\documentclass[a4paper, amsfonts, amssymb, amsmath, reprint, showkeys, nofootinbib, twoside, aps, pre]{revtex4-2}
\usepackage[english]{babel}
\usepackage[utf8]{inputenc}
\usepackage[colorinlistoftodos, color=green!40, prependcaption]{todonotes}
\usepackage{setspace} 
\usepackage[pdftitle={Article}, pdfauthor={Author}]{hyperref} 
\usepackage{xr} 
\usepackage{float}
\usepackage{stackengine}
\usepackage{gensymb} 

\newcommand{\recenter}{\emph{Recenter-COM}} 
\newcommand{\fixingBonding}{\emph{Glued-Monomers}} 
\newcommand{\feneLJ}{\emph{Mutual-Attraction}}
\newcommand{\replicationLike}{\emph{Ladder-like}} 

\newcommand{\singleLooped}{Arc-1-1}

\begin{document}
\title{Kinetics of segregation of topologically-modified ring polymers in cylindrical confinement}
\author{ Harsh Doshi$^1$, Shreerang Pande$^1$, Sathish K. Sukumaran$^{2*}$ and Apratim Chatterji$^1$}
    \email[Correspondence email address: ]{apratim@iiserpune.ac.in, sa.k.sukumaran@gmail.com}
    \affiliation{1. Dept. of Physics, Indian Institute of Science Education and Research, Dr. Homi Bhabha Road, Pune, India-411008}
     \affiliation{2. Graduate School of Organic Materials Science, Yamagata University, 4-3-16 Jonan, Yonezawa 992-8510, Japan}

     \author{}



\begin{abstract}
    
    In {\em Escherichia coli} ({\em E. coli}), entropic repulsion between the two daughter DNA ring polymers under cylindrical confinement is believed to be an important factor governing chromosomal segregation. The repulsion can be enhanced by topological modifications, i.e., by the introduction of internal loops at certain locations along the contour of the circular DNA. However, the effect of topological modifications on the rate of segregation of ring polymers remains unclear. Therefore, we systematically varied the number and the contour length of loops introduced at selected locations by crosslinking monomers. The appropriate crosslinking was motivated by observations that extruded loops are located mainly near the origin of replication (ori-proximal) region of the {\em E. coli} chromosome. This resulted in the chains becoming intrinsically anisotropic. Using Langevin dynamics simulations of these topologically modified bead-spring polymers, we calculated the time required for segregation under cylinder confinement. With certain caveats, we found that increasing the number of loops resulted in a decrease in the time of segregation. In line with past work, we propose that this is due to the increase in the entropic repulsion between the polymers upon increasing the number of loops. In addition to the number of loops, the contour length of the loops and the mutual orientation of the (anisotropic) chains in the initial configurations played a role in determining the time of segregation.
\end{abstract}

\onecolumngrid
\noindent{ %
    \vspace{10ex}
    \fbox{\parbox{0.95\textwidth}{\small\raggedright
    © 2026 American Physical Society. This is the accepted manuscript of the following article: Doshi, Pande, Sukumaran, and Chatterji ``Kinetics of segregation of topologically modified ring polymers in cylindrical confinement'', Phys. Rev. E \textbf{114}, 054403. The abstract of the accepted article can be found at: https://doi.org/10.1103/dx3v-91ry
    
    This manuscript version is posted with permission for non-commercial scholarly use.}
    }
}
\twocolumngrid

\maketitle


\section{Introduction} \label{sec:intro}



Two self-avoiding linear polymers under sufficiently strong cylindrical confinement minimize their overlap by
segregating along the length of the cylinder 
\cite{Teraoka2004,elena1,Arnold2007,Jun2006,Jun2007,Jun2010,Du2018,Jung2010}.
The segregation occurs because the number of chain configurations (microstates) accessible in the segregated state is larger when compared to that in the overlapped state. 
Note that entropic segregation is typically observed when the diameter of the confining cylinder, $D \lesssim 2 R_g$, where $R_g$ is the radius of gyration of the unconfined polymer~\cite{Jun2010}. 

When compared to linear polymers, the entropic force of segregation under cylindrical confinement is higher for a pair of ring polymers~\cite{elena2,ha2_2012,Shin2014}. 
This is because a ring polymer with $N$ monomers under cylindrical confinement can be considered equivalent to two linear polymers 
of $N/2$ monomers each that overlap along the axis of the cylinder~\cite{ha_review,ha2_2012,Polson2014,Polson2018}. Thus two overlapping ring polymers with $N$ monomers each can be considered equivalent to four overlapping linear polymers of $N/2$ monomers, resulting in a greater entropic force for segregation. The origin of the larger entropic force of segregation for ring polymers can be traced to the difference in the free energy between the segregated and the overlapped state being larger for a pair of ring polymers when compared to linear polymers of identical contour lengths. Since the free energy of athermal polymers is dominated by entropy, this implies that the entropy increase due to the segregation of two polymers of $N$ monomers each is significantly larger for rings when compared to linear chains.

Unlike chromosomes in eukaryotic cells, bacterial chromosomes are ring polymers. It has been proposed that the entropic repulsion between ring polymers is utilized by {\em E. coli} to spatially segregate its two daughter chromosomes into separate halves of the cell \cite{Jun2010,Jung2010,Jun2012,dieterloop,frontiers_dekker}). Note that in the {\em E.coli} cell, replication and segregation proceed simultaneously \cite{Wang2006,Cass2016,Kuzminov2013,Badrinarayanan2015,Woldringh_r_2024}. 
Segregation of the daughter chromosomes is known to crucially depend on the presence of loop extruding Structural Maintenance of Chromosomes (SMC) proteins, for e.g., MukBEF \cite{Kleckner2014,smc_org,Boccard2013,Boccard2021,Sazer2017,Sherratt2020,Mkel2021}. However, the exact function of these proteins in ensuring timely segregation during cell division remains to be elucidated. A plausible scenario is that the SMC proteins enable the creation of loops at specific locations along the contour of the circular DNA, thereby enhancing the entropic repulsion between the daughter chromosomes. One of the primary motivations for this work was to investigate whether such additional loops can expedite the segregation of ring polymers as models of {\em E.coli} chromosomes~\cite{chase,Harju2024,Brahmachari24956}. 

In our previous work, we found that certain topological modifications to a ring polymer assisted in the localization of particular portions of the chain to specific locations along the axis of the cylinder \cite{dna2}. This finding, when appropriately adapted to the biological context, suggested a plausible mechanism for the spatio-temporal localization of certain chromosomal loci of {\em E. coli} under both slow and fast growth conditions \cite{dna1,dna3}. The topological modifications were introduced by crosslinking specific monomers of the ring polymer to create additional loops. In other words, the {\em E. coli} chromosome can be considered as a topologically modified (ToMo) ring polymer. 
Such topological modifications, effected by linker proteins or by loop extruding proteins
e.g. MukBEF, have been observed in bacterial chromosomes~\cite{Lioy2018,Boccard2013,Boccard2021,Gaal} as well as eukaryotic 
chromosomes~\cite{DAsaro2024,DiStefano2021,Khanna2019}, and their effects have been investigated theoretically using ideas from polymer physics~\cite{Brackley2013,Brackley2016,Mithun,Mithun2,marko2009linking,Amitai2016,
Bhandarkar2026,Brahmachari2026,Kadam2023,Roychoudhury2025}.

Topological modifications of a ring polymer, such as the introduction of additional loops, can be performed in an inordinate multitude of ways. With the biological context in mind, we have used experimental results to guide the choice of the particular topological modifications to be investigated.  In the {\em E. coli} chromosome for instance, it is known that the additional loops are primarily located near the origin of replication (ori-proximal) region of the chromosome. Therefore, in our past work, we had explored the effect of relatively simple topological modifications, such as introducing additional loops on only one half of the ring polymer, on the spatio-temporal localization of particular chromosomal loci. However, the effect of the particular topological modifications on the kinetics of chain segregation remained unexplored. Therefore, the effect of variables such as the number and the size (contour length) of loops along the chain contour on the rate of segregation forms the focus of this investigation.

In this work, we have systematically investigated the physical mechanisms by which topological modifications, i.e., the introduction of internal loops, to a ring polymer modify the kinetics of segregation from an overlapping state. Extending these ideas to models for bacterial chromosome segregation is left for the future. 
Models for chromosome segregation tend to be complex as they include several biologically relevant aspects that can mutually influence each other. In such models, it is often hard to disentangle the contributions of the different aspects. Hence, we believe that it is useful to supplement investigations of these complex models with investigations that consider the effect(s) of only {\em one} of the several aspects. Our hope is that insights gained from such an investigation can contribute to the fundamental understanding of daughter chromosome segregation in bacteria. Indeed, the results presented here have guided the design and development of subsequent investigations incorporating chromosome replication, loop extrusion, cell elongation, additional topological modifications, and the release of topological constraints. The results of these investigations will be reported in the future. Note that this approach is similar to that adopted previously, where the findings reported in Ref.~\cite{dna2} were subsequently extended in Refs.~\cite{dna1,dna3}. 
 
The primary focus of this work was to uncover emergent phenomena that result from the tuning of effective entropic interactions between polymer segments under equilibrium or near-equilibrium conditions. 
Together with our previous studies \cite{Roychoudhury2025,Bhandarkar2026}, we believe that the work contributes to the elucidation of the fundamental physics of topologically modified polymers. We anticipate that the physical principles uncovered will offer additional insight into how living systems exploit topological modifications, mediated by enhancer-promoter interactions and other DNA-linking proteins, to regulate the organization and dynamics of chromosomes.

 
 Previous studies have shown that incorporation of active mechanisms such as loop extrusion accelerates DNA-polymer dynamics \cite{chan_activity-driven_2024, kadam_dynamical_2025, wutz_pds5_2026, nuebler_chromatin_2018, gabriele_dynamics_2022, ashwin_organization_2019, bailey_loops_2023}. Faster segregation has also been observed, albeit in a computational model of {\em E.coli} chromosomes~\cite{Harju2024}. In contrast, here we establish that the very presence of internal loops can speed up segregation by increasing the entropic repulsion between the two overlapping polymers. We envisage the biological scenario as follows. Extrusion near the oriC proximal regions of the chromosome in {\em E.coli} creates loops along the chromosomal contour and leads to faster segregation even as replication is in progress.

In the absence of biological mechanisms that specifically enable chain crossings, excluded volume (EV) effects in polymers typically prevent them. When chain crossings were disallowed, the protocol for creating overlapping initial configurations used in our earlier work \cite{dna2} often resulted in concatenation(s) between loops belonging to different polymers which prevented chain segregation. Given this, one of the major challenges was to develop protocols for obtaining initial ToMo chain configurations that were unconcatenated but overlapping. Therefore, we have developed several protocols for the preparation of the initial overlapped configurations confined in a cylinder of diameter $D$.  
Starting from such carefully prepared initial configurations, we determined the time required for the segregation of two ToMo polymers under cylindrical confinement for two different conditions: (i) for fixed monomer volume fraction and (ii) for fixed degree of confinement, $2R_g/D$, where $R_g$ is the radius of gyration of the unconfined ToMo polymer. Case (i) is expected to be relevant for bacterial chromosomes. As $R_g$ varies upon modification of the topology, case (ii) enabled comparison under similar degree of confinement for the different topologies investigated.


The total number of monomers in each polymer was fixed at either $N=200$ or $N=500$. We started with a ring polymer of $N$ monomers and modified its topology to create an asymmetry in the simplest manner we could conceive. The modified topology had one large loop with $N/2$ monomers, while the rest of $N/2$ monomers were equally divided between either two, five or ten smaller loops. We have also considered modified topologies with one large and one small loop but the total number of monomers still remained $N$.
Our investigations revealed that, with certain caveats, increasing the number of loops in the cluster decreased the time of segregation. A similar situation was observed upon varying $N$, provided the ratio of the contour lengths of the large and small loops were kept identical.

The manuscript is organized as follows. Following a brief introduction to the model used and the simulation technique, we detail the four initialization protocols that were used to obtain the initial overlapped configurations of a pair of ToMo polymers. Using these initial configurations, we investigated the kinetics of segregation of the two chains with a particular emphasis on the time of segregation. Note that identifying a ``time of segregation'' is not straightforward and we discuss the motivation behind the particular criterion used in this work. Finally, we also discuss the observed kinetics of segregation in light of the particular topological modifications and the different initialization protocols used. 

\section{Model and Methods} \label{sec:model_methods}
\subsection{Polymer Model} \label{sec:model}
We performed Langevin dynamics (MD) \cite{allen_computer_2017} simulations of two unconcatenated polymers confined in a cylinder. 
The polymers were modeled as Kremer-Grest~\cite{kremer_dynamics_1990} chains. In these chains, the excluded volume interactions between any two monomers of mass $m$ were modeled using the Weeks-Chandler-Andersen (WCA) potential~\cite{weeks_role_1971} given by
\begin{equation}
     U_{WCA}(r) = \begin{cases}
        4 \epsilon \: \left[\:\left(\frac{\sigma}{r}\right)^{12} - \: \left(\frac{\sigma}{r}\right)^6 \: \right] + \epsilon \:\:\:\:\: & \forall \: r \leqslant 2^{1/6}\sigma \\
        0 & \forall r > 2^{1/6}\sigma
        \end{cases}
\label{eq:wca}
\end{equation}
where $r$ is the distance between the interacting monomers, $\sigma$ and $\epsilon$ are respectively the length and energy scale in our simulations.
The interaction between the wall of the confining cylinder and the monomers was modeled using the WCA potential such that the bead experienced increased repulsion as the distance of its center from the wall, $r_w \rightarrow \sigma/2$. 

Polymers were created by connecting monomers using an additional finitely extensible nonlinear elastic (FENE) potential~\cite{kremer_dynamics_1990} given by
\begin{equation}
    U_{FENE}(r) = -\frac{1}{2} \: k {R_0}^2\: \text{ln}\left[1 - \left(\frac{r}{R_0}\right)^2 \right]
\label{eq:fene}
\end{equation}
where $r$ is the distance between the two monomers, $k$ determines the stiffness of the potential, and $R_0$ is the maximum possible value of $r$. For all of the simulations, we set $k = 30 \: \epsilon/\sigma^2$ and $R_0 = 1.5 \: \sigma$. The choice of parameters ensured sufficiently close contact between connected monomers and prevented chains crossings.
Note that the combined effect of the WCA and the FENE potentials resulted in a minimum of the bonded potential at $\approx 0.97\sigma$.

The monomers were coupled to a Langevin heat bath at a temperature of $T = \epsilon/k_B$, where $k_B$ is the Boltzmann constant. The equation of motion for the 
$i$-th bead was given by $ m \ddot{\mathbf{r_i}} = - ${\boldmath$\nabla$}$U - \gamma \dot{\mathbf{r_i}} + ${\boldmath$\eta$}, where $\mathbf{r_i}$ is the position vector of the $i$-th bead, $U$ determines the conservative force acting on the bead, $\gamma$ is the friction coefficient, and {\boldmath$\eta$} is the random force satisfying the fluctuation-dissipation theorem.  The equations of motion were integrated using a velocity Verlet scheme with $\gamma= m/\tau_0$, where $\tau_0 = \sigma \sqrt{m/\epsilon}$ and a time step of $\Delta t = 0.005 \tau_0$ during the initialization and $\Delta t = 0.01 \tau_0$ subsequently. All of the simulations were performed using the LAMMPS ($3$ Mar $2020$ version) software package \cite{LAMMPS}.

\subsection{Topological Modifications and their Nomenclature} \label{sec:topology}
\begin{figure*}
        
    \includegraphics[width=0.35\linewidth]{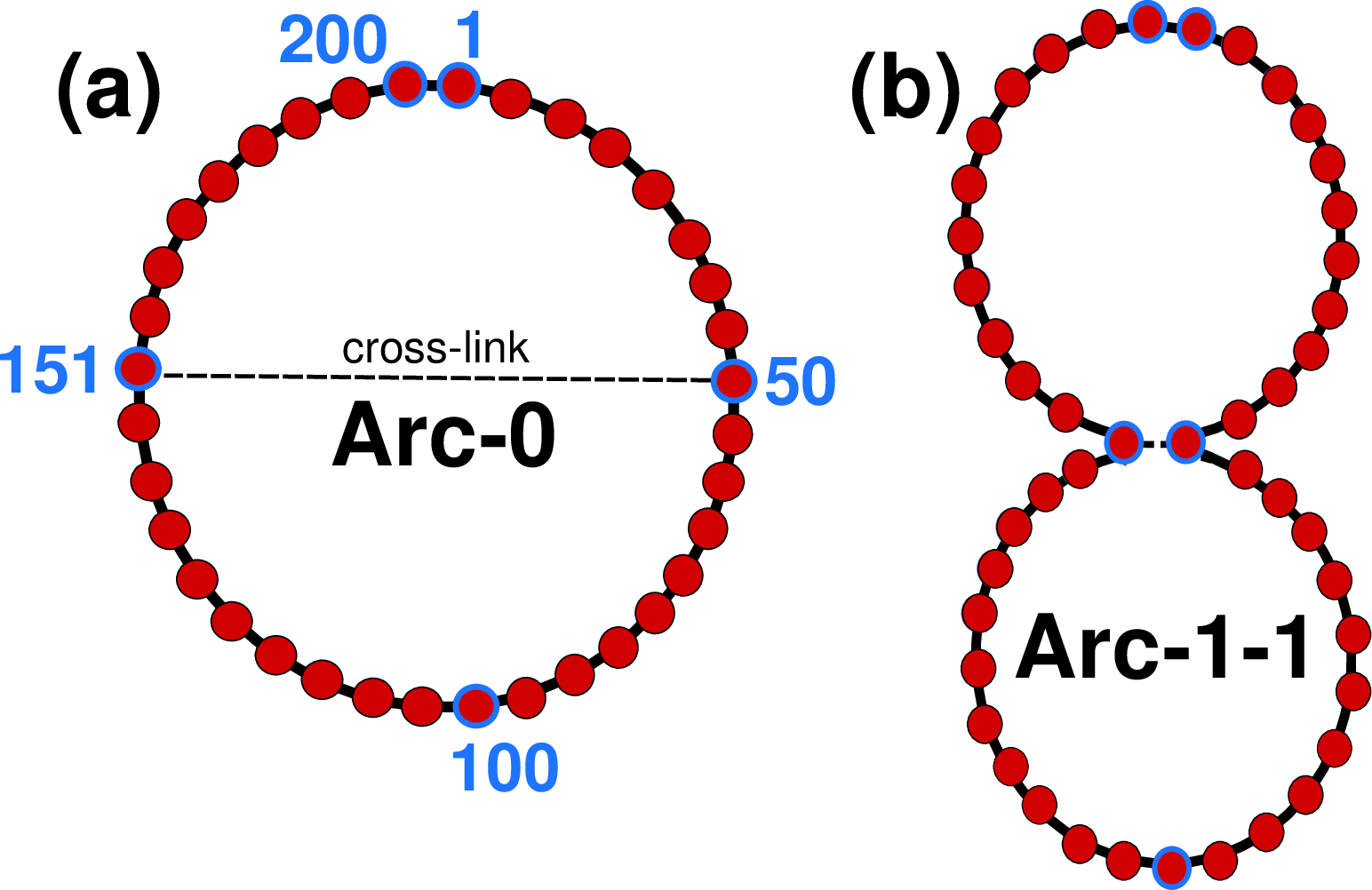}
    \includegraphics[width=0.35\linewidth]{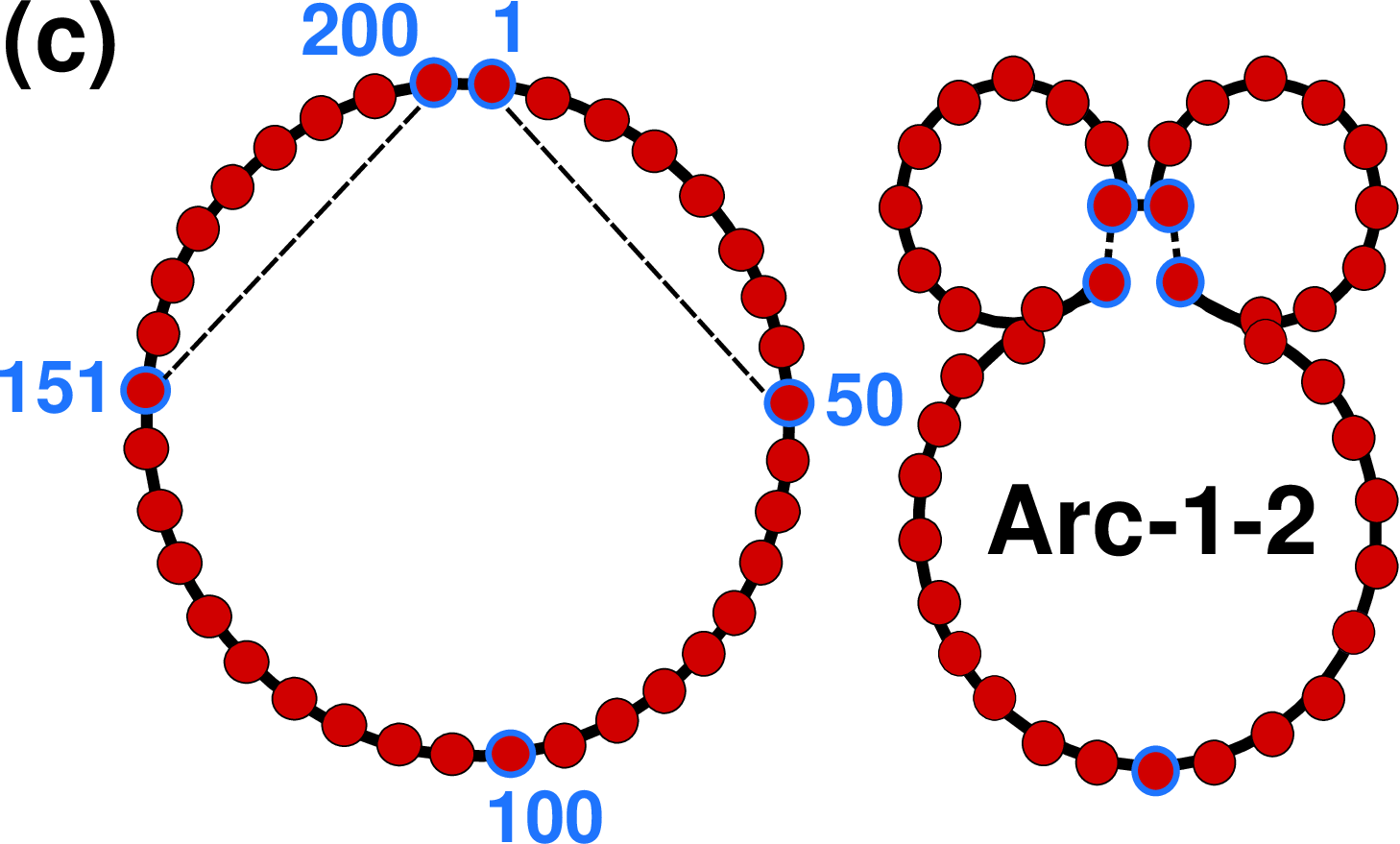}
    \vspace{0.4cm}
    
    \includegraphics[width=0.17\linewidth]{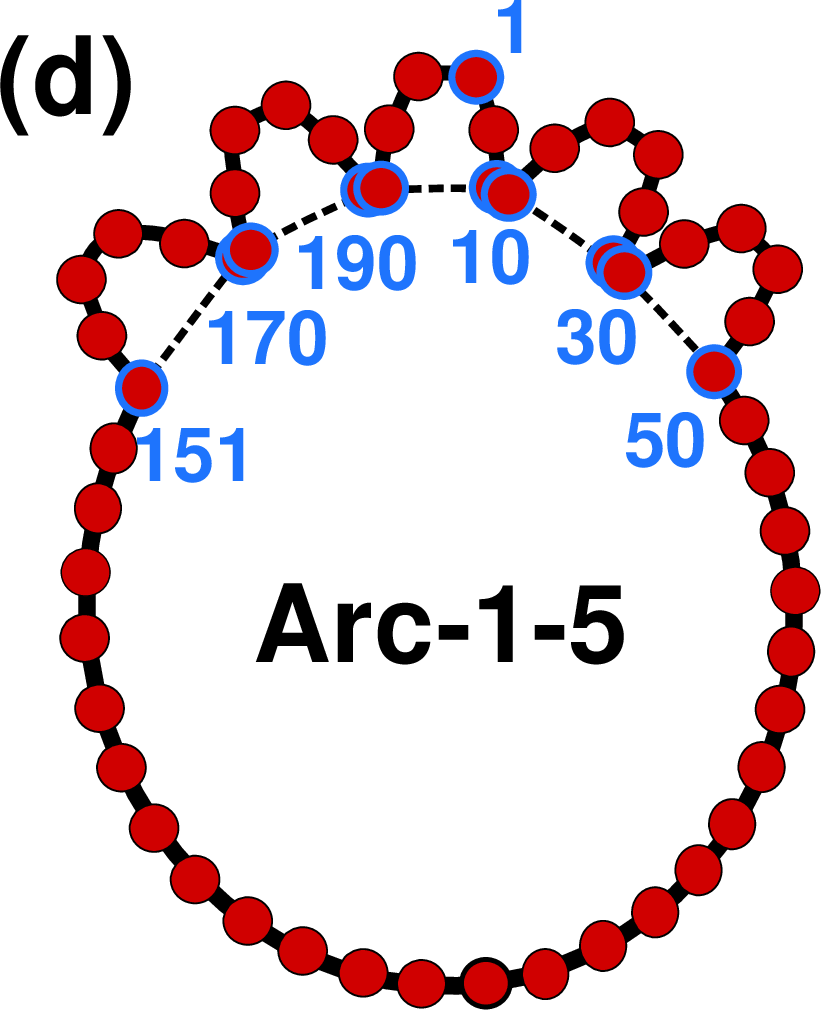}
    \hspace{0.05\linewidth}
    \includegraphics[width=0.14\linewidth]{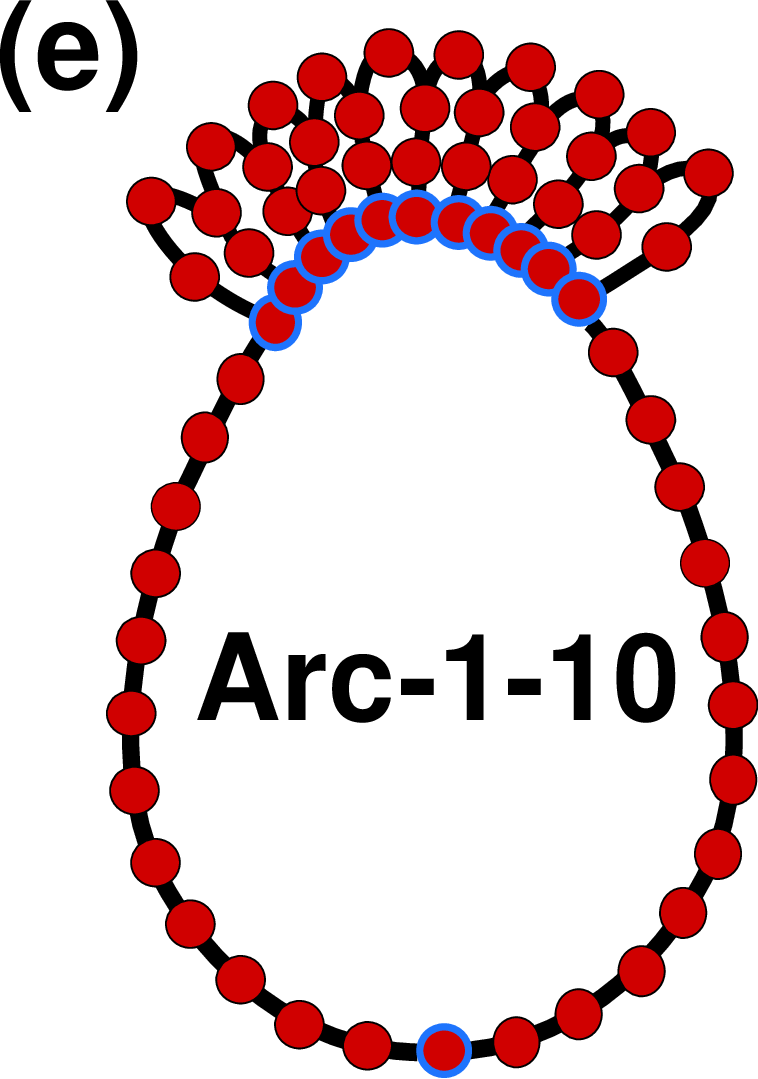}
    \hspace{0.05\linewidth}
    \includegraphics[width=0.40\linewidth]{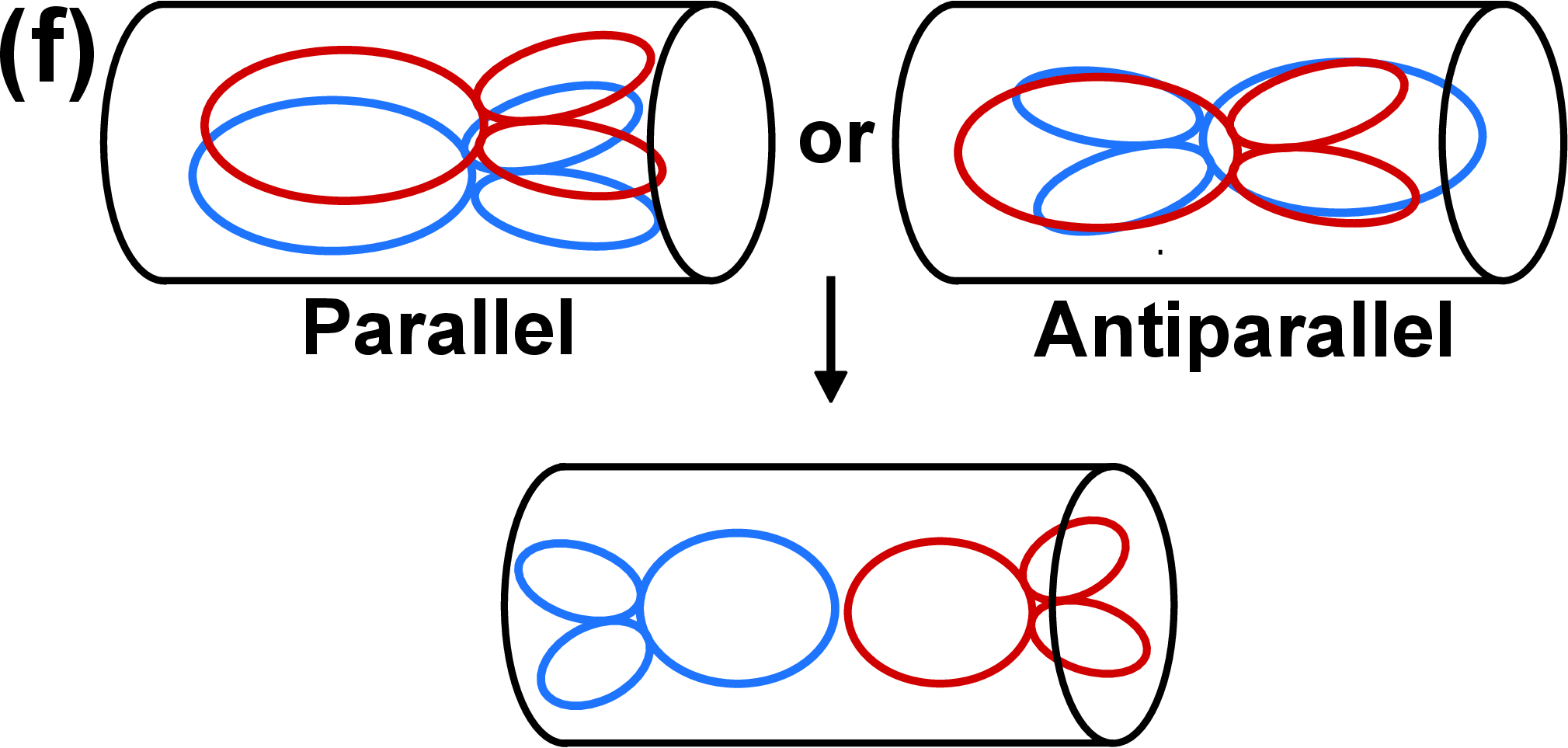}

    \caption{ {\bf Topologically modified polymers}, $N = 200$ (a) Ring polymer. (b) A single crosslink between monomers $50$ and $151$ produced the topology Arc-1-1[100-100]. (c–e) Topologies with multiple loops were generated by introducing $2$, $5$, and $10$ crosslinks resulting in Arc-1-2[100-50], Arc-1-5[100-20], and Arc-1-10[100-10] having one large loop of $100$ monomers and $2$, $5$, or $10$ small loops respectively. The numbers next to the chain contour (in blue) indicate the indices of the crosslinked monomers (indicated by enclosing the monomers within a thin blue circle). In (e), the monomer indices of Arc-1-10 have been omitted. (f) Schematic of the segregation process of two Arc-1-2 polymers. The top panels show two possible overlapped configurations -- parallel and antiparallel -- while the bottom panel depicts the final segregated state.}
    \label{fig:topologies}    
\end{figure*}

We created several ToMo polymers by crosslinking specific monomers along the contour of the ring polymer.
The crosslinking between monomers was implemented using a FENE spring, identical to the one used for the bonds between monomers. Refer figure~\ref{fig:topologies} (a)-(e) for a schematic of the different topologies investigated in this work. A more detailed explanation is provided below.

\emph{(i) Arc-1-1 with one crosslink}: A single crosslink is introduced in a $N=200$ ring polymer
to create two loops of $100$ monomers each.  The number of monomers in each loop is specified in square 
brackets, refer figure~\ref{fig:topologies}(b). We have also investigated Arc-1-1[140-60], where a crosslink was added between monomers $1$ and $60$, such that the polymer had one large loop with $140$ monomers and a small loop with $60$ monomers. Thus topology Arc-1-1 can have two loops of equal size or one large and one small loop within the ring polymer. By modifying the position of the crosslink, we can systematically vary the relative sizes of the two loops.

\emph{(ii) Additional loops}: We introduced additional  crosslinks in the ring polymer to create multiple loops and thereby design different topologies. We reiterate that all of the small loops were located in one half of the ring contour. Refer~\ref{fig:topologies} (c), (d), and (e), for a schematic of ToMo polymers Arc-1-2, Arc-1-5 and Arc-1-10. The crosslinking was done such that the size of the larger loop remained constant across all of the topologies irrespective of the number of the small loops. 
For a polymer with $200$ monomers, the large loop had $100$  monomers, Arc-1-2 had two small loops each with $50$ monomers. Similarly, Arc-1-5 had five small loops each with $20$ monomers and Arc-1-10 had ten small 
loops each with $10$ monomers. In line with the above, Arc-1-10[250-25] had one large loop with $250$ monomers and ten small loops with $25$ monomers. Note that with this nomenclature, the total number of monomers, $N$, in a given polymer can be deduced, i.e, $N=500$. We have determined the time of segregation for a pair of overlapping ToMo polymers for (i) $N=200$ and  (ii)  $N=500$ monomers.


\subsection{Dimensions of the Confining Cylinder} \label{sec:cyl_dim}
We simulated two ToMo polymers of identical topology confined in a cylinder whose dimensions were chosen in two distinct ways. One choice was motivated by the geometry of bacterial cells such as \emph{E.~coli}. The \emph{E.~coli} cell typically exhibits an aspect ratio of $\approx 5$ in cells poised for division \cite{Cass2016, Woldringh_r_2024}. Accordingly, we fixed the aspect ratio of the cylinder of length $L$ by setting its diameter $D = L/5$. The monomer volume fraction was fixed at $0.2$, consistent with estimates of the volume fraction of chromosomes in the range $0.1$–$0.2$ \cite{Carignano2024}. This choice yielded $D = 6.24\,\sigma$ for $N = 200$ and $D = 8.70\,\sigma$ for $N = 500$ monomers. 
The $R_g$ values for the different topologies calculated from simulations of unconfined single chains with $N=200$ and $N=500$ are provided in Section \ref*{SI-sec:cyl-dim-and-rg} of Supplementary Information \cite{supplementary} (SI-\ref*{SI-sec:cyl-dim-and-rg}).

In the second case, we considered an effectively ``infinite'' cylinder, i.e., $L \gg D$. The diameter of the cylinder, $D$, was chosen such that the degree of confinement, $\delta \equiv 2 R_g^{\mathbb{T}}/D = 1.34$, remained constant for all polymer topologies. Here, $R_g^\mathbb{T}$ denotes the radius of gyration of an unconfined single polymer of topology $\mathbb{T}$. This ensured that each topology experienced similar degree of confinement relative to its unconfined size. The reasons for the particular choice of $\delta$ are discussed further below.


\subsection{Generating unconcatenated but overlapped initial configurations}
In this study, we considered two polymers to be ``overlapped'' if their individual centers of mass (COMs) were sufficiently close to each other. Given this, one would expect each polymer to be extended along the axis of the cylinder such that the monomer density remained uniform on average. As segregation progressed, the polymer COMs moved away from each other until the two chains occupied separate regions along the length of the cylinder.

Generating initial configurations of overlapped ToMo polymers that were unconcatenated posed several challenges. Under strong confinement, polymers exhibit a tendency for spontaneous segregation (see Fig.~\ref{fig:topologies}(f) for a schematic). Given this, it was essential to prepare statistically independent overlapped configurations that could serve as unbiased starting points for investigating the kinetics of segregation of two cylindrically confined polymers. To this end, we developed four distinct initialization protocols. For asymmetric topologies, we observed that the polymers in the overlapped configurations predominantly adopted two distinct orientations: \emph{parallel} or \emph{antiparallel} (see Fig.~\ref{fig:topologies}(f)). We found that, in general, the dependence of rate of segregation on the topology remained consistent across the three of the four different initialization protocols, and any deviations could be attributed to differences in the mutual orientation of the polymers in the initial configuration (whether parallel or antiparallel) or the specific constraints imposed during initialization.

To generate overlapped configurations under cylindrical confinement, we began with two polymers that were spatially well separated inside a cylinder of diameter $D_0$ and length $L_0$ such that $D_0 \gg D$ and $L_0 \gg L$. The cylinder dimensions were gradually reduced until the desired confinement was achieved while the chains remained unconcatenated. However, since two polymers under cylindrical confinement tend to spontaneously segregate, additional constraints (described below) were imposed to ensure that the polymers remained overlapped along the axis of the cylinder throughout the initialization process.

\begin{figure}
    \centering
    \includegraphics[width=0.65\linewidth]{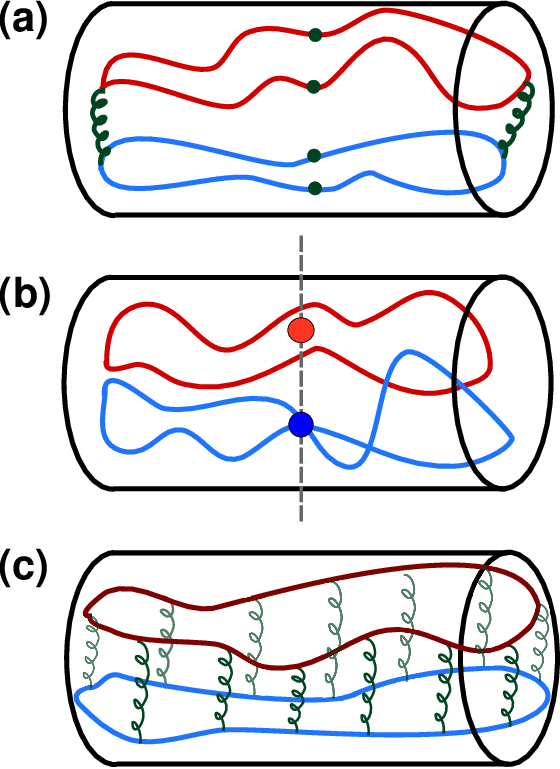}
    \caption{ {\bf Constraint implementation} (a) In the \fixingBonding{} protocol, selected monomers (filled circles) were fixed in space, and harmonic springs were introduced between specific monomers belonging to different polymers. (b) In the \recenter{} protocol, the centers of mass (COMs) of both polymers (filled circles) were constrained to remain in a vertical plane at the center of the cylinder. (c) In the \replicationLike{} protocol, each monomer of one polymer was connected to the corresponding monomer in the other polymer via weak harmonic springs (see text). Note that the ``Mutual Attraction'' constraint implementation is not shown.} 
    \label{fig:constraint}
\end{figure}

\emph{Initialization runs}: We began by gradually shrinking the initially large cylinder from $t = 0$ until the desired confinement was reached at $t = t_0'$. The polymers were further equilibrated until $t = t_0$ while subject to the additional constraints imposed during the four initialization protocols (refer Fig.~\ref{fig:constraint} and see below for additional details). 
The time $t = t_0$ marked the end of the \emph{initialization run} after which the constraints were removed and the polymers allowed to segregate. For each of the four initialization protocols, the entire procedure was repeated to generate $50$ independent unconcatenated overlapped configurations for each topology. Using the overlapped configurations prepared by a particular initialization protocol, the time of segregation for each topology was estimated. As discussed later in the manuscript, we explicitly verified that the polymers remained in an overlapped configuration until $t = t_0$. Below, we describe the additional constraints imposed during the four initialization protocols.

(i) \fixingBonding{}: For two polymers, we selected two monomers positioned diametrically opposite along the $N = 200$ ring contour, for instance, monomers with indices $50$ and $151$, see Fig.~\ref{fig:topologies}(a)). These four monomers were fixed in space on the mid-plane passing through the center of the cylinder and perpendicular to its axis such that the monomers did not overlap. In addition, monomers with indices $1$ and $100$ from one polymer were bonded to their counterparts in the other polymer through a potential given by
\begin{equation}
U_H = \frac{1}{2}\kappa (r - r_0)^2,
\label{eqn:uh}
\end{equation}
with $\kappa = 1000\, \epsilon / \sigma^2$ and $r_0 = \sigma$. For the Arc-1-2 polymer, monomers $1$ and $200$ were crosslinked to monomers $50$ and $ 151$ respectively. Hence, monomers $1$ and $200$ of both polymers were located near the mid-plane. However, the rest of the two polymers were free to stretch out on either side of the mid plane.
For polymers with $N = 500$ monomers, the glued and
bonded monomers were chosen analogously. A schematic of the constraints imposed is shown in Fig.~\ref{fig:constraint}(a). A similar procedure was adopted for the initialization of the other topologies.

(ii) \recenter{}: The centers of mass (COMs) of the two polymers were constrained to lie on the mid-plane passing through the center of the cylinder and perpendicular to its axis (see Fig.~\ref{fig:constraint}(b)). 
The constraint was implemented as follows. After each position update, the COM displacement from the mid-plane was computed for each polymer, and all of the monomers of that polymer were shifted by the negative of this displacement to restore the COM to the mid-plane. This procedure is expected to keep the two polymers overlapped along the axis of the cylinder.

(iii) \replicationLike{}: Each monomer of one polymer was bonded to the corresponding monomer of the other polymer through the weak harmonic potential $U_H$ described above, refer Eq.~\ref{eqn:uh}. In other words, monomer $i$ of one polymer interacted with monomer $i$ of the other polymer for all $1 \leq i \leq N$ (see Fig.~\ref{fig:constraint}(c)). Note however that $\kappa = 10\, \epsilon / \sigma^2$ and $r_0 = 3\,\sigma$ was used. These values allowed spatial fluctuations of the monomers but maintained sufficient overlap between the two polymers.

(iv) \feneLJ{}: As before, monomers belonging to the same polymer experienced the WCA potential. However, interactions between monomers belonging to different polymers were modeled using
\begin{equation}
U_{LJ} = 
\begin{cases}
4\epsilon_{LJ}\left[\left(\dfrac{\sigma_{LJ}}{r}\right)^{12} - \left(\dfrac{\sigma_{LJ}}{r}\right)^{6}\right] + U_0, & r < r_0,\\[6pt]
0, & r \geq r_0,
\end{cases}
\end{equation}
where, $\epsilon_{LJ} = 0.5\,\epsilon$, $\sigma_{LJ} = \sigma$, and the cutoff distance $r_0 = 2.5\,\sigma$. The constant $U_0$ ensured that $U_{LJ}(r_0) = 0$. The value of $\epsilon_{LJ}$ was chosen as above because smaller values were insufficient to maintain overlap for certain topologies, e.g. Arc-1-10.
Note that unlike the WCA potential, which is purely repulsive, this potential is attractive for $2^{1/6} \sigma < r \leq r_0$. 

For further details of the initialization protocol, refer SI-\ref*{SI-sec:tech-initialization}. At the end of each initialization run, we explicitly verified the absence of any concatenation by the methods detailed in 
SI-\ref*{SI-sec:concatenation}.

\begin{figure}
    \centering
    \includegraphics[width=0.9\linewidth]{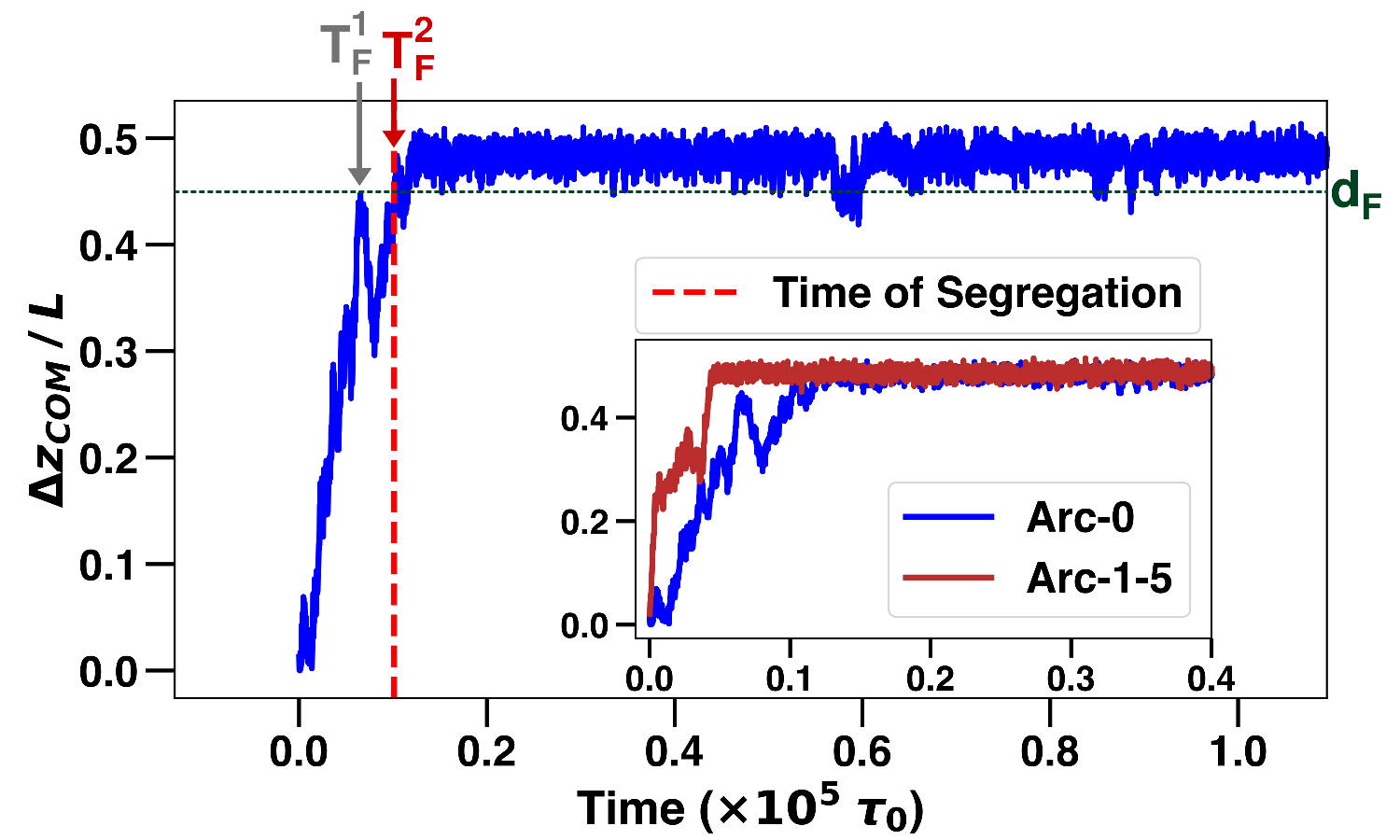}
    \caption{{\bf Determining the time of segregation} Time evolution of the normalized distance between the centers of mass, $\Delta z_{COM}/L$, for one trajectory of two Arc-0 (ring) polymers is shown. Here, $L$ is the length of the cylinder. The increase in $\Delta z_{COM}/L$ indicates progressive segregation of the two Arc-0 polymers from their initial overlapped configuration. The gray arrow marks the time $T_F^1$, corresponding to a temporary decrease in $\Delta z_{COM} / L$ associated with transient remixing. The red arrow and the dotted line denote the time of segregation $T_F^2$, determined according to the criterion discussed in the main text. The inset compares representative $\Delta z_{COM} / L$ trajectories for Arc-0 and Arc-1-5. Data for additional topologies are provided in SI-\ref*{SI-sec:multi-seg-comparison}.
    We have indicated $d_F=0.45L$ for $N=200$ by a dashed line.}
    \label{fig:com-distance}
\end{figure}

\subsection{Segregation Criterion} \label{sec:seg-criterion}
Starting from an overlapped initial configuration at $t = t_0$, we removed the constraints that were imposed during initialization to ensure overlap between the two polymers. In the absence of these constraints, the two polymers began to segregate. The simulation runs were continued for a duration of $10 N^2 \tau_0$. 
A topology-independent criterion was employed to quantitatively determine whether the segregation was complete. This is discussed in greater detail below.

Beginning at time $t_0$, the distance $D_{\mathrm{COM}}$ between the centers of mass (COMs) of the two polymers along the axis of the cylinder was calculated. When $D_{\mathrm{COM}} \approx 0$, the polymers were considered overlapped. As segregation proceeded, $D_{\mathrm{COM}}$ gradually increased and approached $L/2$. Thereafter, $D_{\mathrm{COM}}$ fluctuated around $L/2$, see Fig.~\ref{fig:com-distance}. In this ``steady state'', the polymers occupied different halves, with their COMs located approximately near $L/4$ and $3L/4$ along the axis. Occasional decreases in $D_{\mathrm{COM}}$ (gray arrows in Fig.~\ref{fig:com-distance}) corresponded to transient remixing. 
Ideally, one would like to define the time of segregation as the time beyond which the two polymers remained segregated indefinitely. In microscopic systems, such as the ones under consideration, remixing is possible due to thermal fluctuations given sufficient time. Therefore, in practice, to define the time of segregation, we employed the following two-step criterion:\\
(i) $D_{\mathrm{COM}}$ exceeded the threshold value $d_F$ during the simulation for time $> t_0$, possibly several times, see Fig.~\ref{fig:com-distance}. The time at which the aforementioned condition was satisfied for the $i$-th time was denoted as $T_F^i$ .\\
(ii) Starting from each $T_F^i$, $\langle D_{\mathrm{COM}} \rangle$, the time-average  over a time duration $T_s = 5 N^2 \tau_0$, must satisfy $\langle D_{\mathrm{COM}} \rangle > d_S$. Here $d_S$ is another threshold. In the rare instances where $10 N^2 \tau_0 - T_F^n < 5 N^2 \tau_0$, the time average was calculated for $T_s = 10 N^2 \tau_0 - T_F^n$. The second condition ensured that the polymers remained segregated for a period of time.

If both of the conditions described above were satisfied, the corresponding $T_F^i$ was identified as the time of segregation, $\tau_{seg}$, for that particular simulation run. An illustrative example of the procedure is shown in Fig.~\ref{fig:com-distance}. The first instance when $D_{\mathrm{COM}}$ crossed the threshold $d_F$ was marked as $T_F^1$ (gray arrow in Fig.~\ref{fig:com-distance}). However, $D_{\mathrm{COM}}$ subsequently dropped below $d_S$ and the condition $\langle D_{\mathrm{COM}} \rangle > d_S$ was not satisfied. The next crossing of $d_F$ occurred at $T_F^2$ and the second condition remained satisfied subsequently. Therefore, the time of segregation for this particular simulation run was determined to be $T_F^2$ (red dotted line in Fig.~\ref{fig:com-distance}).

The two threshold values were set at $d_F = 0.45 L$ and $d_S = 0.40 L$ for $N = 200$, and $d_F = 0.48 L$, and $d_S = 0.43 L$ for $N = 500$. The threshold values for $L \gg D$ will be discussed later. The two pairs of thresholds were chosen to be different for different $N$ as the fluctuations of $\Delta z_{COM} / L$ depended on $N$. In several instances, the two conditions were not satisfied till the end of the simulation run. For these runs, the polymers were considered unsegregated and the time of segregation undetermined. Therefore, these runs were not included in the analysis of the statistics of $\tau_{seg}$ for a particular topology.

\begin{figure}[htb]
        \includegraphics[width=0.8\linewidth]{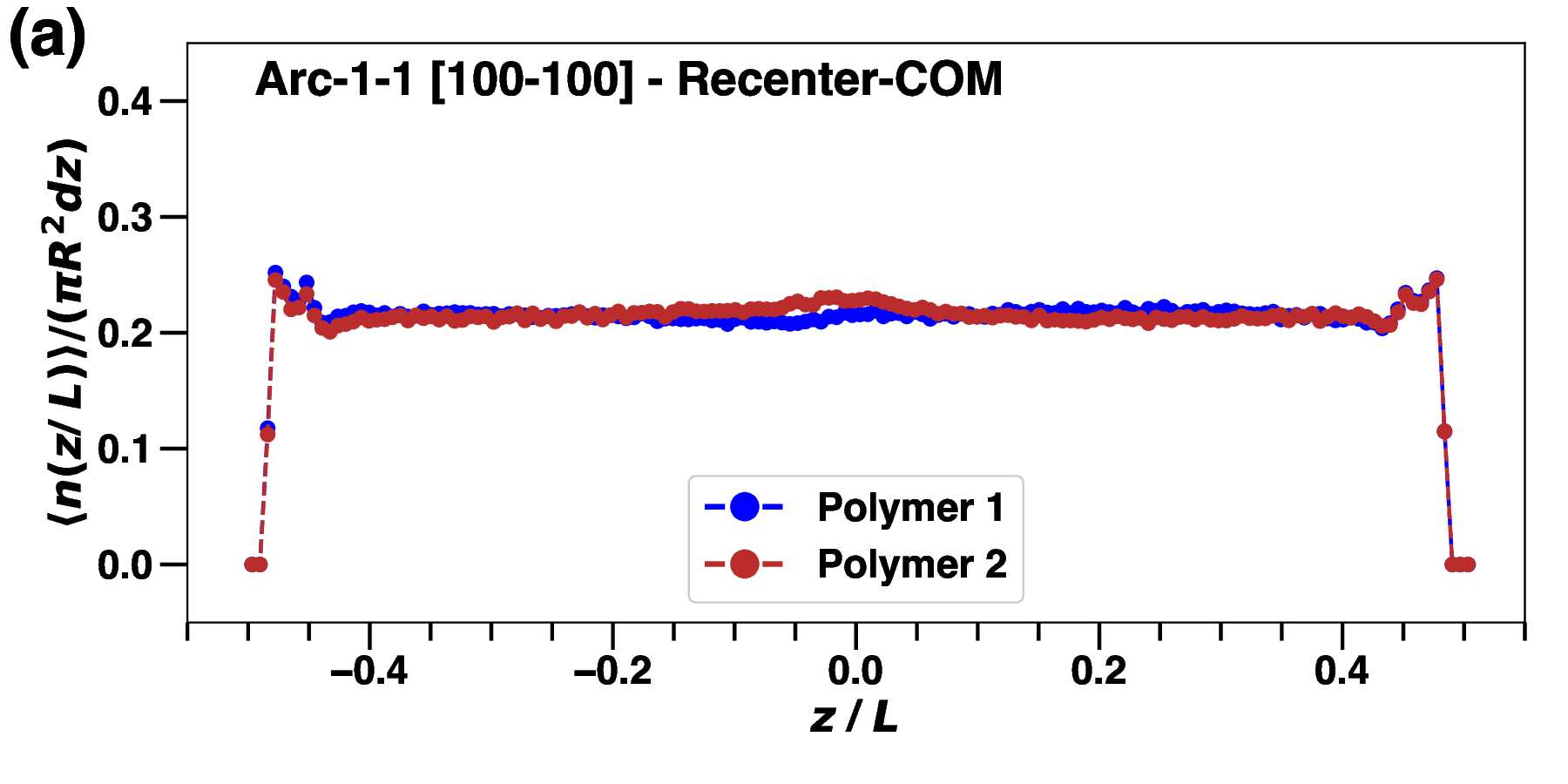}
        \includegraphics[width=0.8\linewidth]{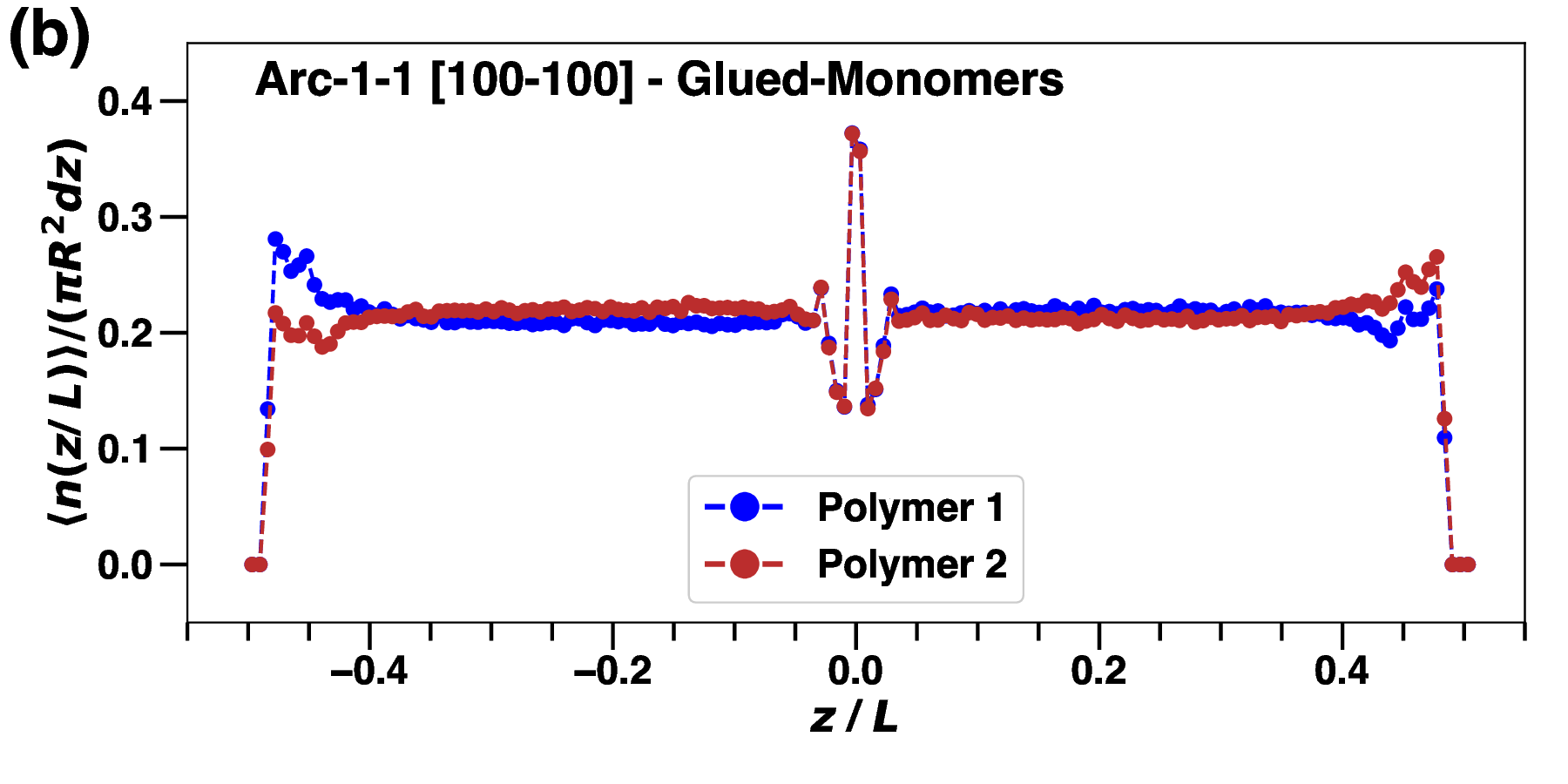}
      \includegraphics[width=0.8\linewidth]{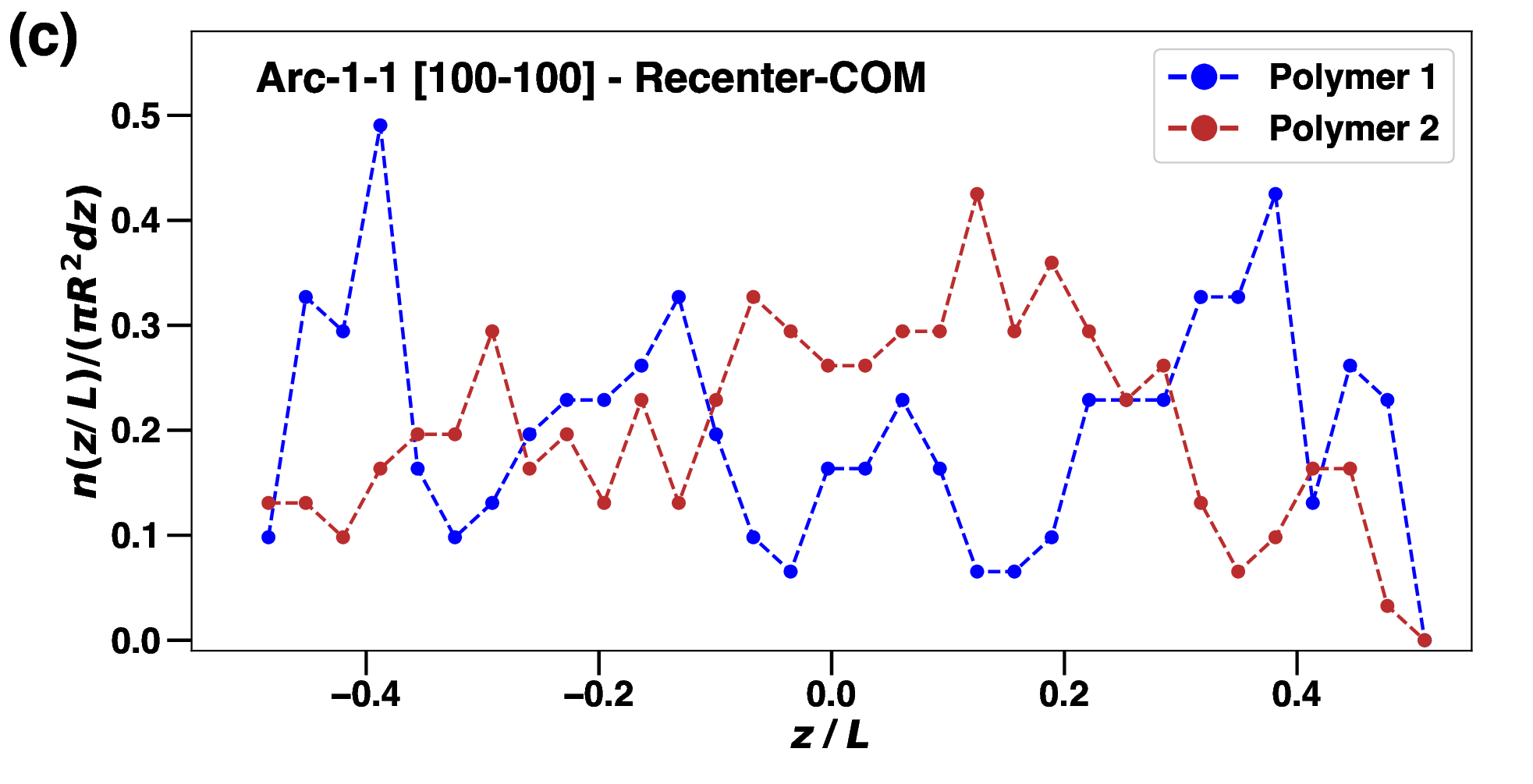}
       \includegraphics[width=0.8\linewidth]{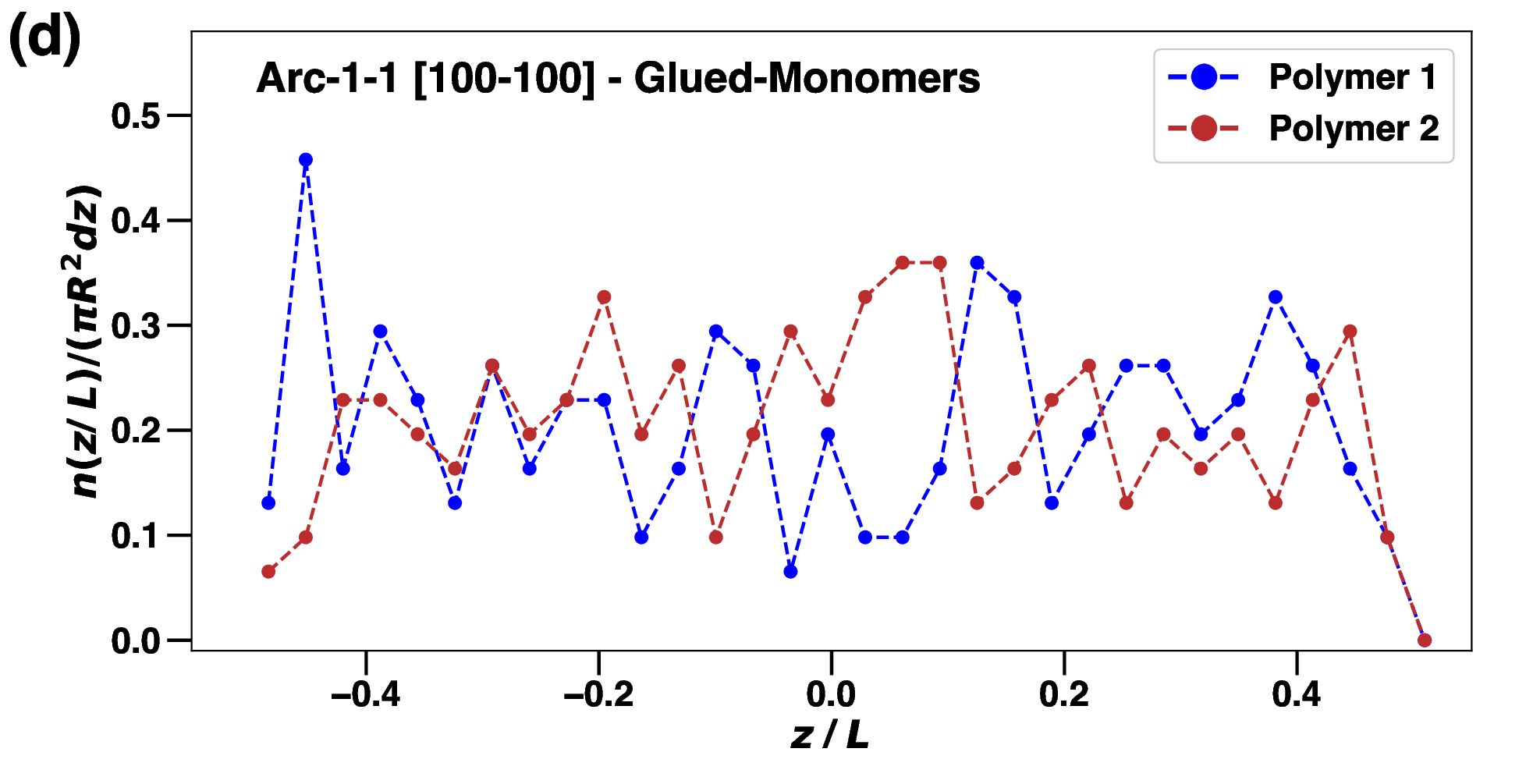}
    \caption{ {\bf Characterizing the initial configurations} Spatial variation of monomer number density of two Arc-1-1[100-100] polymers with $z/L$. Here, $z$ is the displacement from the midpoint of a cylinder of length $L$ along its axis. The monomer number density was calculated by determining the number of monomers, $n(z/L)$, in a disk of diameter $D$ and thickness $dz$ centered at each $z$. Time-averaged monomer number density, $\langle n(z/L) \rangle$, over the initialization run for the two indicated initialization protocols are shown in (a) and (b). For details, see the main text. Instantaneous monomer number density of the configuration at the end of the initialization run for the two  initialization protocols indicated are shown in (c) and (d). Note that for (a) and (b) $dz = 0.2\,\sigma$ while for (c) and (d) $dz = \sigma$. 
    } 
     \label{fig:initialization-long-mon-dens}
\end{figure}


\begin{figure}
        \includegraphics[width=0.9\linewidth]{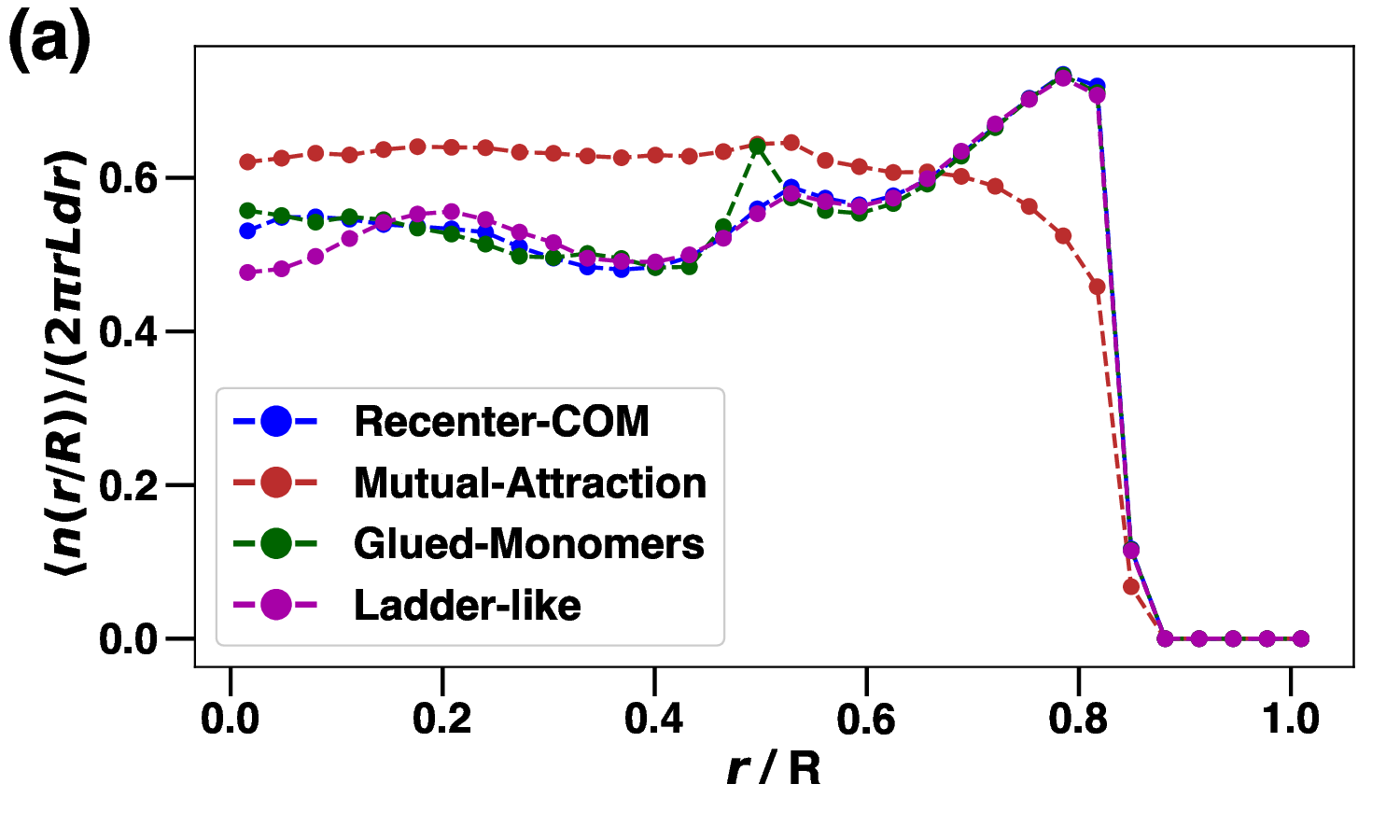}
        \includegraphics[width=0.9\linewidth]{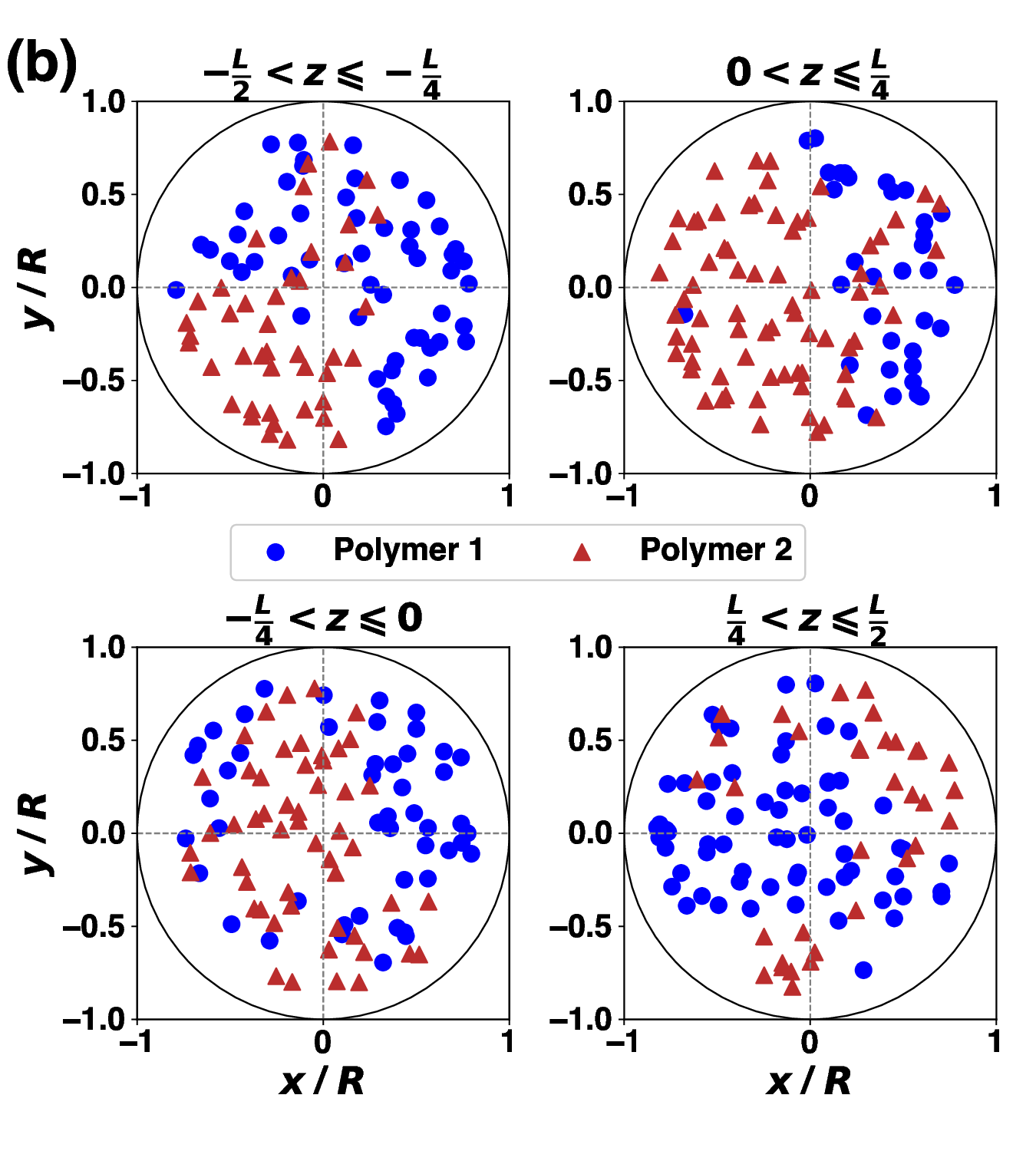}
    \caption{ {\bf Monomer distribution: radial and in the $x-y$ plane} 
    (a) Spatial Variation of time-averaged monomer number density along the normalized radial coordinate, $r/R$, with $R = D/2$ and shell thickness $\text{d}r = 0.1 \: \sigma$. (b) Distribution of monomers projected onto the $x-y$ plane. The cylinder has been divided into four portions along its axis. The monomers of the two polymers are indicated by different symbols and colors. The data shown in (a) and (b) are for two Arc-1-1[100-100] polymers initialized using the \recenter{} protocol.}
    \label{fig:initialization-rad-mon-dens}
\end{figure}


\section{Results and Discussion}\label{sec:results}

\subsection{Characterizing overlapped configurations}\label{sec:mixed-validity}
To verify that the initialization protocol did not introduce artifacts, such as partial segregation 
of the two polymers during the generation of the initial overlapped configurations, we quantified the degree of overlap between the two polymers by calculating several statistical measures. To this end, we calculated the spatial variation of the time-averaged monomer density of two individual Arc-1-1[100-100] polymers with the normalized distance from the midpoint on the axis of the cylinder. This is shown for the initialization protocol \recenter{} in Fig.~\ref{fig:initialization-long-mon-dens}(a) and for \fixingBonding{} in Fig.~\ref{fig:initialization-long-mon-dens}(b). The monomer density was averaged over the configurations obtained during the {\em initialization run} between $t_0'$ and $t_0 = t_0' + 2\times10^5\tau_0$ and sampled every $20\,\tau_0$.

The spatial variation of the time-averaged monomer density with $z/L$ may appear uniform even if the two polymers were actually segregated at each instant but repeatedly switched positions between the two length-wise halves of the cylinder during the initialization run. To check for this possibility, Figs.~\ref{fig:initialization-long-mon-dens}(c) and 
\ref{fig:initialization-long-mon-dens}(d) show the spatial variation of the instantaneous monomer number density with $z/L$ for the final configuration of the {\em initialization run}. The monomers of both polymers spanned nearly the entire length of the cylinder and indicated that they remained extended in the starting configuration.
 Similar data for the other initialization protocols  are provided in
 SI-\ref*{SI-sec:other-mon-dens}. For more complex topologies, such as Arc-1-10, while the spatial variation of the monomer number density with $z/L$ was nonuniform  (see SI-\ref*{SI-sec:other-mon-dens}), monomers of both polymers spanned the entire length of the cylinder.


The time-averaged monomer number densities, refer Fig.~\ref{fig:initialization-long-mon-dens}(a) and (b), remained approximately uniform along the length of the cylinder, with a slight increase near the ends due to the presence of the walls. The pronounced increase at the center of the cylinder, $z/L = 0$, visible in Fig.~\ref{fig:initialization-long-mon-dens}(b) resulted from the four monomers that were permanently fixed at that location.  

To determine the radial distribution of the monomers, the spatial variation of the time-averaged monomer number density with $r/R$ for the four initialization protocols is shown in Figure~\ref{fig:initialization-rad-mon-dens}(a). Here, $r$ is the radial distance from the axis of the cylinder. The density at different $r$ was calculated using concentric shells of length $L$ and thickness $\mathrm{d}r$.   
Except for the \feneLJ{} protocol, peaks can be observed near the walls of the cylinder; see figure~\ref{fig:initialization-rad-mon-dens}(a).  For \feneLJ{}, the attractive part of the potential between the monomers of different chains resulted in a slightly higher density away from the wall. Similarly to Fig.~\ref{fig:initialization-long-mon-dens}(b), the peak observed near $r/R = 0.5$ for the \fixingBonding{} initialization protocol was a result of four of the monomers being held fixed at the mid-plane of the cylinder.

\begin{figure}[ht!]
    \centering
    \includegraphics[width=0.9\linewidth]{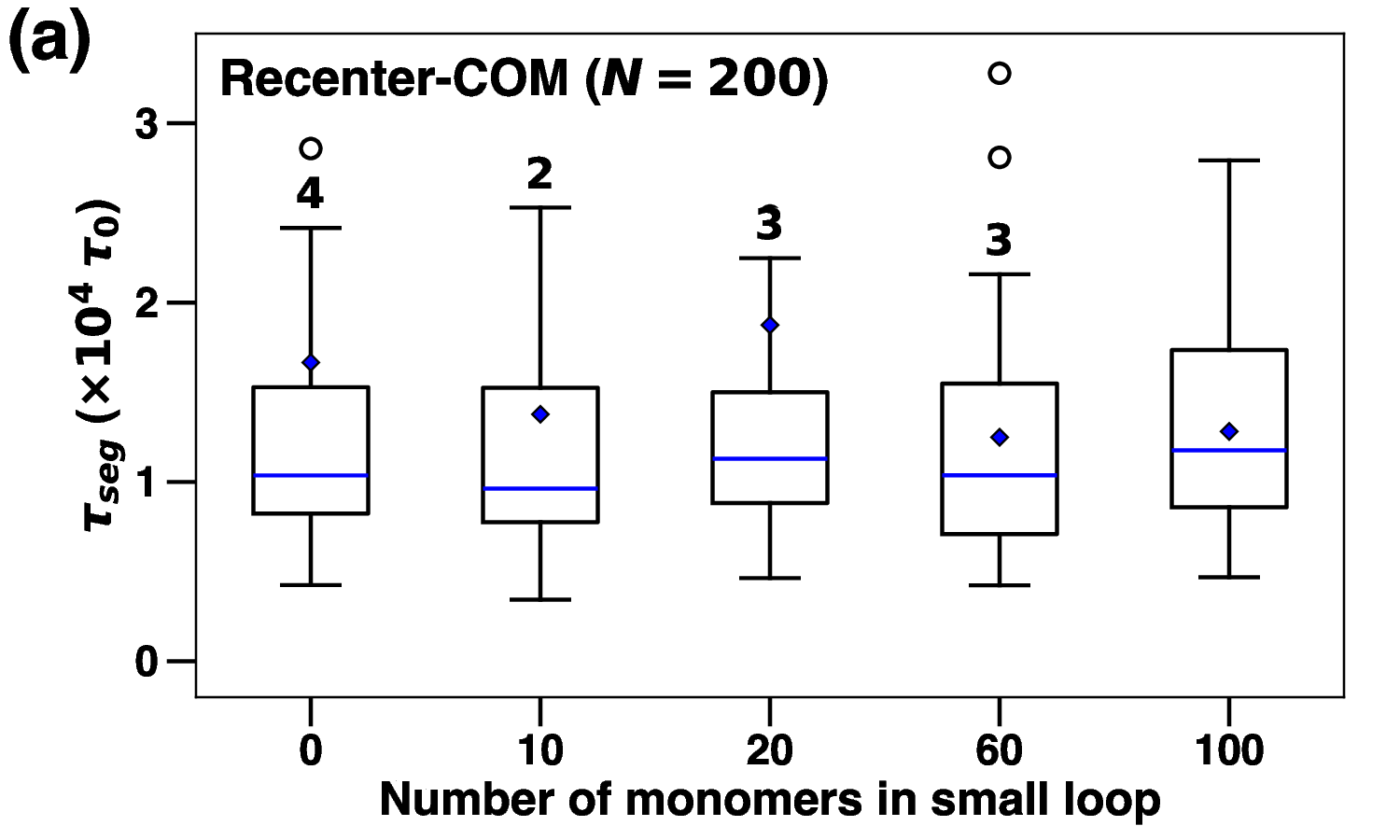}
    \hfill
    \includegraphics[width=0.9\linewidth]{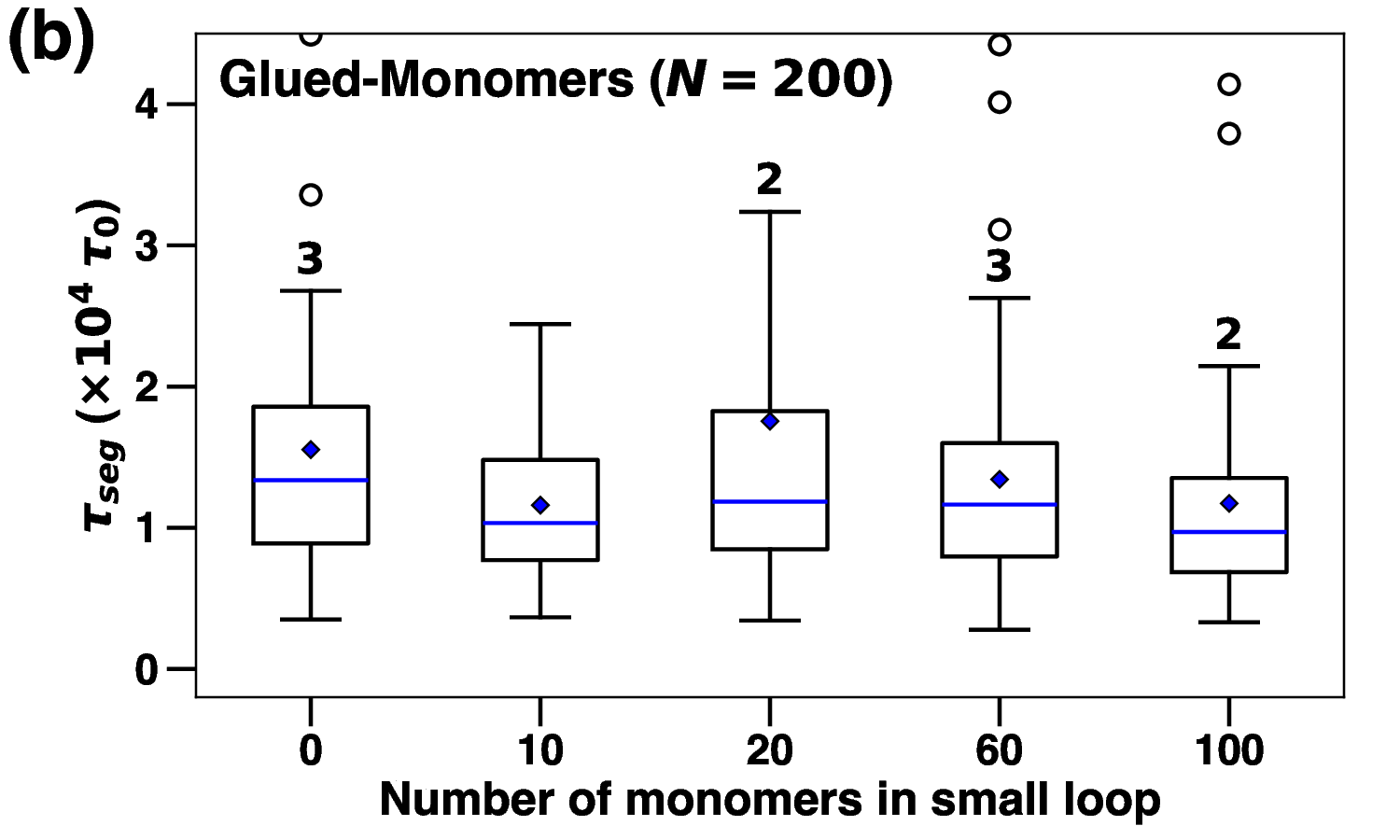}
    \hfill
    \caption{ {\bf Varying loop lengths}. Box plot displaying several summary statistics 
    for the variation of $\tau_{seg}$ with $N_S$, the number of monomers in the small loop,    where the data is for a pair of Arc-1-1 polymers with $N=200$ monomers in each polymer. Mean (blue diamonds), second quartile or the median (Q$_2$, blue line within the box), third quartile (Q$_3$, top edge of the box), first quartile (Q$_1$, bottom edge), and the number of outliers have been provided. The whiskers extend from the edges of the box to data points lying within $1.5 \times \mathrm{IQR}$ where $\mathrm{IQR} \equiv \mathrm{Q}_3 - \mathrm{Q}_1$ is the interquartile range. Data points beyond $1.5\times\mathrm{IQR}$ from the box edges were classified as outliers ($\circ$). The range of the y-axis was selected for ease of comparison across datasets; outliers beyond this range are not shown. 
    Probability distributions of the topologies were compared pairwise using a t-test and the resulting $p$-values are provided in SI-\ref*{SI-sec:si-hist-seg-times}.
    We have performed t-tests and provided $p$-values 
    on comparing a pair of distributions of segregation times at a time, to have a 
    quantitative measure of whether the distributions are significantly different, refer 
     SI-\ref*{SI-sec:si-hist-seg-times}.
}
    \label{fig:phi-single-box-plots}
\end{figure}

\begin{figure*}[ht!]
        \centering
        \includegraphics[width=0.49\linewidth]{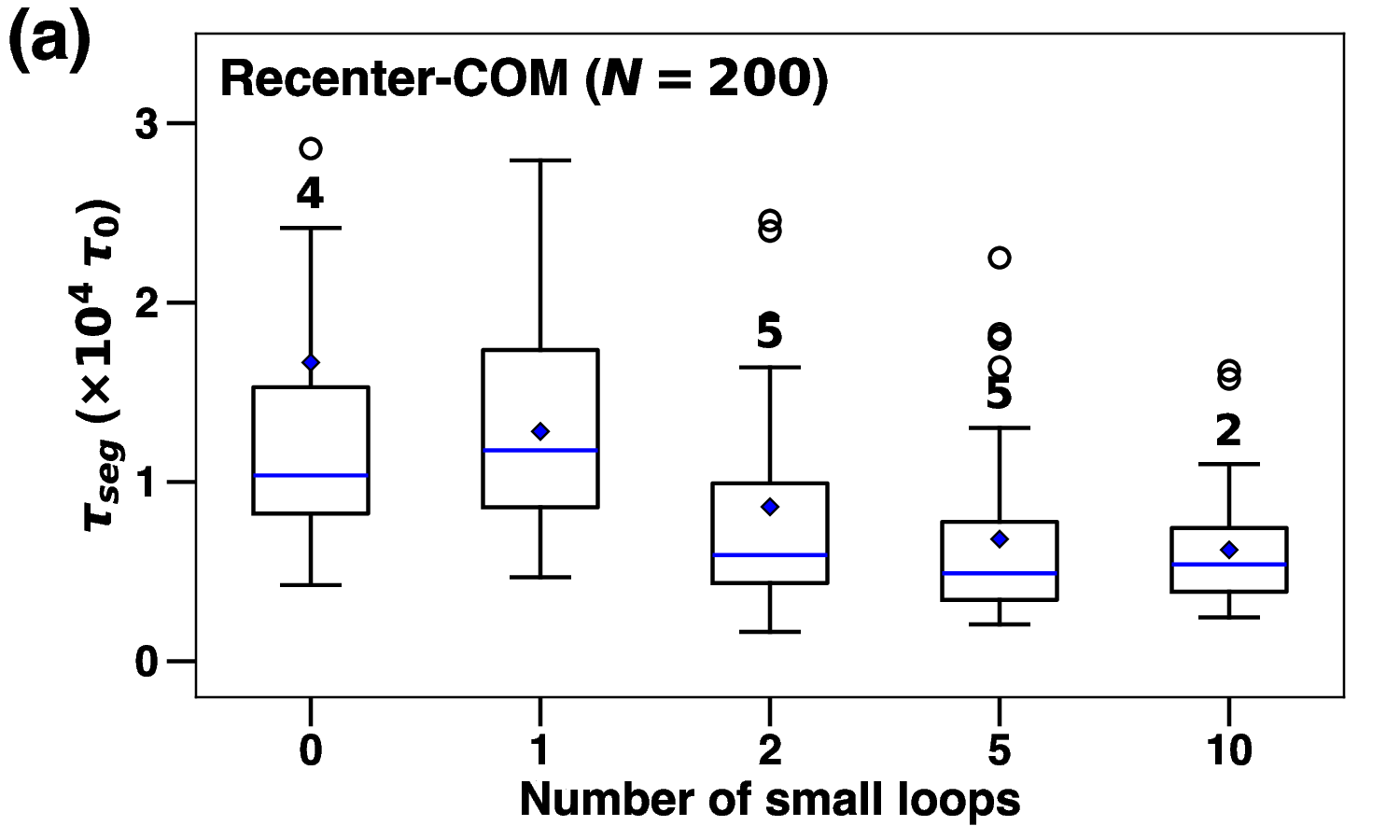}
        \label{fig:rec-multi-box-plot} 
    \hfill
        \includegraphics[width=0.49\linewidth]{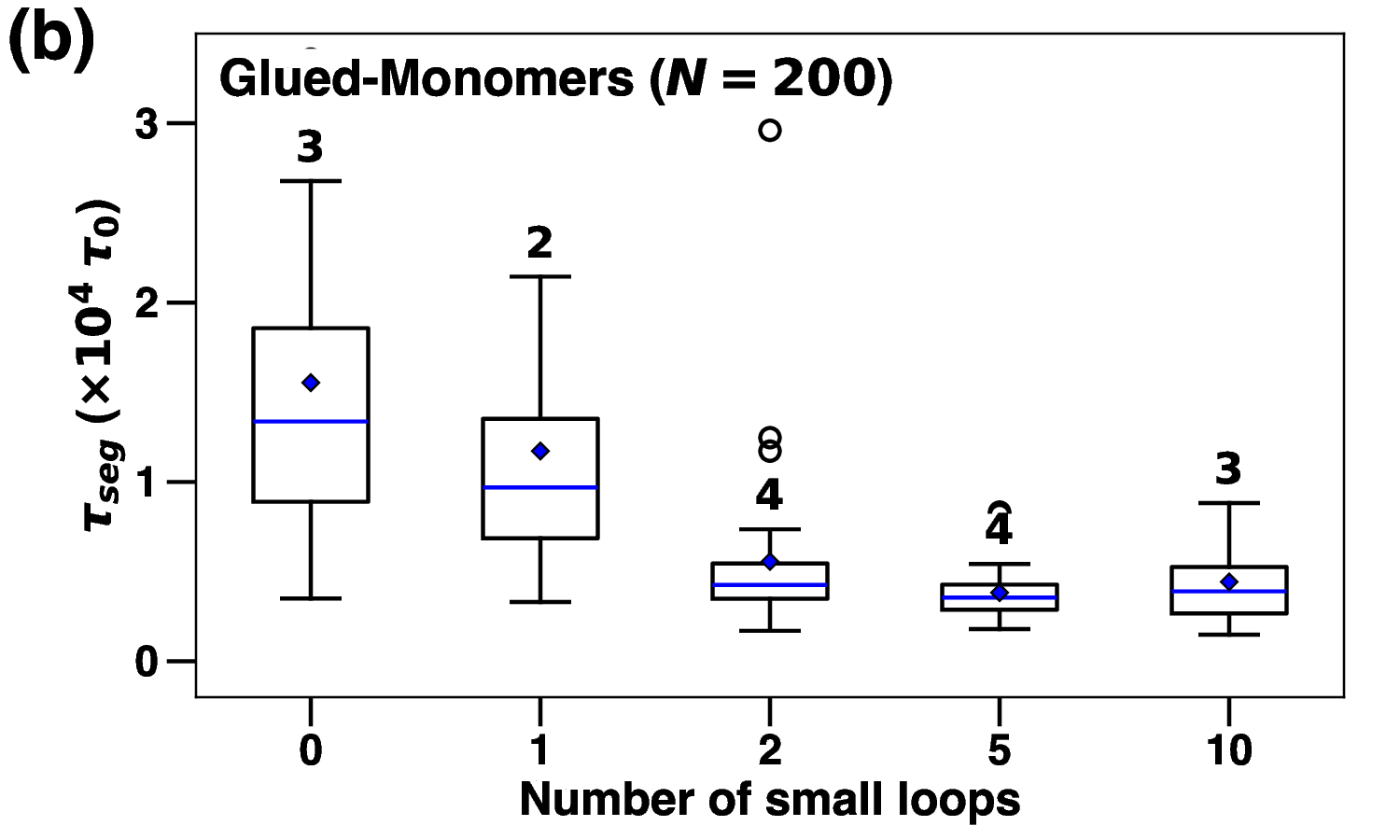}
        \includegraphics[width=0.49\linewidth]{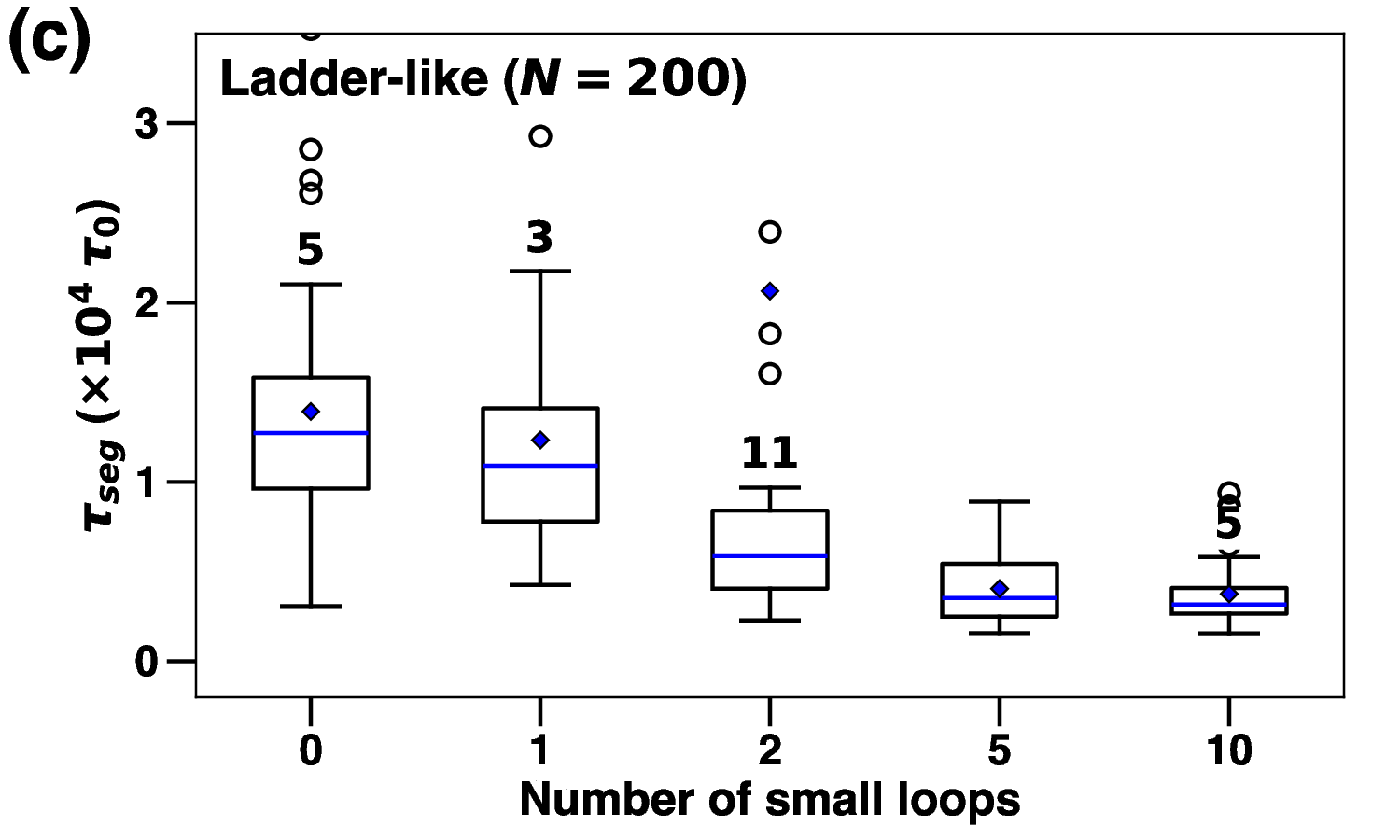}
    \hfill
        \includegraphics[width=0.49\linewidth]{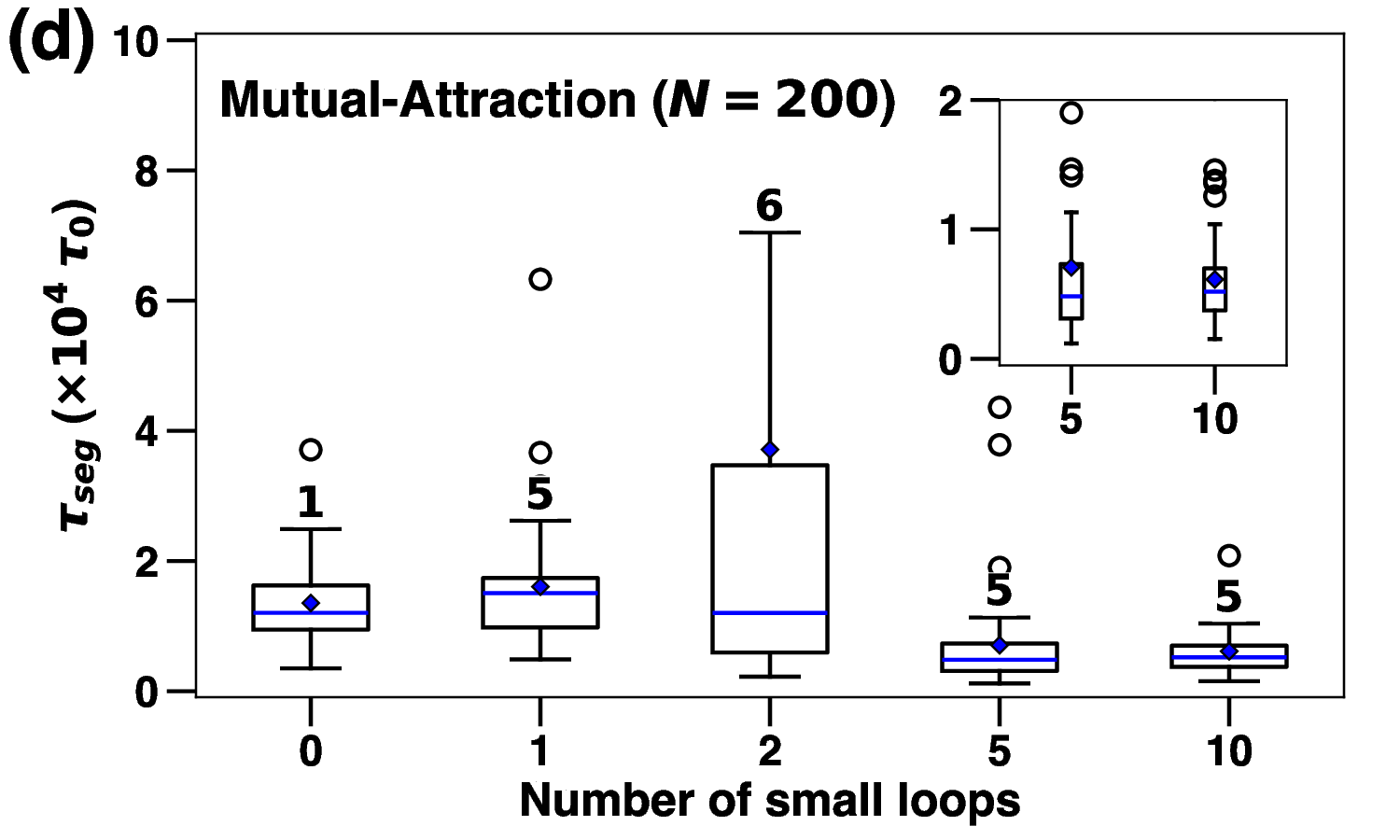}
    \caption{ \textbf{Varying the number of small loops: $N = 200$}. Box plot displaying several summary statistics 
    for the variation of $\tau_{seg}$ with the number of small loops in each polymer for $N = 200$. In all cases, the large loop contained $N/2 = 100$ monomers. Results for the four initialization protocols are provided. For details of the summary statistics and their representation in the box plot, refer Figure~\ref{fig:phi-single-box-plots}.
    Probability distributions of the topologies were compared pairwise using a t-test and the resulting $p$-values are provided in SI-\ref*{SI-sec:si-hist-seg-times}.
    We have performed t-tests and provided $p$-values 
    on comparing a pair of distributions of segregation times at a time, to have a 
    quantitative measure of whether the distributions are significantly different, refer 
     SI-\ref*{SI-sec:si-hist-seg-times}.
}
    \label{fig:phi-multi-box-plots}
\end{figure*}

As before, even if the time-averaged radial distribution was uniform, the instantaneous distribution need not be. Therefore, we visualized the distribution of the monomers along the $x - y$ plane in the configuration at $t = t_0$. For this, we divided the cylinder into four quarters, i.e., $ -L/2 < z  \leqslant -L/4 $, $ - L/4 < z  \leqslant 0 $, $ 0 < z  \leqslant L/4 $, $ L/4 < z  \leqslant L/2 $. The monomer positions within each quarter were projected onto a plane perpendicular to the axis of the cylinder, i.e., in the $x-y$ plane, and shown in Fig.~\ref{fig:initialization-rad-mon-dens}(b) for the \recenter{} protocol. As can be seen, the two polymers exhibited clear separation in the x-y plane in all of the regions considered. The radial segregation could be understood as follows. Two overlapping ring (or linear) polymers confined in a cylinder effectively partition the available cross-section among themselves resulting in confinement blobs of reduced diameter. For further discussion, see~\cite{ha_review, ha2_2012, Jung2010}.  Note however that the polymers remained overlapped along the axis of the cylinder. Similar data for two Arc-1-10[100 10] polymers can be found in SI-\ref*{SI-sec:sectional-mon-dens}, subsection A.
Similar data for Arc-1-1[250-250] and Arc-1-10[250-25] ($N=500$) topologies for two initialization protocols can be found in SI-\ref*{SI-sec:other-mon-dens-500}.

In addition to the radial monomer density and the monomer distribution in the $x-y$ plane discussed above, time-averaged density in different sections of the cylinder is shown in SI-\ref*{SI-sec:sectional-mon-dens}, subsection B. The partitioning into sections enabled a spatially resolved characterization of density variations across different regions of the confining geometry. The analysis was motivated in part by a potential artifact of the shrinking-cylinder protocol. Rapid reduction of the confining volume during the protocol leads to rapid displacement of the boundary walls. This can result in fast accumulation of monomers near the moving walls. Such an accumulation of monomers can be rather long-lived owing to their higher density. 


\subsection{Effect of Topology on $\tau_{seg}$: $L/D = 5$}
At the end of the initialization run, i.e., at $t_0$, the constraints imposed to ensure overlap were removed and the simulations continued. With a few exceptions, the distance between the centers of mass of the two polymers increased with time and approached an asymptotic value, refer Figure~\ref{fig:com-distance}. For each run, the time of segregation, $\tau_{seg}$, was determined using the criterion discussed in Section 2 E.


{\em Varying loop lengths in Arc-1-1: $N = 200$}

Before examining the different topologies, 
we investigated the effect of varying the number of monomers in the small loop, $N_S$, on $\tau_{seg}$ of $N=200$ Arc-1-1 polymers. For example, a polymer with $N_S = 60$, denoted by Arc-1-1[140-60], contains $60$ monomers in the small loop and $140$ in the large loop. 

Preliminary analysis of the data obtained from the different simulation runs suggested that $\tau_{seg}$ exhibited a right-tailed distribution, refer SI-\ref*{SI-sec:si-hist-seg-times}. 
The reader can appreciate that neither the distributions, nor the means and standard deviations facilitate a succinct comparison between the different topologies. \\
Moreover, it is difficult to draw definite conclusions by comparing the distributions for the different topologies and initial conditions considered in this study. Therefore, in addition to the mean of the aforementioned distribution, we have provided box plots with several summary statistics, refer Fig.~\ref{fig:phi-single-box-plots}. Mean (blue diamonds), second quartile or the median (Q$_2$, blue line within the box), third quartile (Q$_3$, top edge of the box), first quartile (Q$_1$, bottom edge), and the number of outliers are shown. The whiskers extended from the edges of the box to the data points lying within $1.5 \times \mathrm{IQR}$, where $\mathrm{IQR} \equiv \mathrm{Q}_3 - \mathrm{Q}_1$ is the interquartile range. Data points beyond $1.5\times\mathrm{IQR}$ from the box edges were classified as outliers ($\circ$).




For the \singleLooped{} topology, Fig.~\ref{fig:phi-single-box-plots}~(a) and (b) show box plots for the variation of $\tau_{seg}$ with $N_S$ for two different initialization protocols. Results for the other two initialization protocols are provided in SI-~\ref*{SI-sec:single-loop-segregation}. All four initialization protocols yielded consistent results for the \singleLooped{} topology which suggest that varying $N_S$ produced no discernible effect on $\tau_{seg}$.

The above result can be rationalized as follows. In the overlapped state, the two Arc-1-1 polymers are likely to be parallel to each other along the axis of the cylinder. Therefore, each polymer can be thought of as occupying one half of the accessible volume and comprising a certain number of blobs. Therefore, to a first approximation, the free energy of the system can be taken to be proportional to the total number of blobs, i.e., twice the number of blobs in each polymer. Upon varying $N_S$ while keeping $N$ fixed, each polymer would continue to occupy approximately one half of the accessible volume within the cylinder and hence, the total number blobs is unlikely to vary significantly. Consequently, upon varying $N_S$, any change in the free energy of the system in the overlapped configuration is likely to be insignificant. As a result, similar values of $\tau_{seg}$ can be expected for different $N_S$. 

{\em Varying the number of small loops}: (a) $N=200$

Unlike the near independence on $N_S$ discussed above, $\tau_{seg}$ varied with the number of small loops in the polymer, see Figure~\ref{fig:phi-multi-box-plots} for $N=200$. Details of the topologies considered can be found in Sec.~\ref{sec:topology}. For all of the initialization protocols, the Arc-0 and the Arc-1-1[100-100] topologies exhibited comparable $\tau_{seg}$. In contrast, for two of the four protocols used, i.e., \recenter{} and \fixingBonding{}, topologies containing two or more small loops displayed smaller $\tau_{seg}$ when compared to Arc-0. The results with  \recenter{} and \fixingBonding{} protocols can be rationalized by considering the change in the entropic cost of overlap of two polymers upon introduction of the small loops~\cite{Bhandarkar2026}.
Consistent with this, previous work~\cite{Bhandarkar2026} has verified that there was negligible difference between the internal energy of the overlapped and the segregated configurations, and hence the segregation is predominantly entropy driven. The difference in the free energy between the overlapped and segregated configurations of a pair of 
ToMo polymers is because of an increase in entropy in segregated state and not due to a decrease in the internal energy.
Similarly to the entropic repulsion between two ring polymers under cylindrical confinement, the small loops also entropically repelled each other. As a consequence, upon increasing the number of small loops in a polymer beyond one, the entropic cost
of overlap for the two polymers also increased. This can be understood as follows: though the individual loops
were smaller, the number of monomers in the cluster remained fixed at N/2 monomers. Moreover, the cluster of
loops have only a limited ability to stretch out and/or explore conformations because of the imposed crosslinking
constraints. The overlap of a cluster of loops with loops of the other polymer is associated with a greater free
energy (entropic) cost, as has been demonstrated previously in Ref.~\cite{Bhandarkar2026}.

The greater effective entropic repulsion between the polymers leads to faster segregation.
The energetic contribution due to excluded volume interactions is likely to be minimal. For instance, even when the value of the diameter, $\sigma$, in the WCA potential was significantly reduced, entropy-driven organization of ToMo polymers was still observed. This was systematically validated in our previous work where we investigated the organization of ToMo polymers in a sphere \cite{Roychoudhury2025}.

Note however that $\tau_{seg}$ did not decrease appreciably when the number of small loops exceeded two, having nearly saturated at the value observed for Arc-1-2. To understand this, we considered the extent of confinement experienced by the small loops. The diameter of the small loops when not under cylindrical confinement, approximated as twice the radius of gyration of the loops in an unconfined single chain, $2R_g^{sl} \approx 7.2\,\sigma, 4.0\,\sigma$ and $2.5\,\sigma$ respectively for Arc-1-2, Arc-1-5, and Arc-1-10. The diameter of the cylinder, $D \approx 6.2\,\sigma$. Since $2R_g^{sl} < D$ for Arc-1-5 and Arc-1-10, it is plausible that increasing the number of small loops beyond two did not significantly increase the entropic cost for segregation further and hence, resulted in the aforementioned near saturation of $\tau_{seg}$. It is germane to note that although the median $\tau_{seg}$ remained comparable for Arc-1-2, Arc-1-5, and Arc-1-10,  the interquartile range was smaller and the number of outliers were fewer for Arc-1-10 when compared to Arc-1-2.

\begin{figure*}[ht!]
        \centering
        \includegraphics[width=0.49\linewidth]{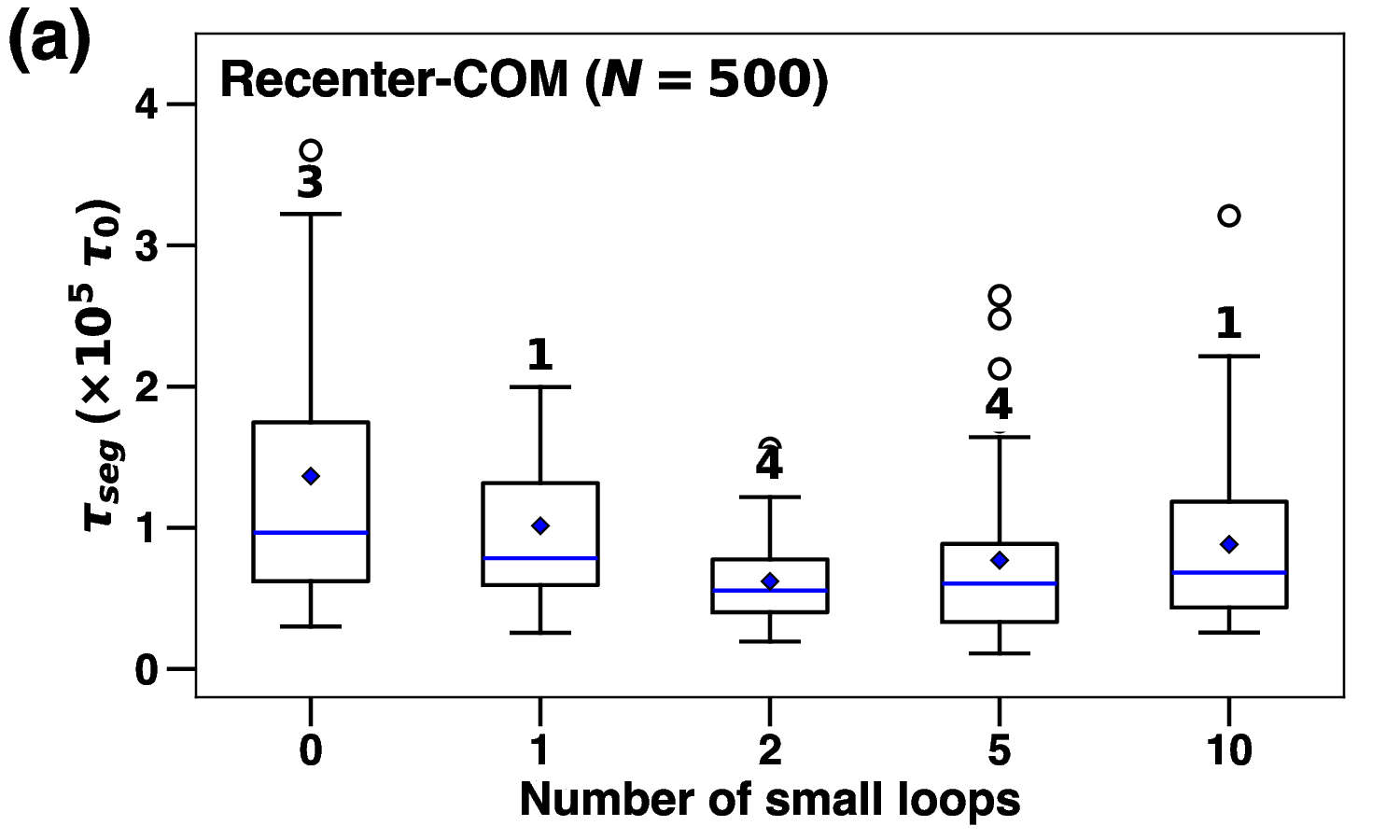}
    \hfill
        \includegraphics[width=0.49\linewidth]{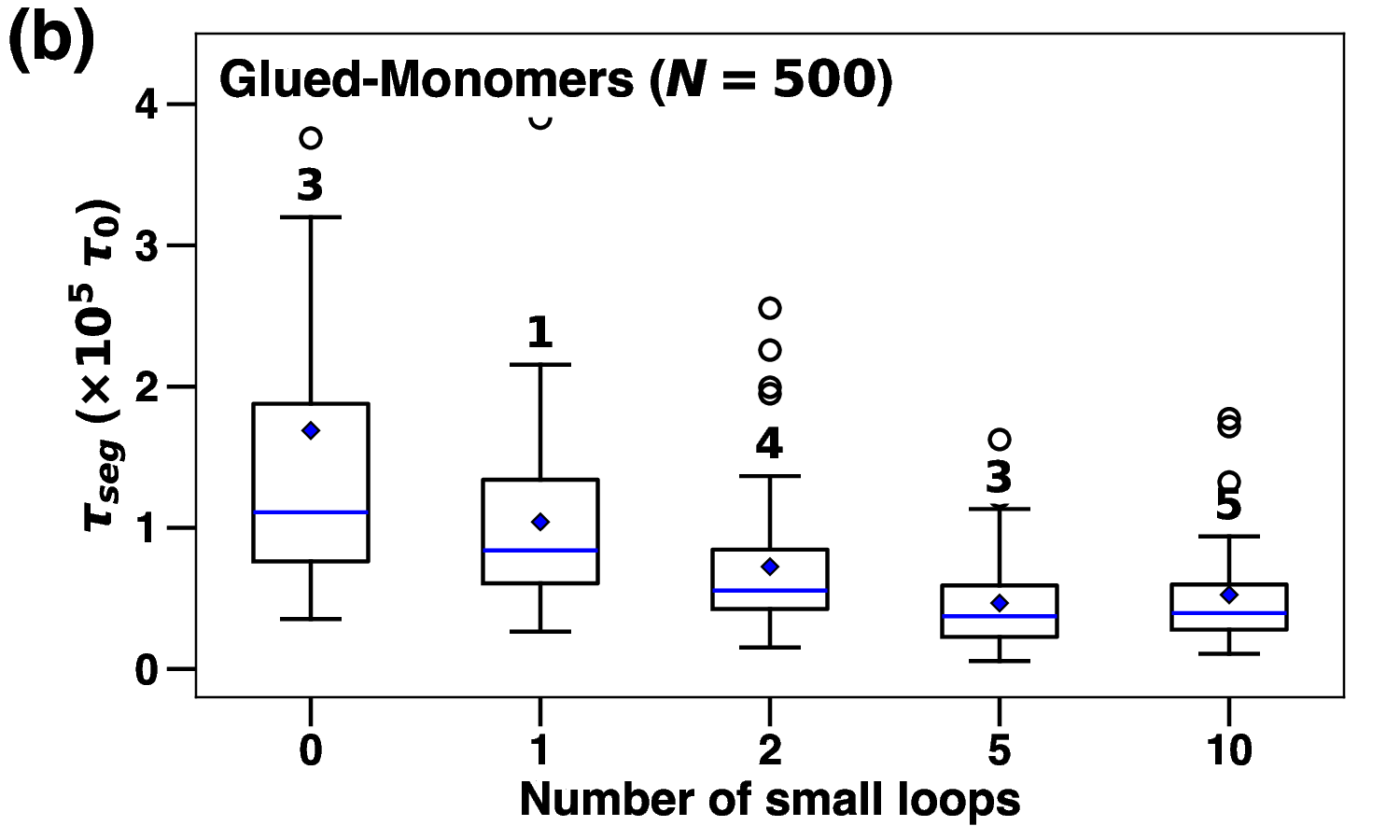}
        \includegraphics[width=0.49\linewidth]{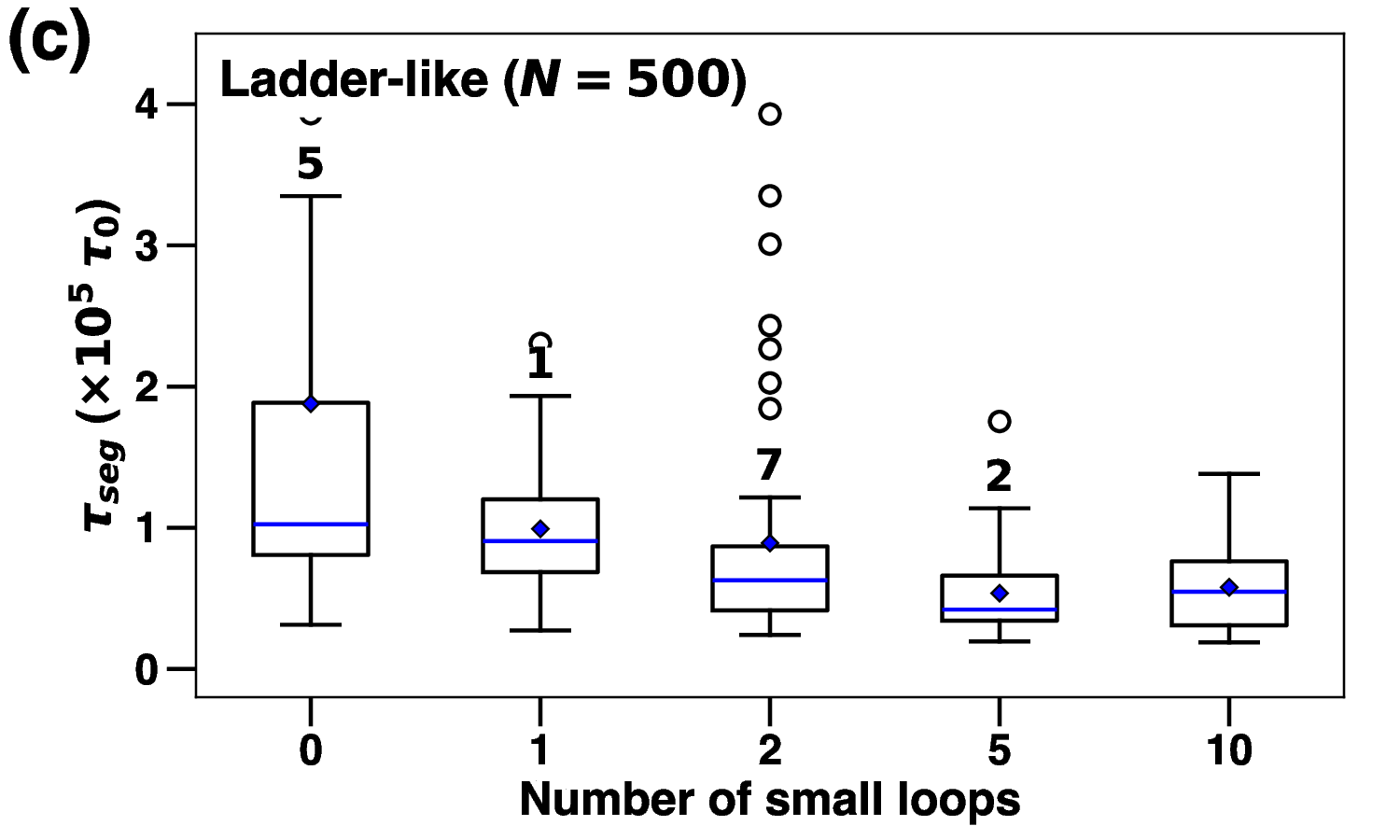}
    \hfill
        \includegraphics[width=0.49\linewidth]{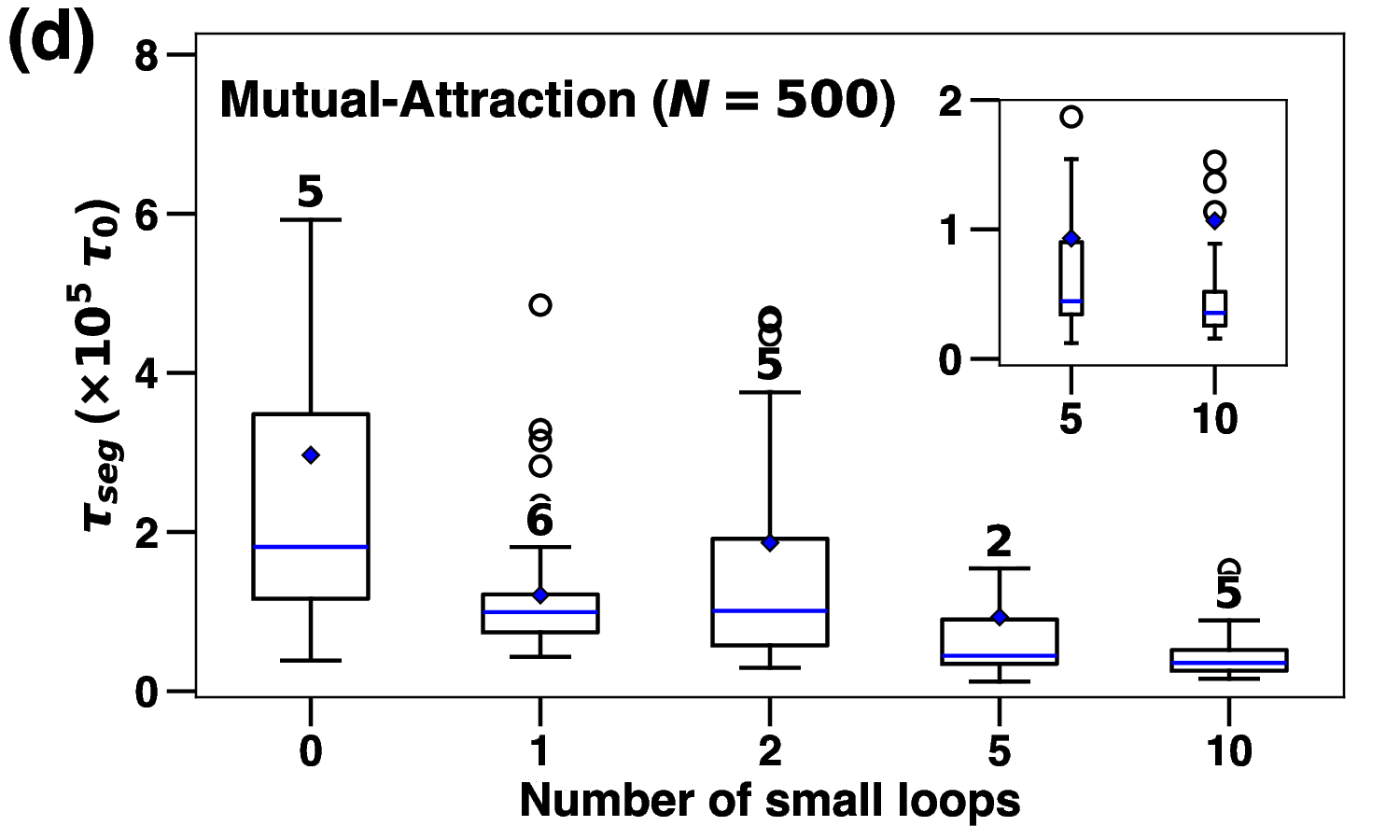}
   
    \caption{
    \textbf{Varying the number of small loops: $N = 500$} Box plot displaying several summary statistics 
    for the variation of $\tau_{seg}$ with the number of small loops in each polymer for $N = 500$. In all cases, the large loop contained $N/2 = 250$ monomers. Results for the four different initialization protocols are provided. For details of the summary statistics and their representation in the box plot, refer Figure~\ref{fig:phi-single-box-plots}.
    Probability distributions of the topologies were compared pairwise using a t-test and the resulting $p$-values are provided in SI-\ref*{SI-sec:si-hist-seg-times}.
    We have performed t-tests and provided $p$-values 
    on comparing a pair of distributions of segregation times at a time, to have a 
    quantitative measure of whether the distributions are significantly different, refer 
     SI-\ref*{SI-sec:si-hist-seg-times}.}
    
    \label{fig:phi-500-multi-box-plots}
\end{figure*}

While the variation of $\tau_{seg}$ with the number of small loops observed for \recenter{} and \fixingBonding{}  protocols were similar, the results for the \replicationLike{} and \feneLJ{} protocols were different from the other two. In these cases, $\tau_{seg}$ did not decrease for Arc-1-2, but did for Arc-1-5 and Arc-1-10, refer Figure~\ref{fig:phi-multi-box-plots} (c) and (d). Arc-1-2 exhibited several outliers, i.e., an unusually large distribution of $\tau_{seg}$. For \feneLJ{}, the reason for this anomalous behavior could be traced to the mutual interpenetration of the two polymers. The configurations generated using the \feneLJ{} protocol often contained mutually interpenetrating chains, where one polymer {\em threaded} into the region spanned by the other. The threadings were the result of the Lennard-Jones attraction between the monomers of the different polymers. The attraction was introduced during initialization to maintain the overlap between the two polymers and was switched off at $t_0$, i.e., before the segregation of the two polymers was investigated. As entropic repulsion is expected to induce segregation of the two polymers, spontaneous formation of new threadings after $t_0$ were deemed unlikely. Therefore, examination of the configurations at $t_0$ obtained from the different initialization protocols sufficed. We found that Arc-1-2 exhibited a higher incidence of threadings which suggested a plausible reason for the observed high value of $\tau_{seg}$. For the Arc-1-5 and Arc-1-10, the small loops with only $20$ and $10$ monomers each respectively, impeded significant threading. This can be seen in the few representative snapshots, from $50$ independent runs, shown in SI-~\ref*{SI-sec:threading-snapshots}. 
For the \replicationLike{} protocol also, we found threaded loops in the system with Arc-1-2 polymers, which again is an artefact of holding together the monomers from different polymers with springs. Smaller  segments of the polymers can (probabilistically) twist around each other as in a DNA-double helix, and this can easily lead to threading.  Thus the  higher values of $\tau_{seg}$  statistics for Arc-1-2 is an 
artefact due to threading. We explicitly confirmed that while the other two initialization protocols generated overlapped configurations, the configurations did not exhibit threadings to any significant degree. The general  trend remained that an increase in the 
number of loops led to  faster segregation.

\textit{Varying the number of small loops}: (b) $N = 500$.

Starting from initial configurations where the chains were overlapped, the analysis described above was performed also for polymers with $N = 500$. For the unconfined Arc-1-2, Arc-1-5, and Arc-1-10 topologies, the approximate diameter of the small loops was, $2R_g^{sl} = 12.7\,\sigma, 7.3\,\sigma$ and $4.6\,\sigma$, respectively. The diameter of the cylinder, $D = 8.7\,\sigma$. For these topologies, the $\tau_{seg}$ are shown in Fig.~\ref{fig:phi-500-multi-box-plots}. For the \fixingBonding{} and \replicationLike{} protocols, $\tau_{seg}$ decreased with increasing number of small loops, consistent with that observed for $N=200$. However, $\tau_{seg}$ obtained when the initial configurations were generated using the \recenter{} and the \feneLJ{} protocols deviated from this trend, with Arc-1-5 and Arc-1-10 exhibiting unexpectedly large $\tau_{seg}$ in the \recenter{} case. Examination of the initial configurations, at $t_0$, revealed that the orientation of the two polymers was {\em antiparallel} to each other, unlike the {\em parallel} orientation observed for the other two initialization protocols. In such antiparallel configurations, the cluster of small loops of one polymer overlapped with the large loop of the other. In parallel configurations, in contrast, the cluster of small loops of one polymer overlapped with the cluster of small loops in the other. Segregation from an antiparallel configuration likely involves crossing a higher free-energy barrier 
 (or relatively smaller gradient in free energy) when compared to parallel configurations with the result being larger $\tau_{seg}$. The existence of a free-energy barrier (or a smaller free energy gradient) is discussed in the next section based on the analysis of the onset of segregation from {\em Parallel} and {\em Antiparallel} configurations.

Therefore, we identified the number of {\em parallel} and {\em antiparallel} initial configurations generated by the \recenter{} and  the \feneLJ{} initialization protocols for $N=200$ and $N=500$, see SI-\ref*{SI-sec:orientation}. We found that as the number of small loops increased, the frequency of obtaining an {\em antiparallel} initial configuration overwhelmingly exceeded that of obtaining a {\em parallel} configuration. 
For the other two initialization protocols, the particular implementation of the constraints ensured that the two polymers were parallel in the overlapped configuration. Similar details regarding the initial configurations for $L/D \gg 1$, to be discussed later in the manuscript, is also provided in 
SI-\ref*{SI-sec:orientation}. 

In the initial configuration, the propensity of the two overlapped polymers to adopt one of the two orientations  depends on their topology. Arc-1-2 polymers possess relatively large loops, and the entropic cost associated with the overlap of clusters of small loops is significantly less as compared to that for Arc-1-5 or Arc-1-10 architectures. Consequently, Arc-1-2 polymers can be found in either orientation, with a higher probability of being {\em antiparallel}. In contrast, a pair of polymers with the Arc-1-5 and Arc-1-10 topologies preferentially adopt the {\em antiparallel} orientation in which the clusters of small loops belonging to the two polymers can remain spatially separated along the axis of the cylinder.


Furthermore, for initialization using the \feneLJ{} protocol and consistent with the observations for $N=200$, $\tau_{seg}$ of $N = 500$ Arc-1-2 also exhibited a larger interquartile range (IQR) when compared to Arc-1-5 and Arc-1-10. For $N=200$, we had attributed this to the mutual chain interpenetration observed in the initial configurations prepared using the \feneLJ{} protocol. This remains true for $N=500$ polymers as well. A visual overview of the segregation pathways using snapshots of the chain configurations is provided in SI-~\ref*{SI-sec:orientation-and-seg}. In the next subsection, we discuss the kinetics of segregation and the associated mechanisms.

\subsection{Kinetics of Segregation}
\begin{figure}[h!]
     \centering
        \includegraphics[width=0.8\linewidth]{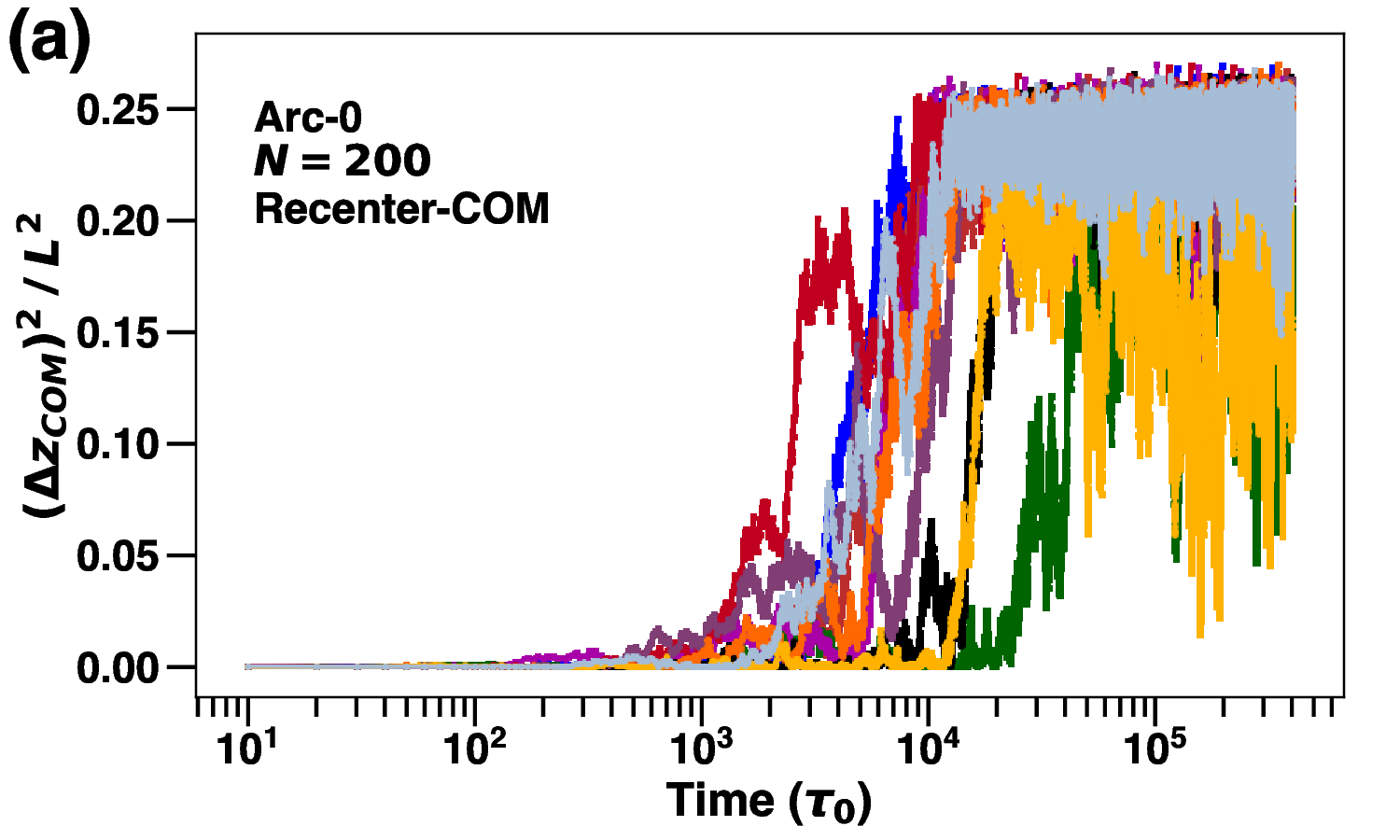}
        \includegraphics[width=0.8\linewidth]{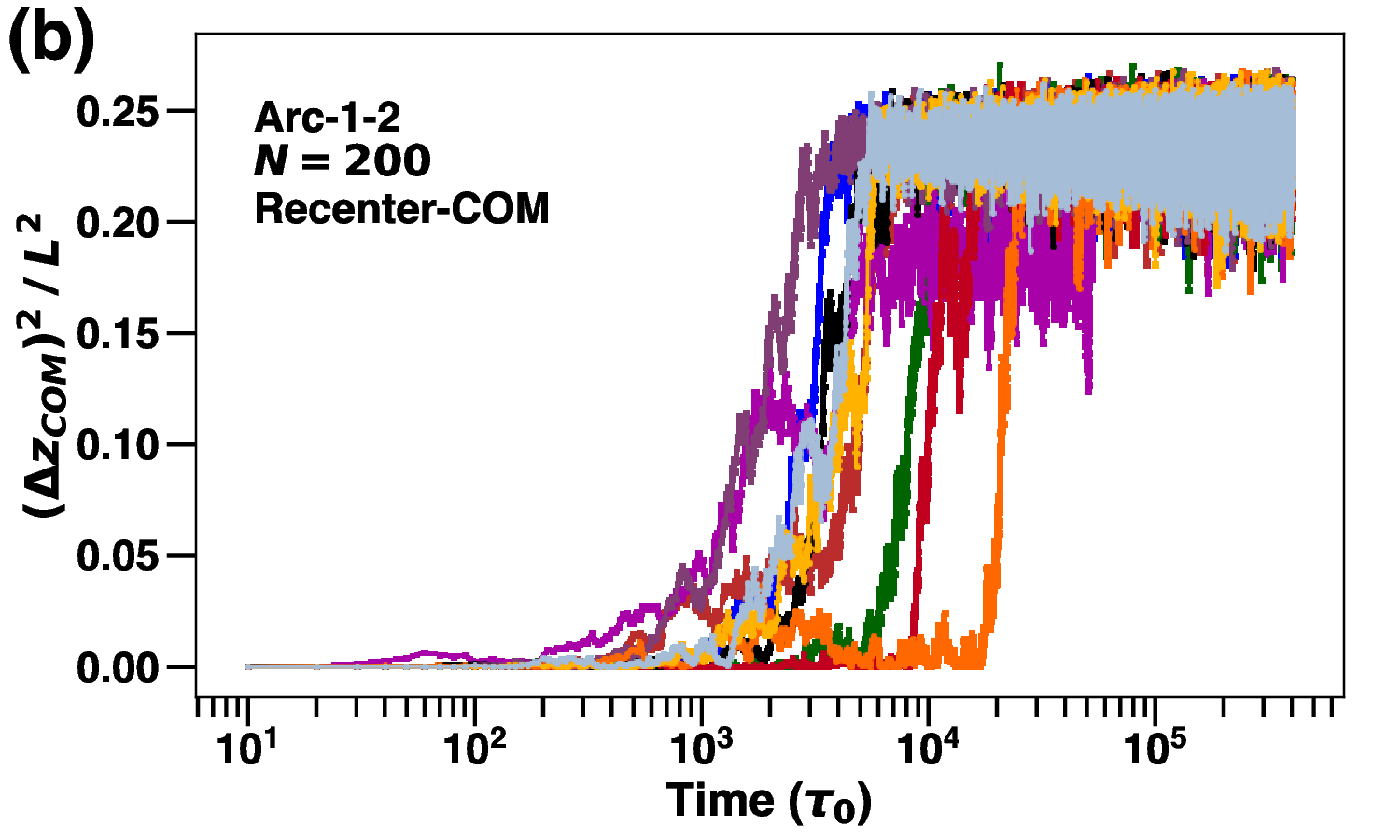}
        \includegraphics[width=0.8\linewidth]{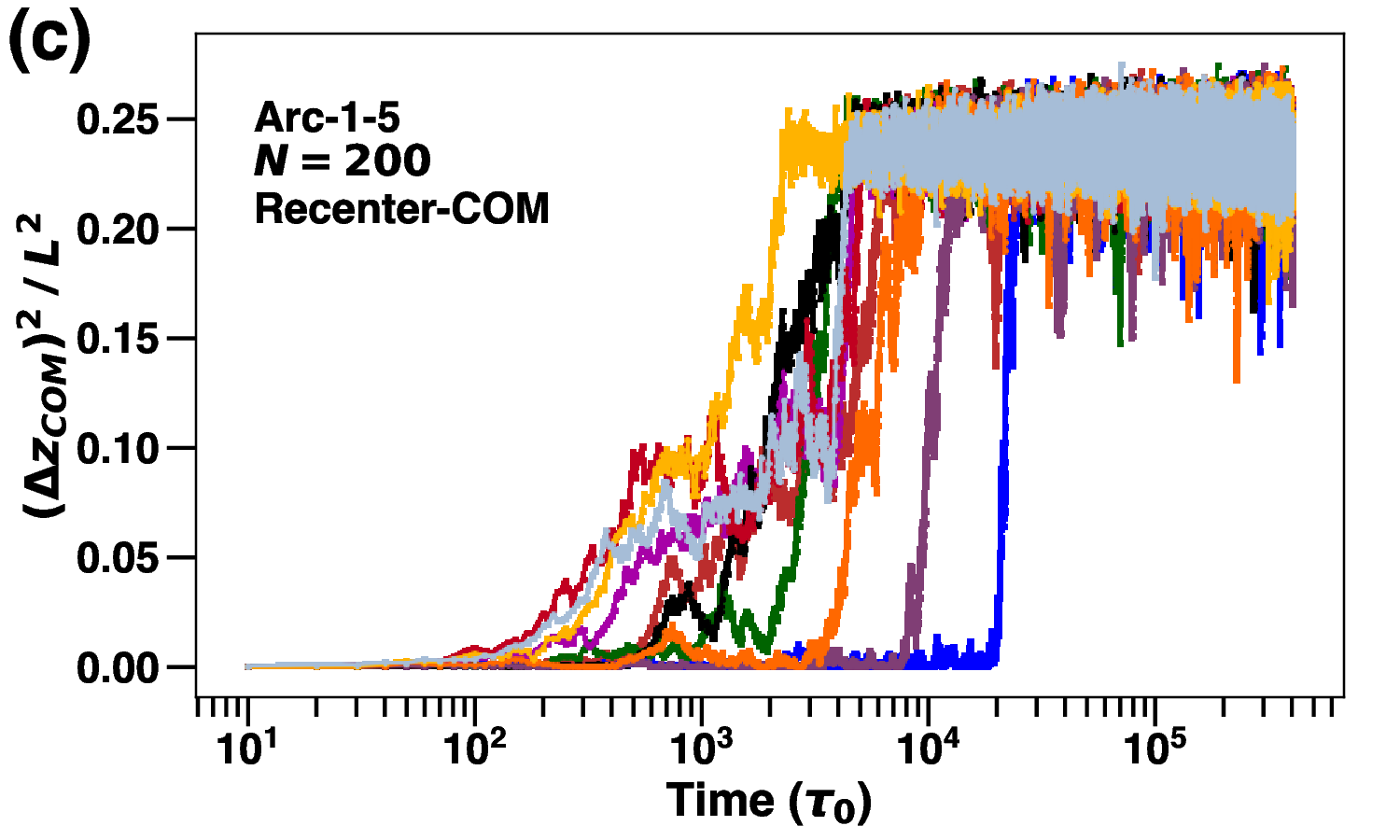}
        \includegraphics[width=0.8\linewidth]{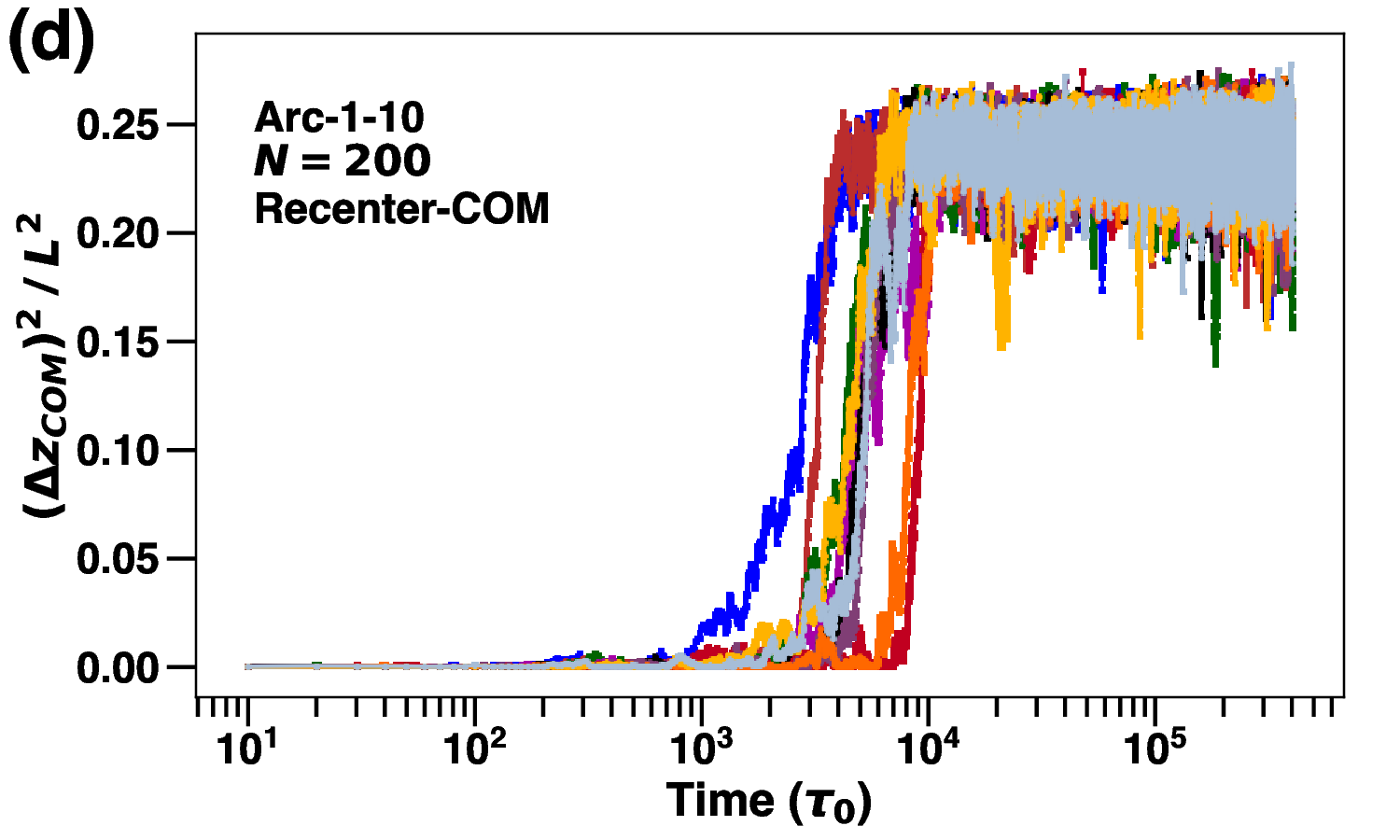}
    \caption{Time evolution of the squared separation along the axis of the cylinder between the centers of mass of two $N = 200$ polymers, $(\Delta z_{COM})^2$, for $L/D = 5$. Ten independent trajectories for (a) Arc-0, (b) Arc-1-2, (c) Arc-1-5, and (d) Arc-1-10 initialized using the \recenter{} protocol are shown.}
    \label{fig:sq-com-runs-rec}
\end{figure}

Results presented earlier clearly indicated that both the median and the IQR of $\tau_{seg}$ were strongly affected by the topological modifications and the nature of the overlapped initial configuration which depended on the initialization protocol. Below, we examine the kinetics of segregation in more detail. 

To this end, starting from $t_0$, we examined the individual trajectories of two $N = 200$ Arc-0, Arc-1-2, Arc-1-5, and Arc-1-10 ToMo polymers. Figure~\ref{fig:sq-com-runs-rec} exhibits the time evolution of the squared separation along the axis of the cylinder between the centers of mass (COMs) of the two polymers, $(\Delta z_{\text{COM}})^2$, normalized by the squared length of the cylinder, $L^2$. Data for $10$ (out of $50$) independent trajectories initialized using the {\em Recenter-COM} protocol are shown. Data for Arc-1-1 is not presented as it exhibited behavior qualitatively similar to that of Arc-0.
For all of the topologies examined, the two polymers were overlapped at $t_0$ such that $(\Delta z_{\text{COM}})^2 / L^2 \approx 0$. For time $> t_0$, $(\Delta z_{\text{COM}})^2 / L^2$ increased with time until it reached an asymptotic plateau value near $(0.5L)^2 / L^2$ corresponding to complete segregation. However, as discussed below, the time evolution for the different topologies exhibited qualitative differences. 

{\em Arc-1-2, Arc-1-5: trajectories fall into two groups} 

For the two topologies, we observed significant heterogeneity in the time evolution of $(\Delta z_{\text{COM}})^2 / L^2$ between the different trajectories. 
Note that for many of the trajectories the onset of segregation was observed at time $\approx 5 \times 10^2 \, \tau_0$ as indicated by an increase in $(\Delta z_{\text{COM}})^2 / L^2$. In contrast and especially for Arc-1-5, the onset of segregation occurred as early as time $\approx 10^2 \, \tau_0$ or was delayed until time $\approx 10^4 \, \tau_0$. The aforementioned observations suggest that the trajectories can be divided into two groups. For the first group, segregation started at earlier times, slowed down between $10^3 \, \tau_0$ and $5 \times 10^3 \, \tau_0$, before reaching completion well before $10^4 \, \tau_0$. For the second group however, a pronounced delay in the onset of segregation was observed. However, once initiated, the segregation proceeded rapidly towards completion with no apparent slowing down at intermediate times. 

{\em Arc-0 and Arc-1-10:} 

Figure~\ref{fig:sq-com-runs-rec} indicates that there was noticeable heterogeneity in the onset of segregation 
(time $\approx 10^3 - 10^4$) also for Arc-0. Moreover, the increase in $(\Delta z_{\text{COM}})^2 / L^2$ was rather intermittent for all of the trajectories. As a consequence, the time at which 
$(\Delta z_{\text{COM}})^2 / L^2$ reached its asymptotic value also exhibited considerable heterogeneity. Therefore, in contrast to the trajectories of Arc-1-2 and Arc-1-5 polymers discussed above, 
these trajectories could not be neatly categorized into two distinct groups on the basis of their time evolution characteristics. 
In the case of Arc-1-2 and Arc-1-5, the overlap of the loops in the internal cluster of one polymer with the loop(s) of the other polymer is expected to increase the entropic drive for segregation compared to that for Arc-0. This is analogous to the entropic drive for segregation of a pair of overlapping polymers being stronger for rings compared to linear chains~\cite{elena2,ha2_2012,Shin2014}. This difference between the trajectories of Arc-0 and Arc-1-5 is known to arise from the free energy landscape of ToMo polymers, as discussed in detail in \cite{Bhandarkar2026}.  The relatively slower start and the gradual and intermittent segregation of Arc-0 polymers are likely consequences of a lower entropic drive for segregation relative to that for Arc-1-2 and Arc-1-5, combined  with the absence of {\em parallel} and {\em antiparallel} initial configurations for Arc-0.

The time evolution of $(\Delta z_{\text{COM}})^2 / L^2$ for Arc-1-10, however, was markedly 
different from both Arc-0 and the qualitatively similar Arc 1-2 and Arc-1-5. For Arc-1-10, the typical onset
of segregation occurred after time $\approx 10^3 \, \tau_0$. After the onset of segregation, 
$(\Delta z_{\text{COM}})^2 / L^2$ increased rapidly leading to the completion of segregation before time 
$\approx 10^4 \, \tau_0$. For Arc-0, as well as for Arc-1-2 and Arc-1-5, the onset of segregation was well
before $10^3 \tau_0$ in most of the cases. 
The delay in the onset of segregation for Arc-1-10 followed by a rapid increase in
$(\Delta z_{\text{COM}})^2 / L^2$ suggested the existence of a single dominant kinetic barrier,
arising from free-energy differences, that must be overcome before the onset of segregation can occur.
The rapid rate of segregation once the barrier has been crossed suggests a relatively large free-energy difference between the overlapped and the segregated states. Estimates for this free-energy difference for a pair of Arc-1-10 polymers with $N=200$ and $N=500$ can be found in \cite{Bhandarkar2026}. 
Upon onset of segregation, no additional significant free-energy barriers were evident, as no slowing down of segregation was observed at later times.

\begin{figure}
    \centering
    \includegraphics[width=0.9\linewidth]{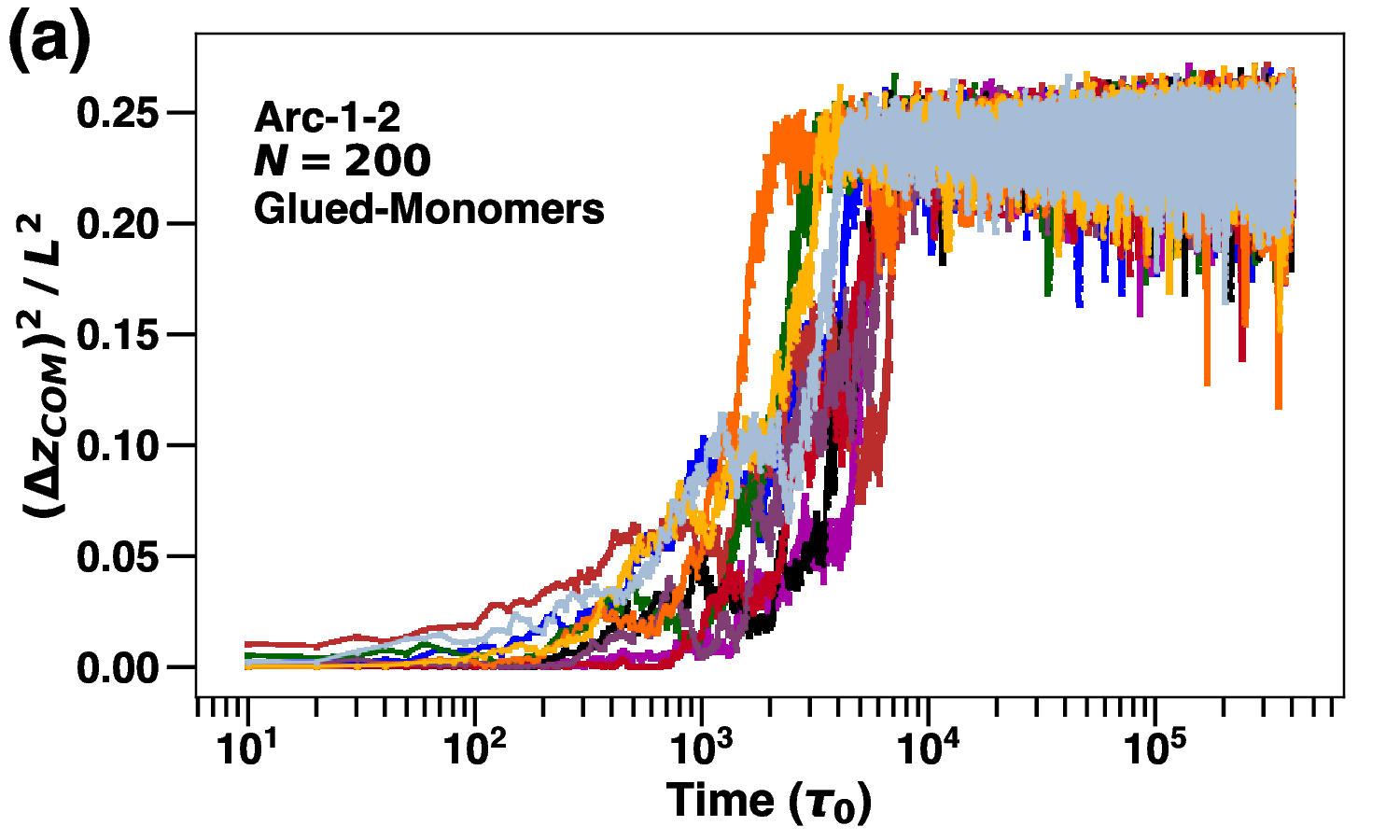}
    \includegraphics[width=0.9\linewidth]{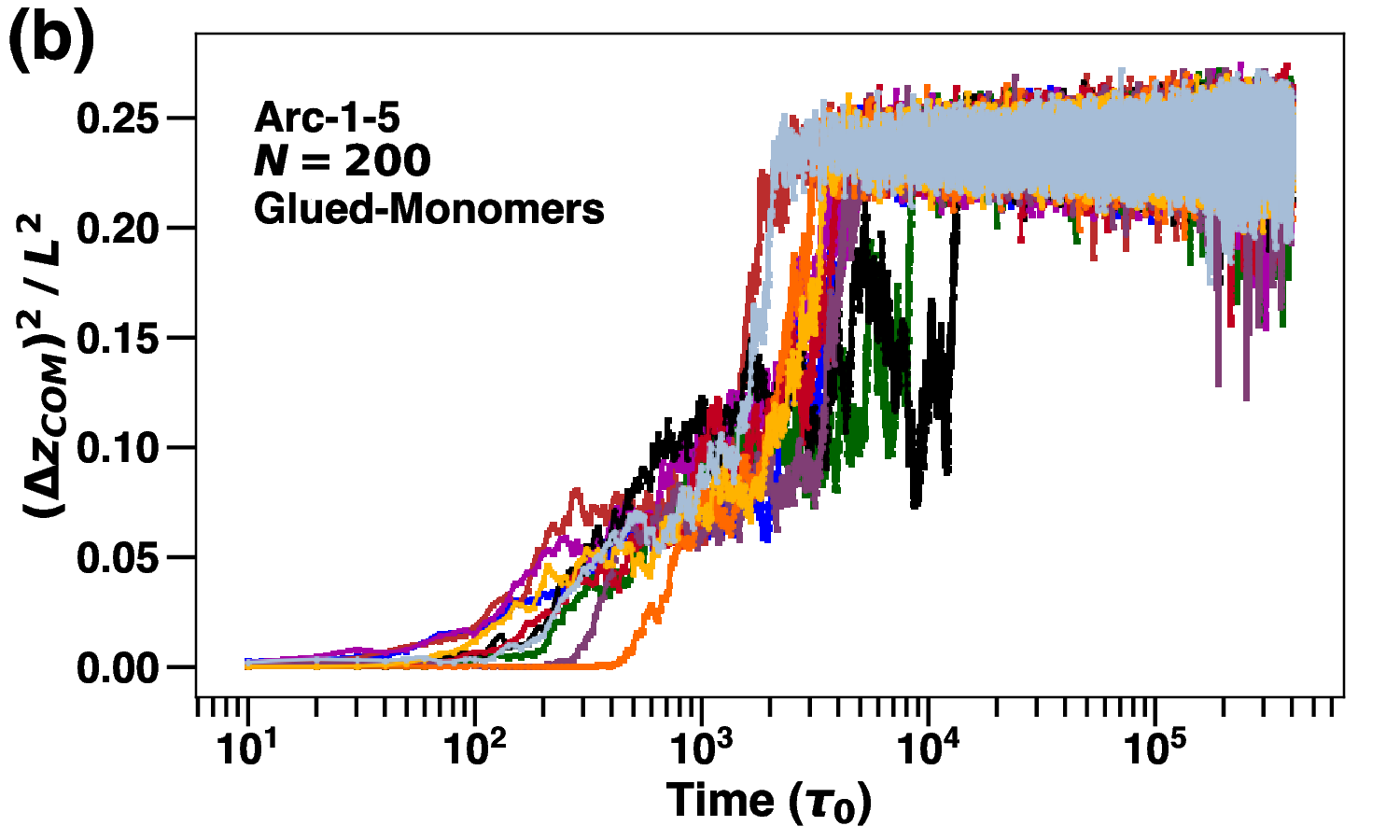}
    \caption{Time evolution of the squared separation along the axis of the cylinder between the centers of mass of two $N = 200$ polymers, $(\Delta z_{COM})^2$, for $L/D = 5$. Ten independent trajectories for (a) Arc-1-2 and (b) Arc-1-5 initialized using the \fixingBonding{} protocol are shown. 
    }
    \label{fig:sq-com-runs-fbsr}
\end{figure}

In Fig.~\ref{fig:sq-com-runs-fbsr}, the time evolution of $(\Delta z_{\text{COM}})^2/L^2$ for Arc-1-2 and Arc-1-5 initialized using the \fixingBonding{} protocol are shown. Several of these trajectories exhibited earlier onset of segregation as compared to those corresponding to the \recenter{} protocol. The  early onset of segregation indicates that the system spontaneously started lowering its free energy.
However, for the \fixingBonding{} protocol, nearly all of the trajectories exhibited considerable slowdown in the 
increase of $(\Delta z_{\text{COM}})^2/L^2$ at time $\approx 10^3 \, \tau_0$. Note also that the slowdown was
more pronounced for Arc-1-5 than for Arc-1-2. These two observations clearly suggest the existence of either a free
energy barrier or a lower free energy gradient at intermediate times. The relation of this to the chain topology
will be discussed below. Note that, for nearly every trajectory, segregation was complete by $10^4 \, \tau_0$. 

In contrast, for the {\em Recenter COM} initialization protocol, all of the simulation runs exhibited a pronounced delay before the onset of segregation for a pair of Arc-1-10 polymers. This observation is consistent with the presence of a free-energy barrier that must be overcome for the segregation to commence. If the delay was solely due to a smaller free energy gradient, one can expect that at least some of the trajectories would have exhibited onset of segregation with little delay. 
However, such behavior was not observed.

A key distinction between the \fixingBonding{} and the \recenter{} initialization protocols was the mutual orientation of the overlapped ToMo polymers. The \fixingBonding{} (and the \replicationLike{}) protocol generated only {\em parallel} configurations, whereas the \recenter{} (and the \feneLJ{}) initialization yielded both {\em parallel} and {\em antiparallel} configurations, refer Fig.~\ref{fig:topologies}(f)). Free-energetically, it is favorable for a cluster of small loops of one polymer to overlap with the more deformable large loop of the other polymer -- an arrangement corresponding to the {\em antiparallel} configuration~\cite{Bhandarkar2026}. For the Arc-1-5 initialized using the \recenter{} protocol, a significantly larger fraction of {\em antiparallel} initial configurations were obtained. For the \fixingBonding{} protocol, only one of the two kinds of configurations was possible depending on the choice of the monomers between which the harmonic constraints were imposed. In this work, we chose to impose constraints during the \fixingBonding{} protocol such that only {\em parallel} configurations resulted. 
In Sec.~\ref{SI-sec:orientation-seg-times} of the SI, we have presented $\tau_{seg}$ data, analogous to that in Fig.~\ref{fig:phi-500-multi-box-plots}, but in separate box plots for {\em parallel} and {\em antiparallel} initial configurations of Arc-1-2, Arc-1-5, and Arc-1-10 polymers for $N=200$  and $N=500$. The dependence of $\tau_{seg}$ on the mutual orientation of the two polymers at the end of the \recenter{} and \feneLJ{} initialization protocols is clear.

\begin{figure}
    \centering
    \includegraphics[width=0.9\linewidth]{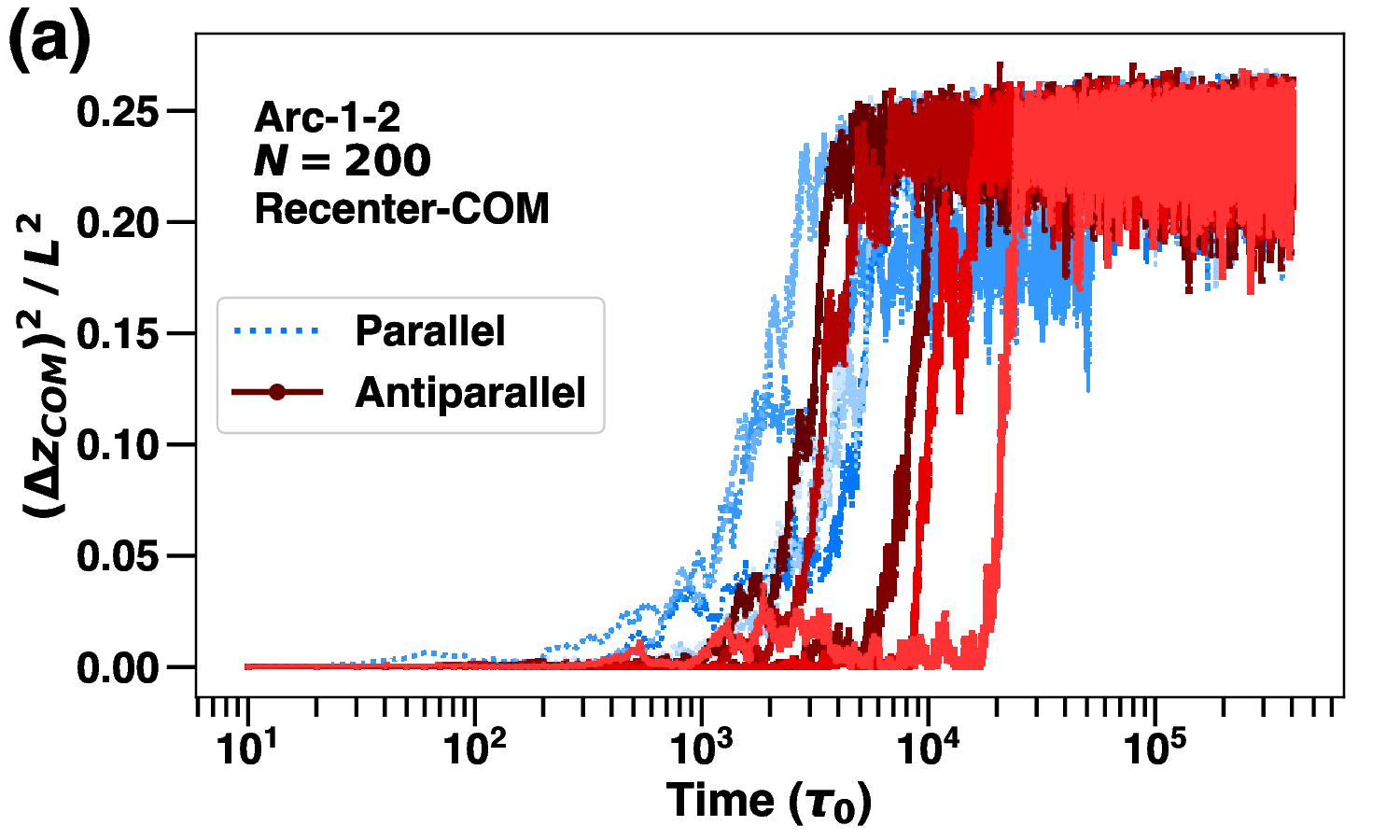}
    \includegraphics[width=0.9\linewidth]{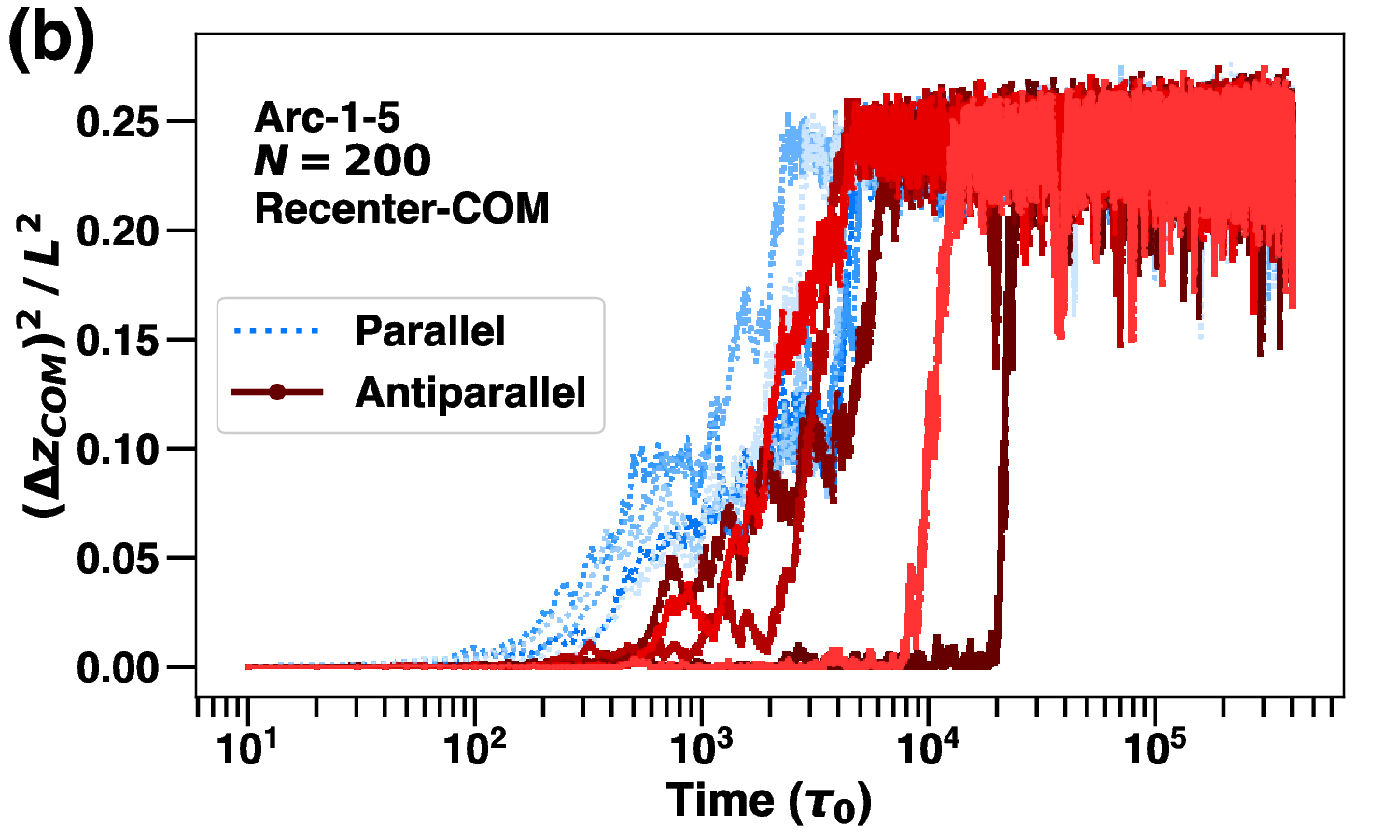}
    \caption{Time evolution of the squared separation along the axis of the cylinder between the centers of mass of two $N = 200$ polymers, $(\Delta z_{COM})^2$, for $L/D = 5$. Ten independent trajectories for (a) Arc-1-2 and (b) Arc-1-5 initialized using the \recenter{} protocol are shown. While the trajectories shown here are identical to that in Fig.~\ref{fig:sq-com-runs-rec}, whether the overlapped initial configuration was parallel or antiparallel has been indicated.
    In SI-\ref*{SI-sec:orientation-and-seg}, we have provided analogous data for the Arc-1-2 and Arc-1-5 topologies for $N=500$. 
    }
    \label{fig:orientation-grouped-sq-com}
\end{figure}

\begin{figure}
    \centering
    \includegraphics[width=0.8\linewidth]{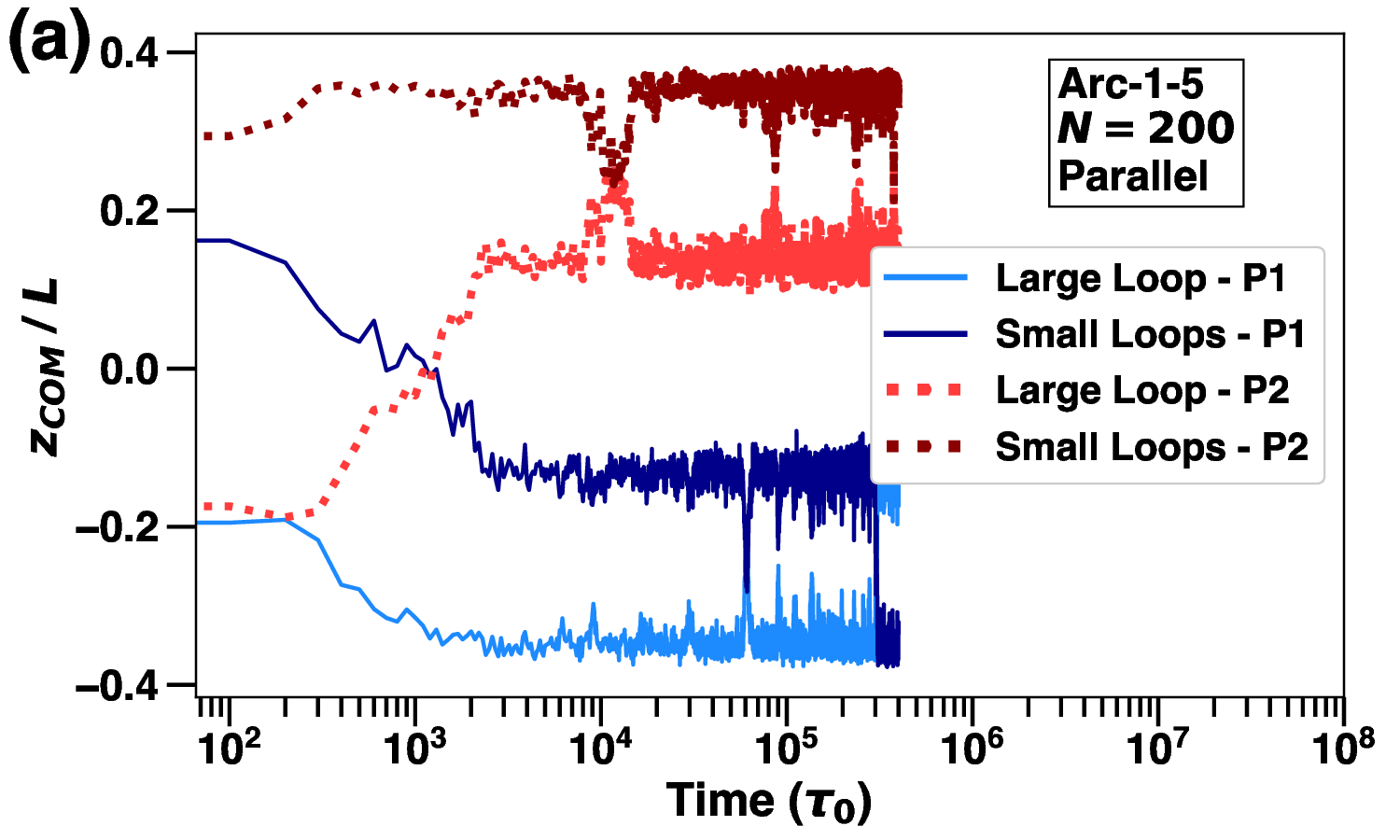}
    \includegraphics[width=0.8\linewidth]{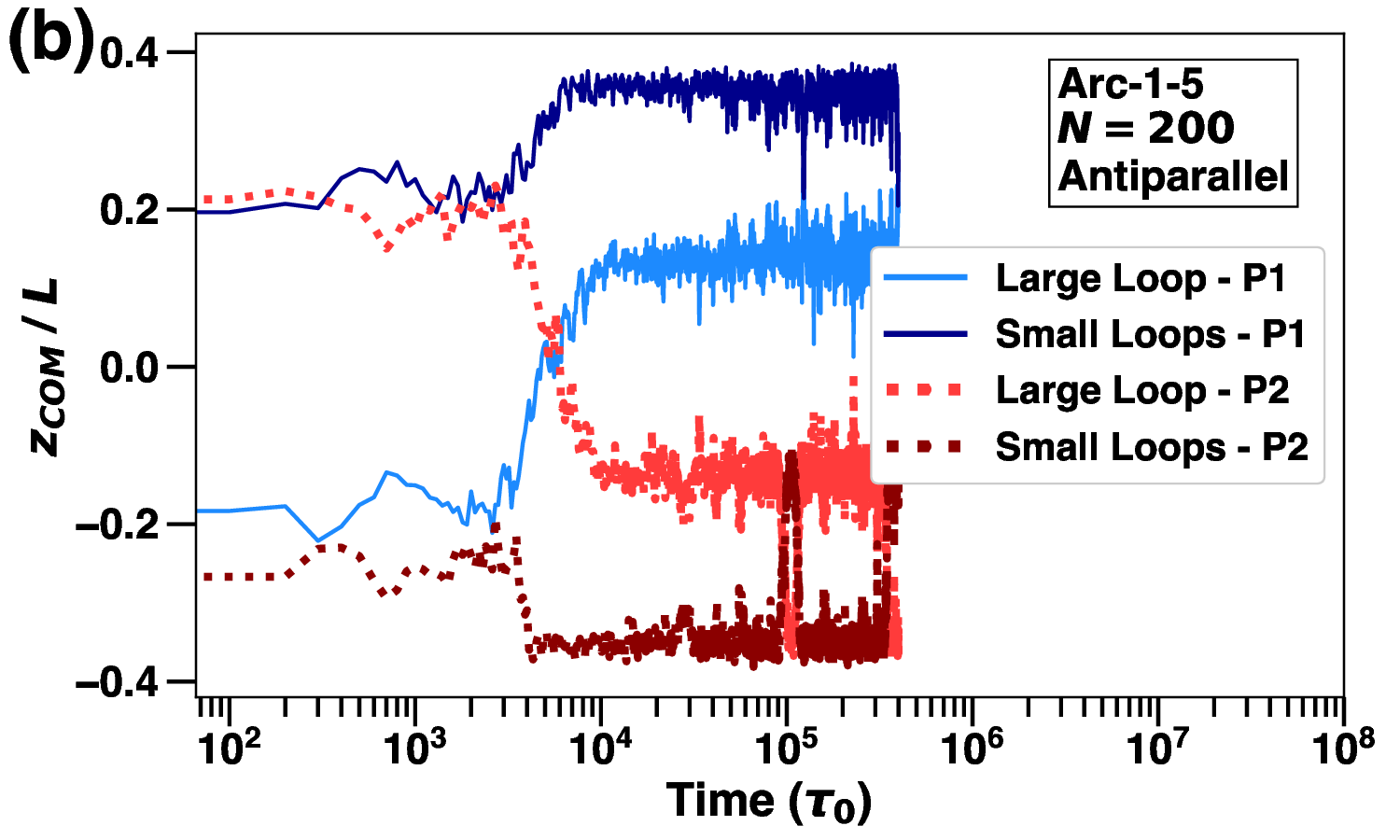}
    \caption{Time evolution of the $z$-coordinate of the center of mass of the large loop and the clusters of small loops of two $N = 200$ Arc-1-5 polymers (P1, P2). (a) From a parallel initial configuration, the polymers segregated via a crossing between the cluster of small Loops of P1 and the large Loop of P2. (b) From an antiparallel initial configuration, the polymers segregated via two crossings between the cluster of small Loops of one polymer and the large Loop of the other, and one crossing between the large Loops of the two polymers.}
    \label{fig:region-com-timeseries}
\end{figure}

In Fig.~\ref{fig:orientation-grouped-sq-com}, the trajectories of Arc-1-2 and Arc-1-5 whose configurations at $t_0$ were either {\em parallel} (blue dots) or {\em antiparallel} (red lines) are shown. Antiparallel initial configurations, in general, exhibited later onset of segregation when compared to parallel initial configurations. For Arc-1-10 initialized using the \recenter{} protocol, we have verified that all of the $50$ initial configurations at $t_0$ were antiparallel. 
Note that while the initial configuration played a role, the particular kinetic pathway towards segregation remained stochastic. 


{\em Kinetic pathways: a closer look}

To better understand the kinetic pathways of segregation, especially in light of whether the initial configuration was parallel or antiparallel, the time evolution of the $z$-coordinates of the centers of mass (COMs) of distinct sections of each polymer is shown in Fig.~\ref{fig:region-com-timeseries}. Examining the COMs of the cluster of small loops and the large loop of each polymer individually enabled us to discern the behavior of the individual sections and identify overlaps during the different stages of segregation. The characteristics of the motion of the clusters of small loops (or occasionally the large loops) toward the poles of the cylinder enabled us to distinguish between the different kinetic pathways.

Figures~\ref{fig:region-com-timeseries} (a) and (b) respectively show representative trajectories for polymers initially in parallel and antiparallel configurations. In the parallel configuration, the COMs of the clusters of small loops in each polymer were located near one pole of the cylinder, while the COMs of the large loops were positioned closer to the other pole. For convenience, we refer to the two polymers as P1 and P2. For time $> t_0$, i.e., once the constraints imposed during initialization were removed, the two clusters of small loops started to move apart from each other. The small loops of P2 migrated towards the nearest pole and remained there. However, the small loops of P1 and the large loop of P2 needed to cross each other to achieve segregation. This involved surmounting a free energy barrier. 
As can be seen from Fig.~\ref{fig:region-com-timeseries}(a), this crossing occurred at time $\approx 10^3\tau_0$ after which the segregation proceeded to completion. 
As noted previously~\cite{Bhandarkar2026}, at equilibrium the clusters of small loops in Arc-1-5 and Arc-1-10 tended to reside near the poles of the cylinder. In line with this expectation, the orientation of P1 flipped by $180^\circ$ at time $\approx 4\times 10^5\tau_0$ such that its small loops were closer to the poles of the cylinder. This rearrangement allowed the more deformable large loops of the two polymers to remain closer to the center of the cylinder and even occasionally overlap, thereby increasing configurational entropy.

\begin{figure}[htbp!]
        \centering
        \includegraphics[width=0.8\linewidth]{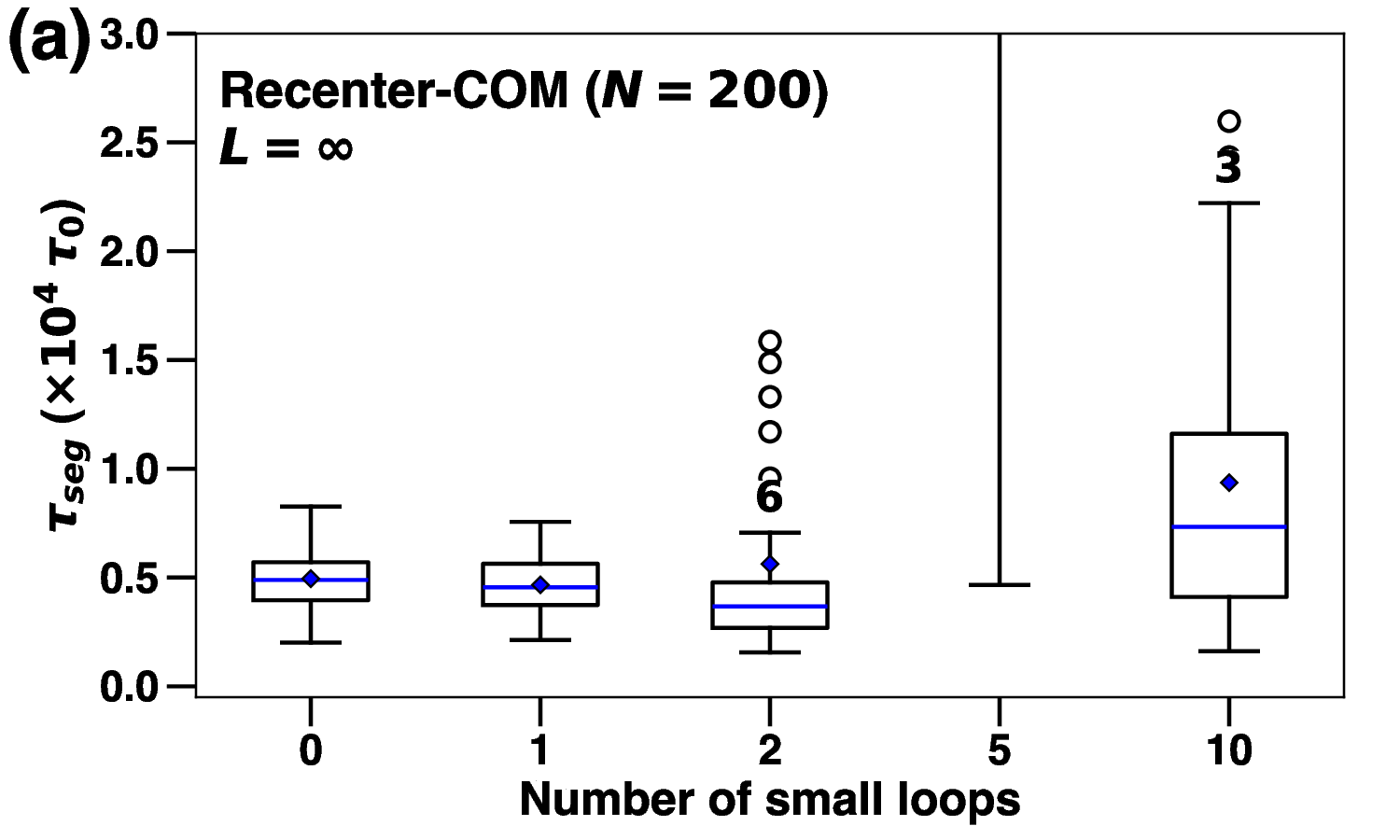}
        \includegraphics[width=0.8\linewidth]{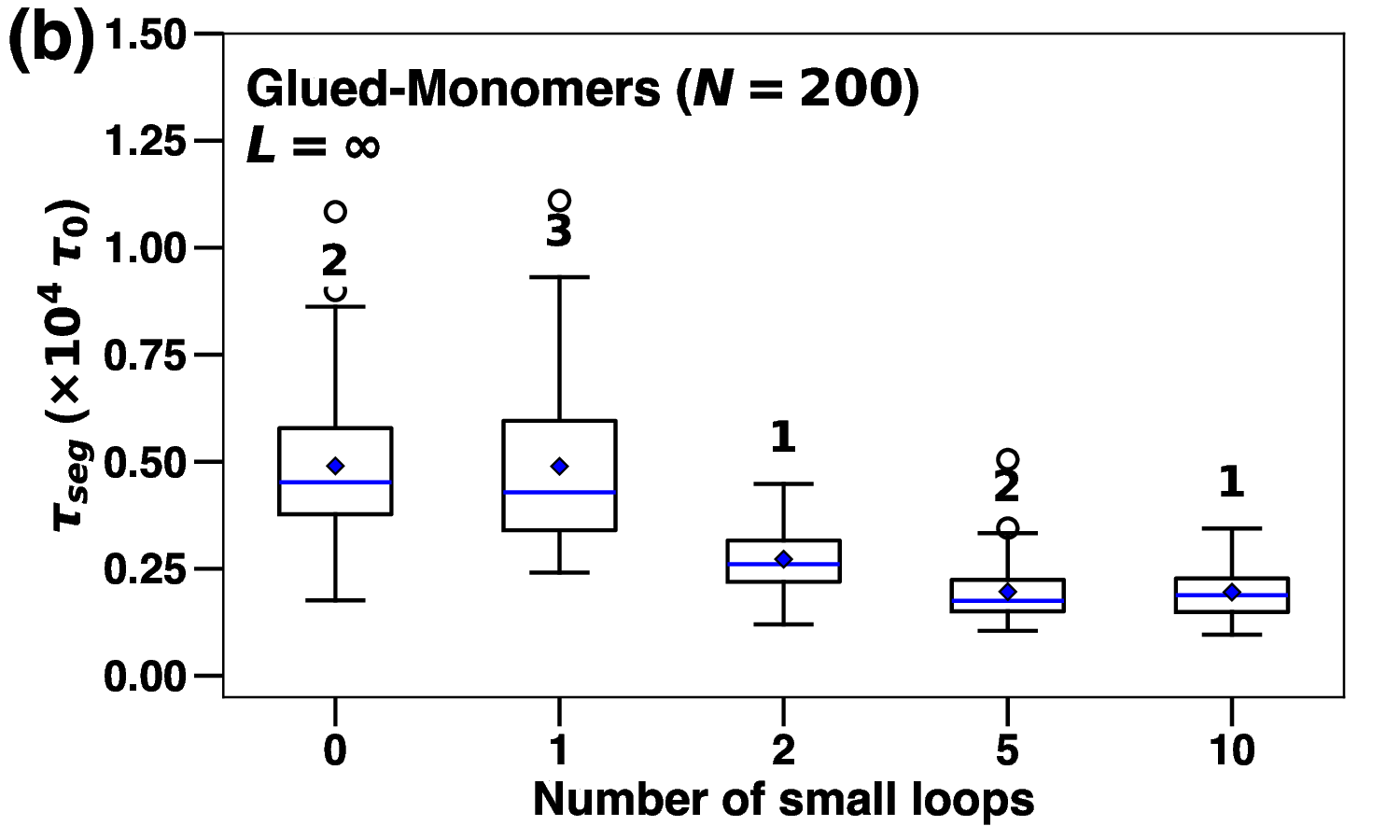}
        \includegraphics[width=0.8\linewidth]{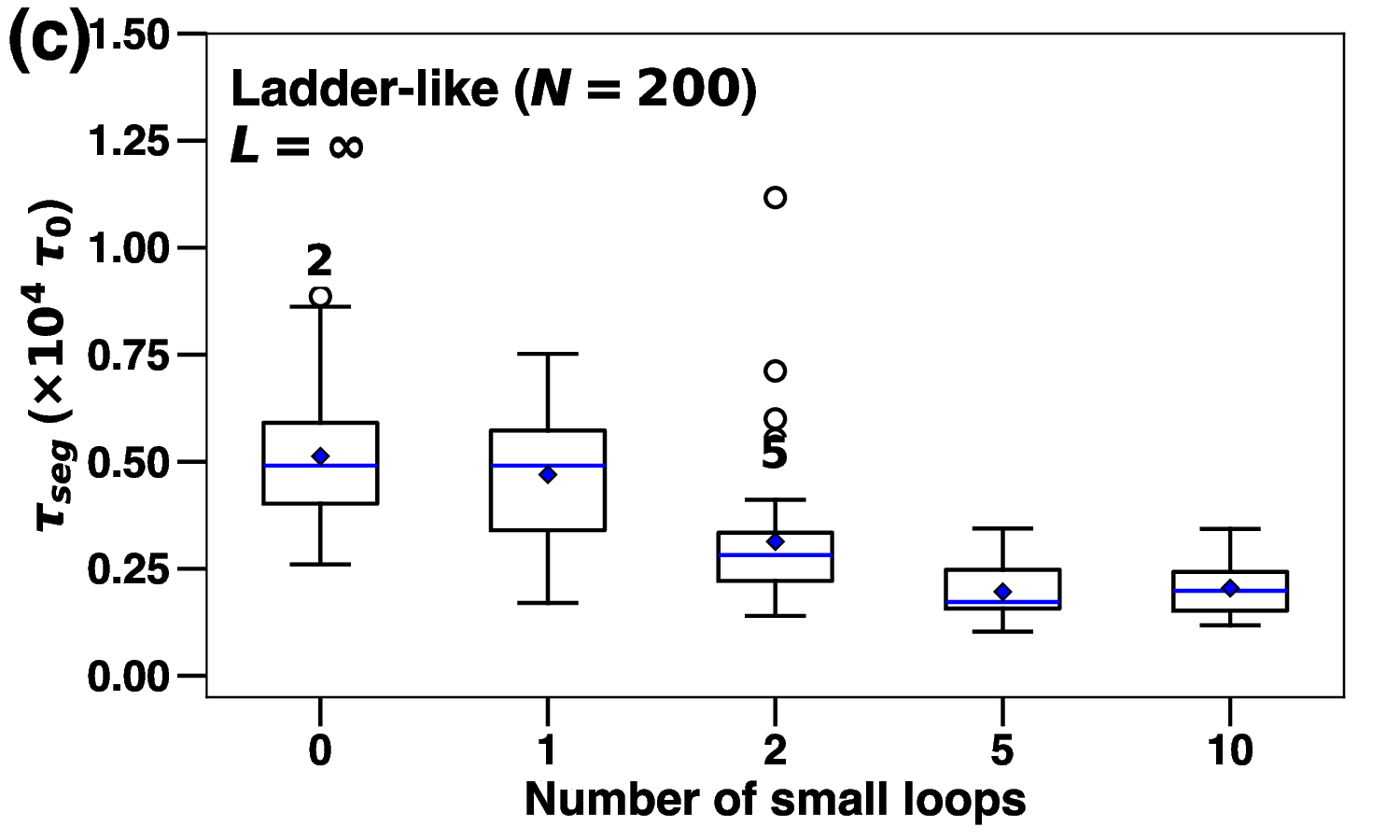}
        \includegraphics[width=0.8\linewidth]{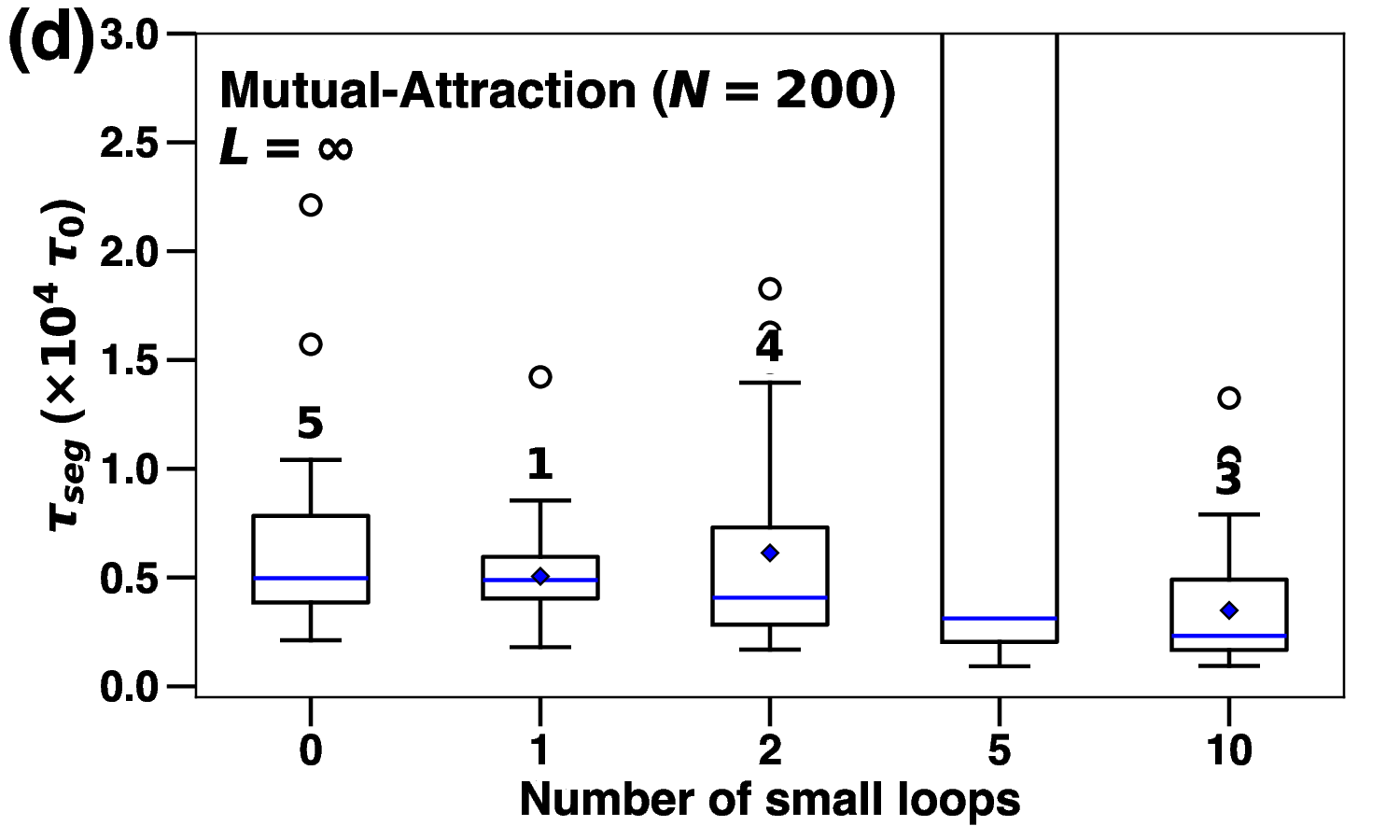}
    \caption{Box plot displaying several summary statistics (refer section III B) for the variation of $\tau_{seg}$ with the number of small loops in each polymer. Each polymer was composed of $N = 200$ monomers and the large loop contained $N/2 = 100$ monomers. The diameter of the cylinder, $D$, was chosen such that the degree of confinement, $\delta \equiv 2 R_g(\mathbb{T})/D \approx 1.34$ while the length of the cylinder, $L$, was chosen to ensure $L \gg D$. Results are shown for different initialization protocols: (a) \recenter{}, (b) \fixingBonding{}, (c) \replicationLike{}, and (d) \feneLJ{}. Each box plot summarizes the data obtained from $50$ independent simulation runs. For details of the summary statistics and their representation in the box plot, refer Figure~\ref{fig:phi-single-box-plots}.
    Probability distributions of the topologies were compared pairwise using a t-test and the resulting $p$-values are provided in SI-\ref*{SI-sec:si-hist-seg-times}.
    We have performed t-tests and provided $p$-values 
    on comparing a pair of distributions of segregation times at a time, to have a 
    quantitative measure of whether the distributions are significantly different, refer 
     SI-\ref*{SI-sec:si-hist-seg-times}.}
    \label{fig:inf-200-multi-box-plots}
\end{figure}




\begin{figure}[!htbp]
    \centering
    \includegraphics[width=0.8\linewidth]{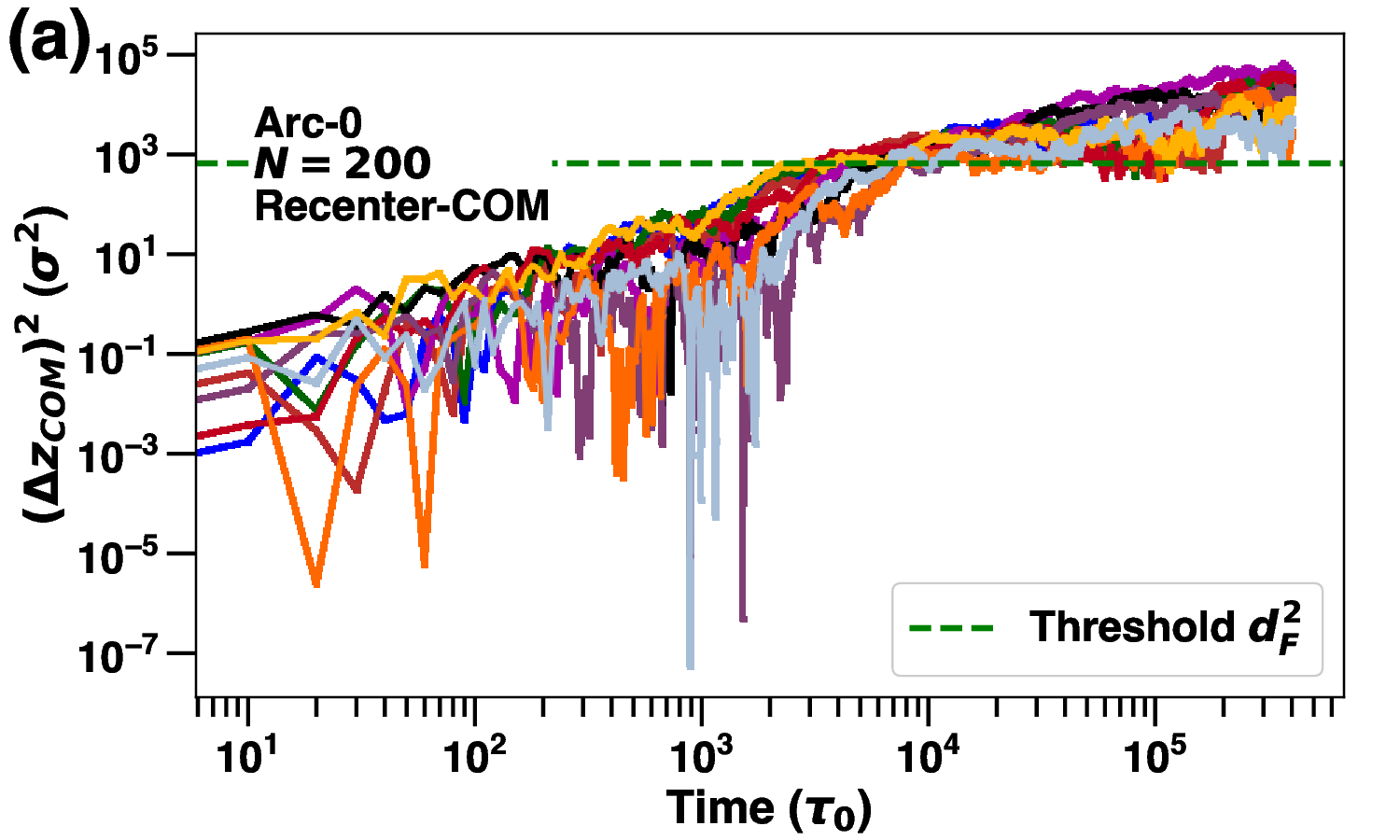}
    \includegraphics[width=0.8\linewidth]{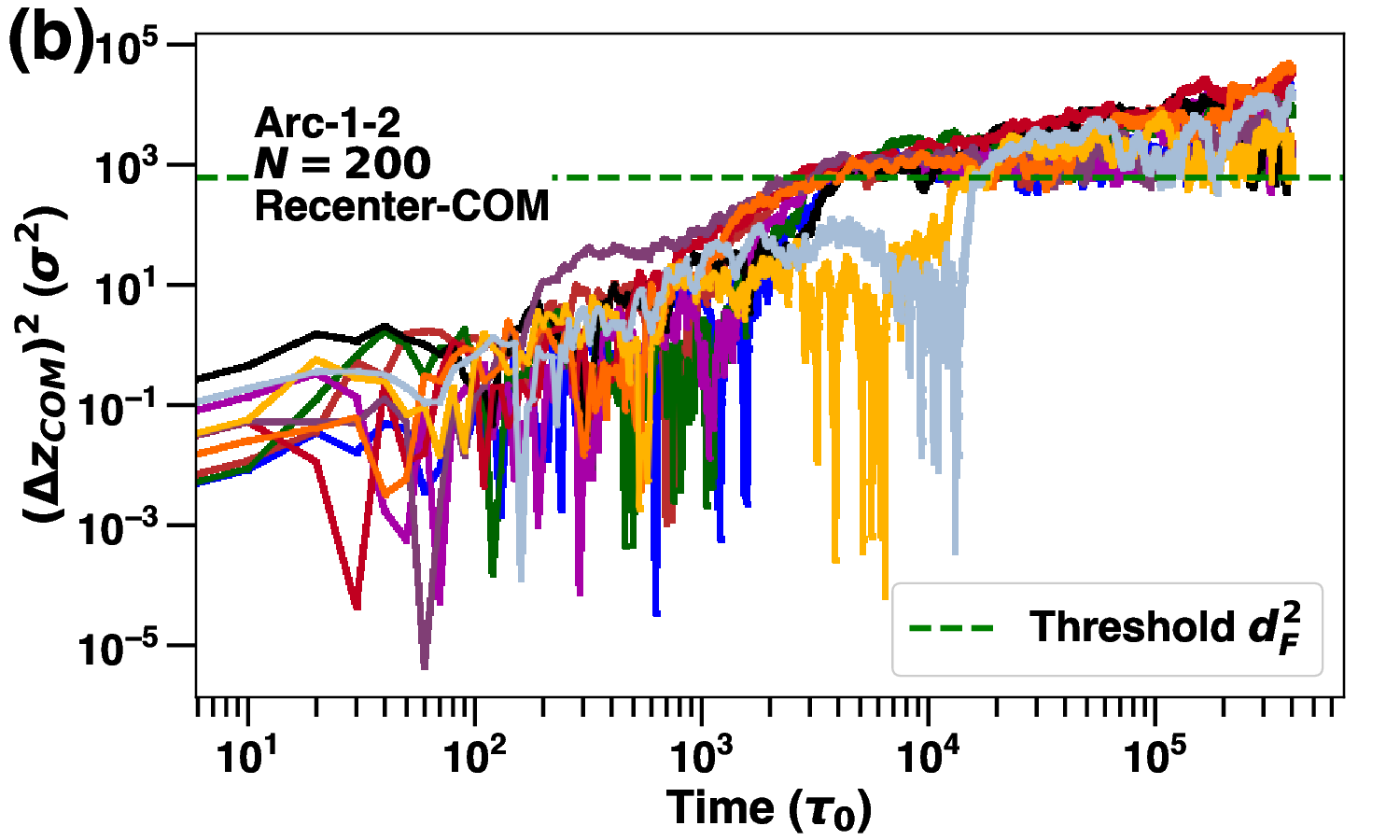}
    \includegraphics[width=0.8\linewidth]{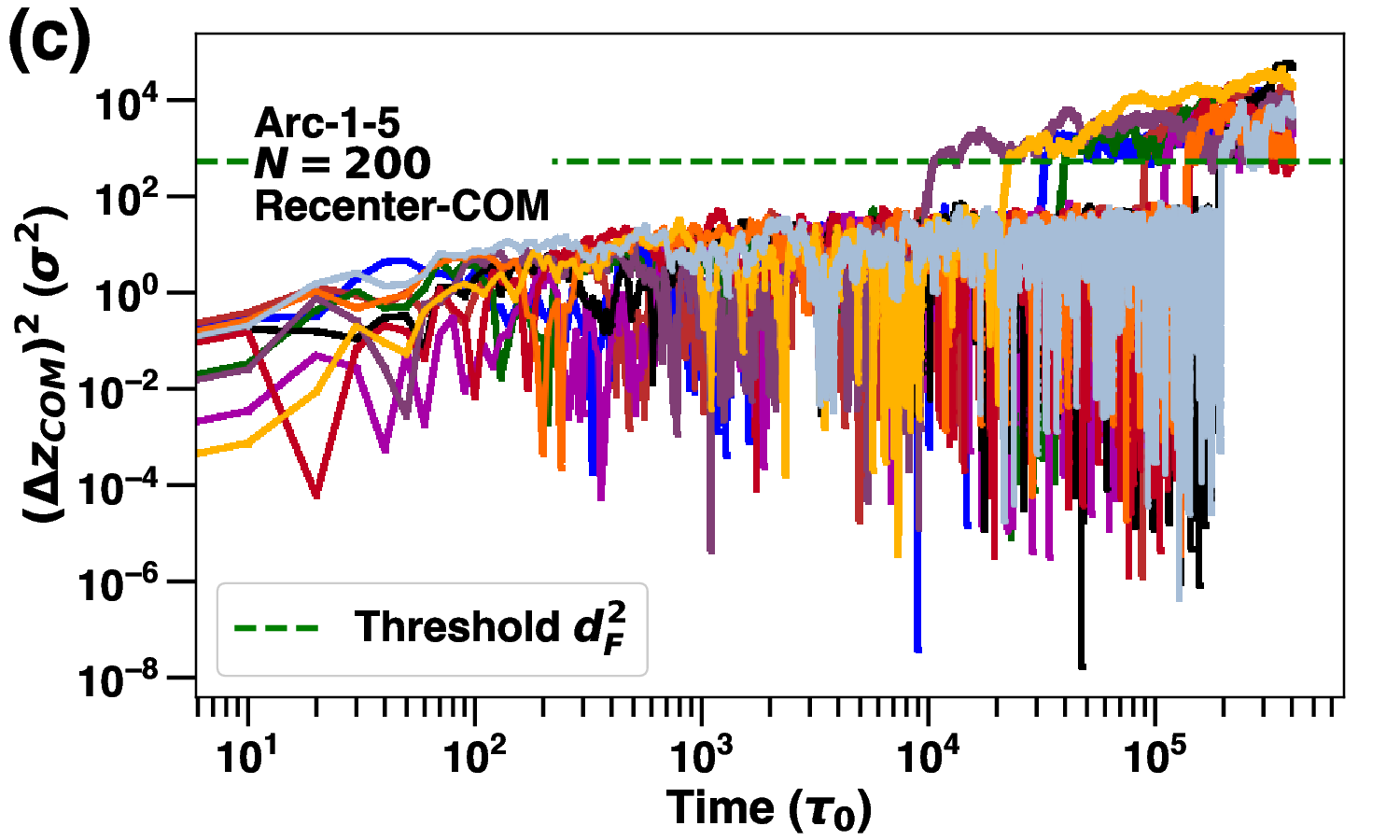}
    \includegraphics[width=0.8\linewidth]{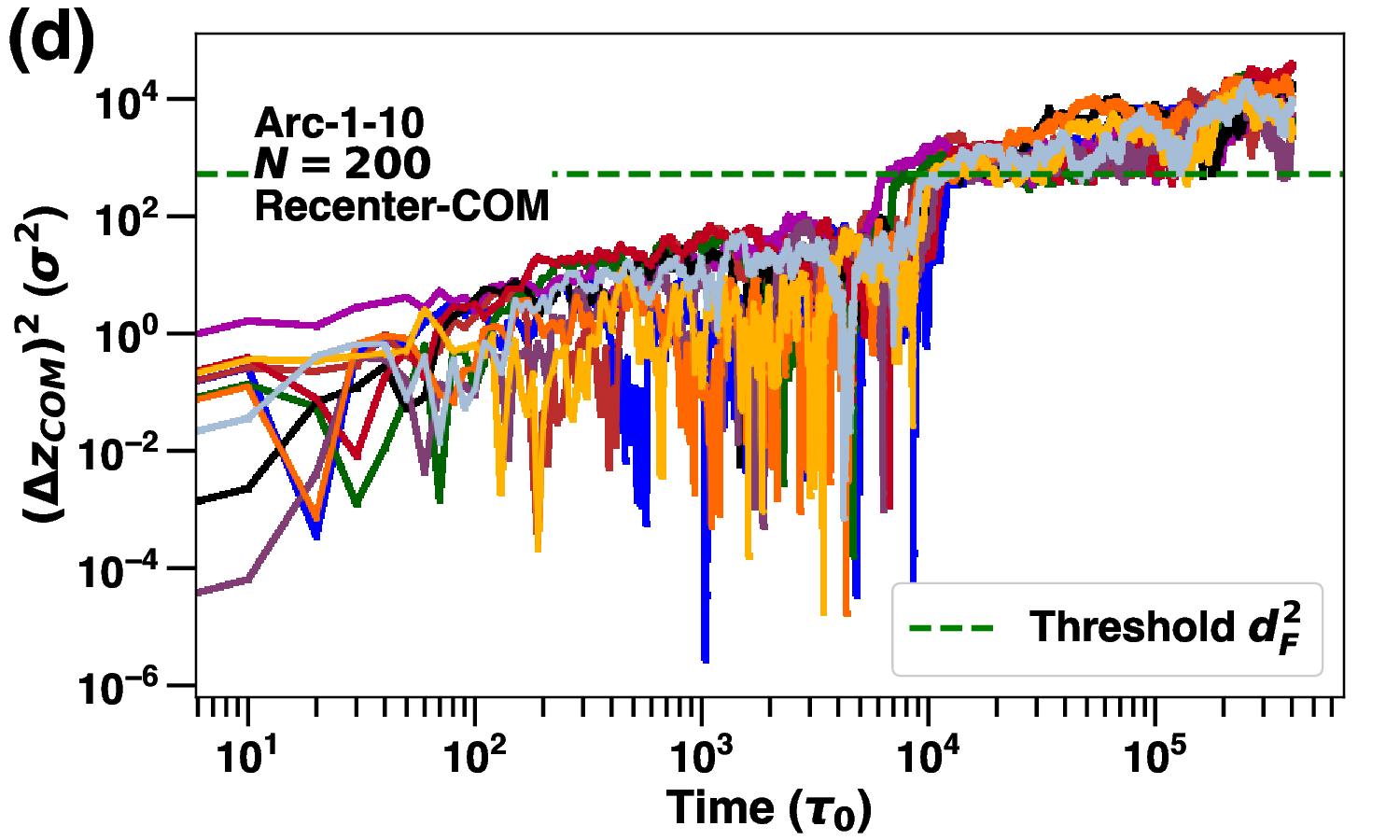}
    \caption{Time evolution of the squared separation along the axis of the cylinder between the centers of mass of two $N = 200$ polymers, $(\Delta z_{\mathrm{COM}})^2$, for $L/D \gg 1$. For reference, the threshold value for segregation, $d_F^2$, is indicated by a dashed line. The initialization was performed using the \recenter{} protocol. For all of the topologies, an initial period was observed where $(\Delta z_{\mathrm{COM}})^2 < \sigma^2$ indicating that the two polymers remained overlapped. Beyond this, $(\Delta z_{\mathrm{COM}})^2$ increased and the increase was more gradual at longer times. For Arc-0 and most of the Arc-1-2 trajectories, polymer segregation was nearly complete by $ \approx 3 \times 10^3\,\tau_0$. In contrast, Arc-1-5 and Arc-1-10 exhibited longer induction times, i.e., the polymers remained overlapped for a longer period of time before undergoing segregation relatively rapidly.}
    \label{fig:sq-com-runs-inf-rec}
\end{figure}

In the antiparallel configuration, the COMs of the cluster of small loops of P1 (or P2) and the large loop of P2 (or P1) were located close to each other. Upon removal of the constraints imposed during initialization, the cluster of small loops began moving towards the closest pole. During the motion of the small loops of say, P2, the large loop of P1 moved towards the center of the cylinder and the overlap with the small loops of P2 was reduced. In a similar fashion, the large loop of P2 reduced its overlap with the small loops of P1. Such a reduction of overlap required time and corresponded to the near plateau seen until time $\approx 10^3 \tau_0$. Thus, polymers initially in the antiparallel configuration spent a considerable amount of time in the overlapped state before significant COM motion occurred. Beyond time $\approx 10^3 \tau_0$, the crossing between the two large loops occurred after which the segregation proceeded to completion. This was associated with overcoming  another free energy barrier or a smaller free energy gradient at the intermediate stages of segregation, corresponding to the crossing of the two large loops. 

In summary, the results discussed above suggest that two key factors affected the variability between the different trajectories during the segregation process. The first was the topology of the polymers. For topologies containing a cluster of small loops, the two overlapped ToMo polymers experienced stronger entropic repulsion. Hence, both the initiation and the completion of segregation tended to occur at shorter times. The second factor was the mutual orientation of the polymer chains in the initial configuration. This depended on the particular initialization protocol used. As discussed above, polymers in a parallel configuration likely encountered a lower kinetic barrier when compared to those in antiparallel configurations. Consequently, polymers initialized in a parallel configuration typically began and completed segregation earlier. For initialization protocols that allowed both parallel and antiparallel configurations, such as \recenter{}, the existence of rather distinct segregation pathways for the two different initial configurations resulted in a broader distribution of $\tau_{seg}$.

\subsection{Effect of Topology on $\tau_{seg}$: $L/D \gg 1$} 
Motivated by the segregation of daughter chromosomes in bacterial cells, we had hitherto considered the segregation of polymers under cylindrical confinement at a fixed monomer volume fraction. In line with this, the length $L$ of the cylinder depended on $N$ but was independent of the particular polymer topology under investigation. 
However, modifying the topology, $\mathbb{T}$, of a polymer chain is likely to change its radius of gyration, $R_g(\mathbb{T})$. Thus, if $L/D$ is determined by the monomer volume fraction, different topologies may experience different degrees of confinement, $\delta \equiv 2R_g(\mathbb{T})/D$.
To elucidate the effect of cylindrical confinement on the kinetics of segregation, we also investigated ToMo polymers at fixed $\delta$. 
To this end, we fixed $\delta \approx 1.34$ and determined $D$ for each ToMo polymer, and used a value of $L$ such that $L \gg D$. The resulting values are tabulated in SI--\ref*{SI-sec:cyl-dim-and-rg}.
Note that $\delta \approx 1.34$ used above is lower than the value for $L/D =5$ at fixed monomer volume fraction discussed earlier. For example, for an unconfined Arc-0 polymer, $2 R_g \approx 16.74\,\sigma$. For $L/D =5$ at fixed volume fraction, we had found that $D \approx 6.24\,\sigma$ which implied that $\delta \approx 2.68$. 

The choice of a lower value of $\delta$ for $L \gg D$ was necessitated by the following. As discussed above, for $L/D = 5$ and $D=6.24\,\sigma$, segregation was observed for all of the ToMo polymers. However, for the same $D$ but $L/D \gg 1$, the antiparallel configurations of Arc-1-5 and Arc-1-10 did not exhibit segregation even at times significantly longer than that for $L/D =5$. A plausible reason for the difference could be traced to the polymer configurations that were observed. For $L \gg D$, at the end of the initialization at $t_0$, Arc-1-5 and Arc-1-10 existed in antiparallel configurations with the monomers in the large loops extending beyond and overhanging the cluster of small loops, refer Fig.~\ref*{SI-fig:inf-anti-reg-single-dist} in SI-\ref*{SI-sec:mix-seg-states-inf}. It is therefore plausible that the free energy barrier for the crossing of the large loop of P1 (P2) and the cluster of small loops of P2 (P1)  was larger for such configurations and effectively suppressed the segregation of the two polymers. Lowering $\delta$ to the value mentioned above enabled segregation to be observed.


Having fixed $\delta$, we investigated the segregation of two polymers in the same manner as before. First, we prepared overlapped initial configurations using one of the four initialization protocols. After verifying that the two polymers exhibited adequate overlap, we simulated segregation and calculated $\tau_{seg}$. The criterion for segregation we used was similar to that for $L/D = 5$ but with an effective cylinder length, $L_{eff}(\mathbb{T}, D) = 2 \times \langle R_{||}(\mathbb{T}, D) \rangle$, where $\langle R_{||}(\mathbb{T}, D) \rangle$ is the average longitudinal extent of one polymer of topology $\mathbb{T}$ confined in a cylinder of diameter $D$. Refer SI-\ref*{SI-sec:tech-initialization} for additional details about initialization and SI-\ref*{SI-sec:cyl-dim-and-rg} for the values of $\langle R_{||}(\mathbb{T}, D) \rangle$ for several $\mathbb{T}$ (and $D$).
Using the $L_{eff}$ defined above, the criterion of segregation discussed in section \ref{sec:seg-criterion} but with threshold values $d_F = 0.48 L_{eff}$ and $d_S = 0.43 L_{eff}$ was used to determine $\tau_{seg}$. It is germane to note that while $L_{eff}$ was used for the determination of $\tau_{seg}$, the actual length of the cylinder in which the simulation was performed was significantly longer.

The $\tau_{seg}$ for the different ToMo polymers prepared using the four initialization protocols are shown in Fig.~\ref{fig:inf-200-multi-box-plots}. For the \fixingBonding{} and the \replicationLike{} initialization protocols, the $\tau_{seg}$ were similar when the number of small loops was either $0$ or $1$, a clear decrease could be discerned for number of small loops $\geqslant 2$. However, for the \recenter{} and \feneLJ{} protocols, the Arc-1-5 topology showed an unusually large $\tau_{seg}$ when compared to the other topologies. The $\tau_{seg}$ of Arc-1-10 was either similar to or larger than the values for topologies with two or fewer loops. This trend was similar to the case of fixed monomer volume fraction with $500$ monomer polymers initialized with the \recenter{} protocol. 

To better understand these observations, we analyzed the individual trajectories for the various topologies initialized using the \recenter{} protocol. Time evolution of the squared separation between the centers of mass, $(\Delta z_{COM})^2$, is shown in Figure \ref{fig:sq-com-runs-inf-rec}. Similar data for $L/D = 5$ was provided in Fig.~\ref{fig:sq-com-runs-rec}. For all of the topologies, $(\Delta z_{COM})^2$ appeared to increase with time while being subject to significant fluctuations. For time $\leqslant 10^2 \tau_0$, $\Delta z_{COM}^2 \leqslant \sigma^2$, indicating that the polymers exhibited substantial overlap.
The increase in $(\Delta z_{COM})^2$ with time indicated the progress of segregation. Nearly all of Arc-0 and most of Arc-1-2 achieved segregation by time $\approx 3 \times 10^3 \tau_0$. The segregated polymers continued to diffuse away from each other, as indicated by the gradual increase of $(\Delta z_{COM})^2$. Trajectories of Arc-1-5 and Arc-1-10, and a few of Arc-1-2, exhibited a longer induction time before initiation of segregation. Arc-1-5 and Arc-1-10 exhibited rapid progression of segregation after the initial induction period.

\begin{figure}[!]
    \centering
    \includegraphics[width=0.8\linewidth]{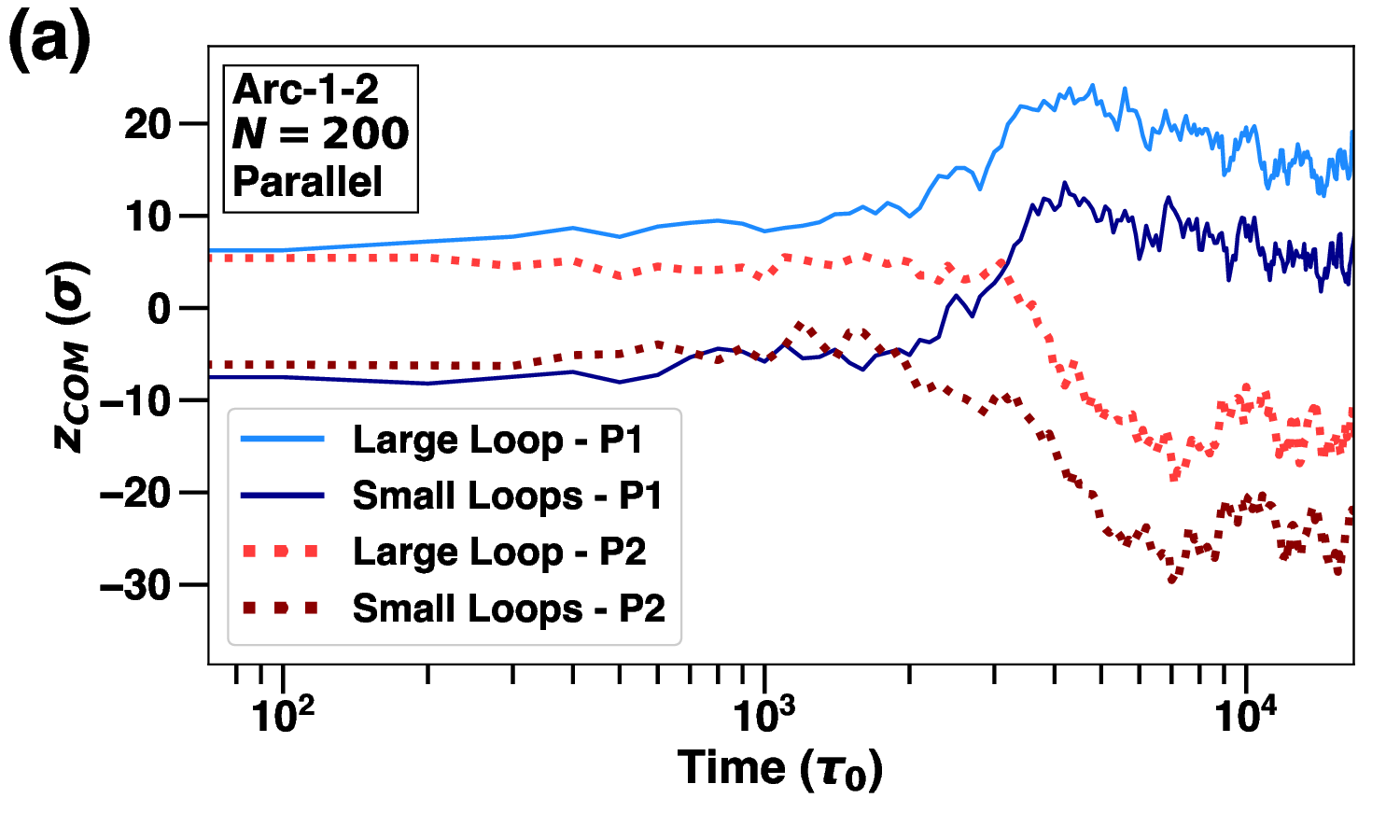}
    \includegraphics[width=0.8\linewidth]{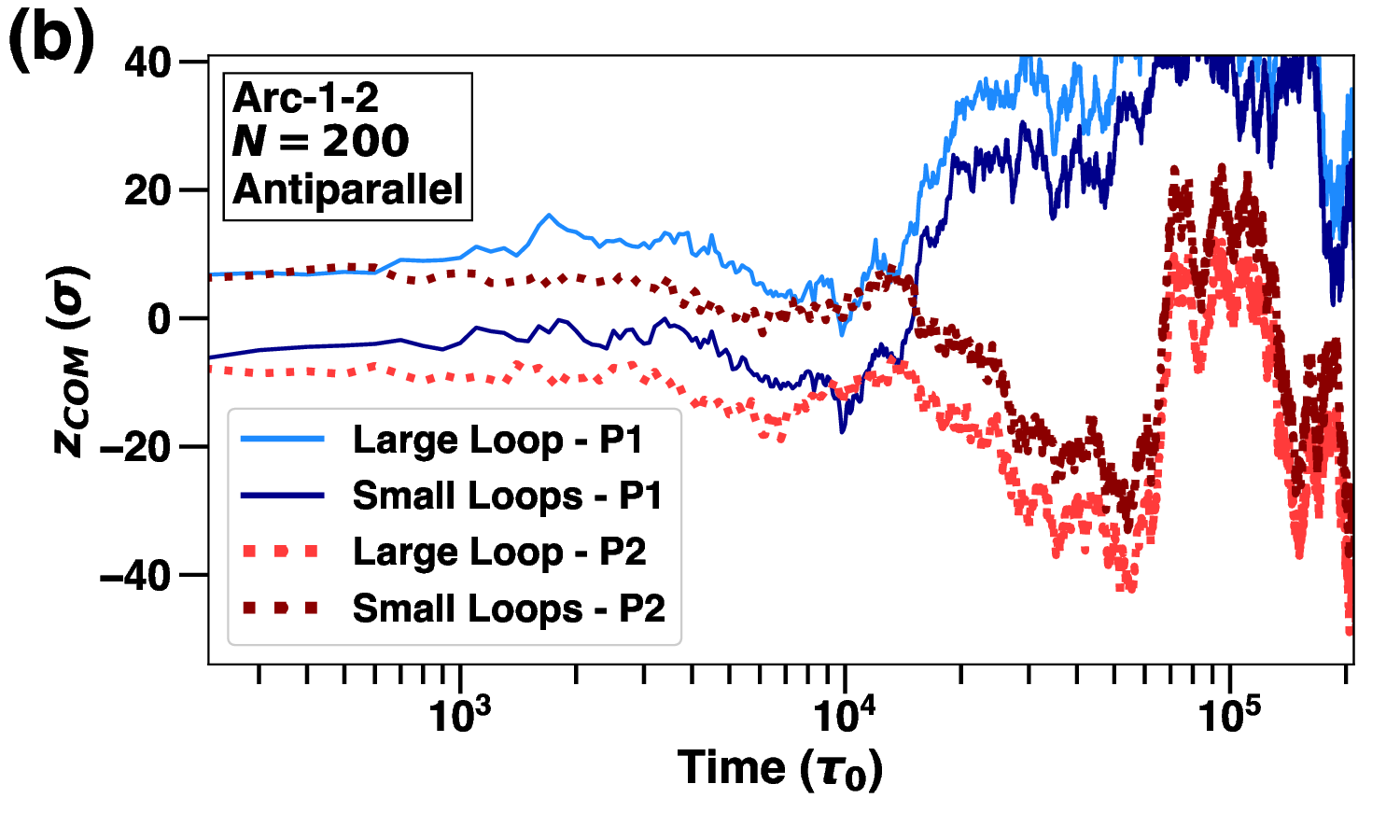}
    \includegraphics[width=0.8\linewidth]{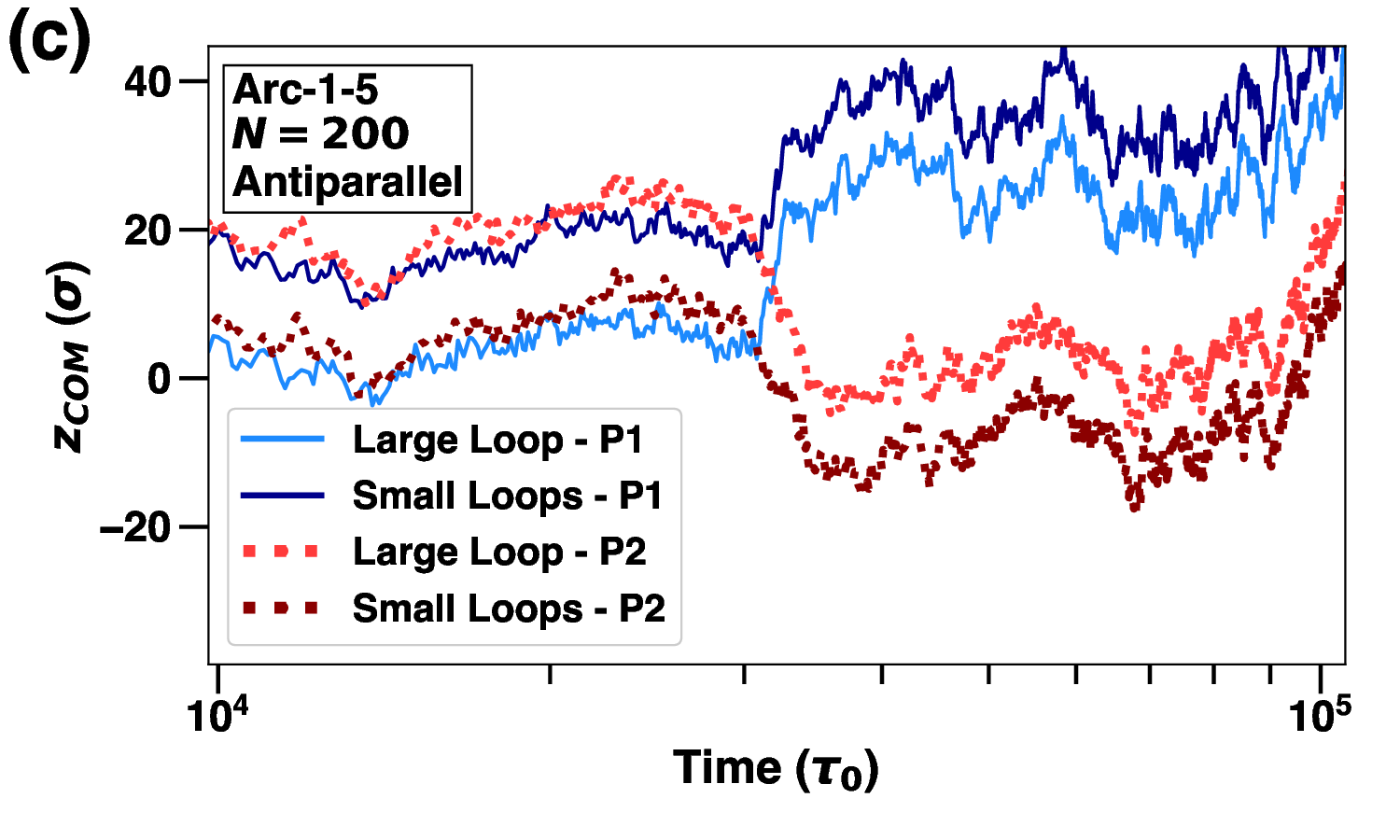}
    \includegraphics[width=0.8\linewidth]{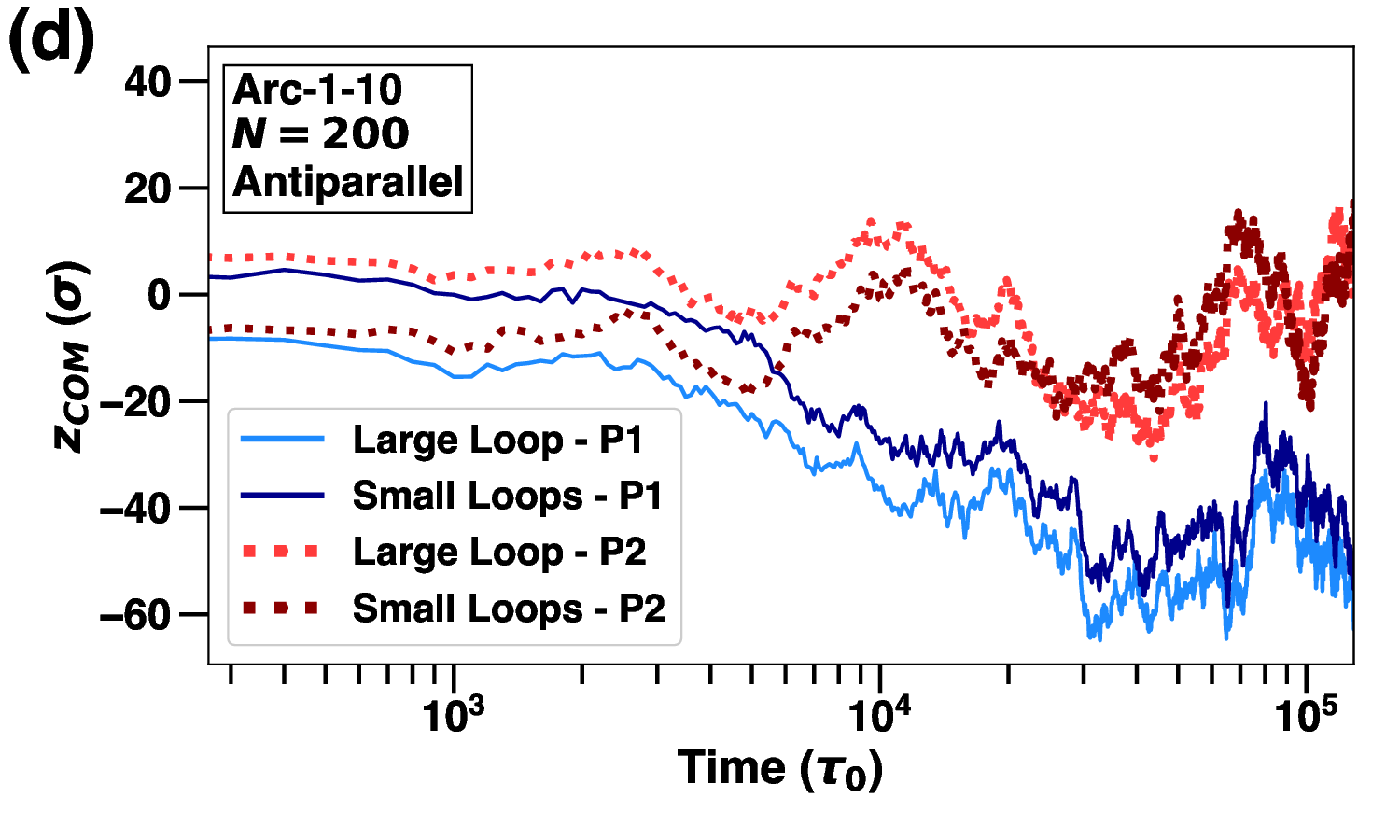}
    \caption{Time evolution of the $z$-coordinate of the center of mass of the Large loop and the cluster of Small Loops of each polymer during segregation for $L/D \gg 1$. The overlapped initial configurations were generated using the \recenter protocol. (a) Segregation of Arc-1-2 starting from a parallel initial configuration involved the crossing between the Large Loop of one polymer and the cluster of Small Loops of the other. (b) Segregation of Arc-1-2 starting from an antiparallel configuration occurred via the clusters of the Small Loops of each polymer crossing each other. This is in contrast to that for $L/D = 5$ discussed earlier, refer Fig.~\ref{fig:region-com-timeseries}. (c) Segregation of Arc-1-5 starting from an antiparallel configuration occurring via the crossing of the two Large Loops. This resulted in the induction period becoming particularly long. (d) Similarly to (b), segregation of Arc-1-10 starting from an antiparallel configuration occurred via the crossing of the clusters of Small Loops of each polymer.}
    \label{fig:inf-region-com-timeseries}
\end{figure}

For all of the topologies initialized using the \fixingBonding{} protocol, time evolution of $(\Delta z_{COM})^2$ is provided in SI-\ref*{SI-sec:mix-seg-states-inf}. From Fig.~\ref*{SI-fig:sq-com-runs-inf-fbsr} of SI, it can be seen that 
the increase in $(\Delta z_{COM})^2$ was closer to monotonic as the effect of fluctuations was weaker. This suggested that the delayed segregation for Arc-1-5 and Arc-1-10 initialized using the \recenter{} and the \feneLJ{} protocols was due to the mutual orientation of the polymers in the initial configurations at $t_0$ being antiparallel. We have explicitly confirmed that this was indeed the case.

Note that for $L/D = 5$, the larger $\tau_{seg}$ exhibited by Arc-1-5 and Arc-1-10 for two of the initialization protocols (refer Fig.~\ref{fig:orientation-grouped-sq-com}) can be attributed to the antiparallel configurations at $t_0$. Clearly, the observations discussed above for $L \gg D$ are consistent with that for $L/D =5$.
As discussed earlier, while $\delta \approx 2.68$ was used for $L/D =5$, lowering $\delta$ to $\approx 1.34$ for $L \gg D$ was necessary so that $\tau_{seg}$ could be determined. This was essentially due to the antiparallel configurations not exhibiting segregation within the time scale of the simulations unless $\delta$ was reduced to $1.34$.

The aforementioned discussion suggests that the kinetic barrier for segregation of the antiparallel configurations was larger for $L \gg D$ than for $L/D =5$. This is because in the antiparallel configurations, the two polymers were overlapped in such a way that the small loops of one polymer were usually nested within the large loop of the other polymer. This led to ``overhangs'' of the large loops on either side of the span of the monomers along the axis of the cylinder. For $L/D = 5$, spatial constraints imposed severe restrictions on the existence of such overhangs. However, for $L/D \gg 1$, the overhangs were free to extend further along the longitudinal axis of the cylinder. Hence, the fraction of monomers in the ``overhangs'' was larger. 
For the initial configurations of two $N = 200$ Arc-1-10 polymers, the axial variation of the monomer density of the large loops and the clusters of small loops is provided in SI-\ref*{SI-sec:mix-seg-states-inf}. This can be used to identify overhangs of the large loops.

It is germane to highlight another difference in the segregation process that depended on whether the initial configuration was parallel or antiparallel. The difference can be understood using the time dependence of $z_{COM}$ for the large loop and the cluster of small loops for each of the two polymers as shown in Fig.~\ref{fig:inf-region-com-timeseries}. In Fig.~\ref{fig:inf-region-com-timeseries} (a), corresponding to the initial configuration of two Arc-1-2 polymers being parallel, the segregation occurred by the displacement of polymer P1 such that its large loop was ahead of the small loops while polymer P2 moved in the opposite direction with the small loops ahead of the large loop. Unless a parallel to antiparallel flip occurred at the very early stages of the segregation, this is essentially the only way the segregation can occur when the starting configuration is parallel. 

In contrast, two distinct scenarios unfolded when the starting configuration was antiparallel. For Arc-1-2 (Fig.~\ref{fig:inf-region-com-timeseries} (b)) and Arc-1-10 (Fig.~\ref{fig:inf-region-com-timeseries} (d)), the polymers moved away from each other such that the large loop was ahead and the small loops trailed behind for majority of the antiparallel simulations. For Arc-1-5 (Fig.~\ref{fig:inf-region-com-timeseries} (c)), conversely, the small loops were ahead and the large loop trailed behind for all of the antiparallel simulations. As the crossings necessary for the segregation to proceed were rather different in each of the two cases, the kinetic barriers, and consequently, $\tau_{seg}$, could be rather different. This is discussed further below.

For $L/D \gg 1$, Arc-1-5 preferentially exhibited antiparallel configurations when initialized using the \recenter{} protocol. We believe that this was a contributing factor to the large $\tau_{seg}$ observed. Note that Arc-1-5 initialized using the \feneLJ{} protocol also exhibited similarly large $\tau_{seg}$. In this case however, we found that mutual interpenetration, i.e., the reciprocal threading of the two polymers similarly to that in fixed monomer density case, was a contributing factor. However, for both of the initialization protocols, $\tau_{seg}$ of Arc-1-10 was substantially smaller than that of Arc-1-5. The reason for this could be traced to the following. As the  small loops in Arc-1-10 were smaller than that of Arc-1-5, they were more compact and hard sphere-like while those in Arc-1-5 were bulkier. 
Therefore, segregation starting from an antiparallel configuration might involve overcoming larger kinetic barriers in the case of Arc-1-5 when compared to Arc-1-10 and thereby, resulted in the observed difference in $\tau_{seg}$. This however is speculation at this stage.

\section{Summary}

We have investigated the effect of introducing topological modifications to polymers under cylindrical confinement on $\tau_{seg}$. For $L/D =5$ and fixed $N$, $\tau_{seg}$ of \singleLooped{} polymers was not significantly affected when the size of the two loops was varied. This can be understood using a blob picture which suggests that the entropic penalty under cylindrical confinement is weakly dependent on the size of the loops. 
In contrast, increasing the number of small loops reduced $\tau_{seg}$ for \fixingBonding{} and \replicationLike{} initialization protocols. Note that the two initialization protocols yielded only parallel configurations at $t_0$. 
The difference in $\tau_{seg}$ can be understood in terms of the relevant free energy differences~\cite{Bhandarkar2026}.

Overlap of two polymers under cylindrical confinement results in a significant entropic penalty. Hence, to maximize the entropy of the system, two polymers spontaneously tend to segregate to different halves of the cylinder. In a ToMo polymer, increasing the number of small loops is expected to increase the entropic penalty in the overlapped state and result in a corresponding decrease in $\tau_{seg}$. However, $\tau_{seg}$ did not continue to decrease indefinitely upon increasing the number of small loops but appeared to reach saturation when the number of small loops increased beyond five. This may be an indication that the entropic penalty does not increase indefinitely upon increasing the number of small loops.

The situation for the other two initialization protocols, \recenter{} and \feneLJ{}, was markedly different as both parallel and antiparallel initial configurations were possible at $t_0$. However, the proportion of parallel and antiparallel initial configurations depended on the particular topology, refer~Tables in SI-\ref*{SI-sec:orientation}. As discussed above, antiparallel configurations resulted in significantly larger $\tau_{seg}$ when compared to parallel configurations. As discussed earlier, past work~\cite{Bhandarkar2026} suggests that this was due to the larger kinetic barriers in the antiparallel case. An additional complication was found in the case of Arc-1-2 initialized using the \feneLJ{} protocol. Here, the two polymers exhibited interpenetration with one polymer threading the region spanned by the other and resulted in larger $\tau_{seg}$. 

For $L \gg D$, $\delta \approx 1.34$ was used to ensure that segregation could be observed during the course of the simulation. Note that this $\delta$ is smaller than that for $L/D =5$, where $\delta \approx 2.68$ was used. Here, both \fixingBonding{} and \replicationLike{} protocols exhibited a decrease in $\tau_{seg}$ upon increasing the number of small loops to five beyond which $\tau_{seg}$ exhibited near saturation. For the \recenter{} protocol, in addition to the possibility of both parallel and antiparallel initial configurations, ``overhangs'' observed at both ends of the polymers also contributed to an increase in $\tau_{seg}$.

In bacterial chromosomes, small loops are extruded on the ori-proximal side by MukBEF proteins. Given that the ori-proximal region is replicated first, small loops exist primarily on one side of the bacterial chromosome. Motivated by this, we modified the topology of the polymer in specific ways and investigated the effect on $\tau_{seg}$. Our findings suggest a plausible route by which the cell can control the speed of segregation through the introduction of small loops at appropriate locations along the replicating chromosome. In the biological context however, segregation proceeds in parallel to replication. Therefore, two daughter DNAs being completely overlapped, the situation found at the end of the initialization protocols, is rather unlikely. Furthermore, the topological constraints due to entanglements can be eliminated through chain crossings (facilitated by topoisomerses) which can affect the speed of segregation. We have shown previously that such topological constraint release leads to faster segregation \cite{dna2}.




\section{Acknowledgements}
The authors acknowledge useful discussions with Arieh
Zaritsky and Conrad Woldringh, which contributed to the design of the project. AC, with DST-SERB (IN) identification No. SQUID-1973-AC-4067, acknowledges funding by DST-SERB (IN)  project CRG/2021/007824. AC is grateful to Biosantexc research and mobility grants which enabled him to visit ENS-Lyon and ENS-Paris Saclay for useful discussions. AC is also grateful for a Fulbright-Nehru grant from USIEF which enabled him to host and extensively discuss with Prof. Jie Xiao. AC also acknowledges several discussions in the meetings organized by ICTS, Bangalore, India, and the use of the PARAM-BRAHMA computing facilities. SKS acknowledges financial support from the Japan Society for the Promotion of Science through Grant-in-Aid for Scientific Research (C)(23K03586). SKS is also grateful to AC and the Department of Physics for facilitating a visit to IISER Pune where part of this manuscript was written.
\section{Data Availability}
The data presented in this work were generated and analyzed using LAMMPS and other homegrown codes that are freely available \cite{harsh_kinetics_data}.

\section{Author Declarations}
\subsection{Conflict of Interest}
The authors have no conflicts to disclose.

\subsection{Author Contributions}
SP, AC and SKS designed and supervised the research project. HD performed the simulations and analysis with input from the other authors. All of the authors contributed to the interpretation of the results. HD and AC wrote the initial draft which was extensively revised by AC and SKS.
\nocite{supplementary}

\bibliographystyle{apsrev4-1}
%

\end{document}